\documentclass[11pt, leqno]{article}
\input amssym.def 
\input amssym.tex
\include{showlabels}
\def\utw{\smash{\rlap{\lower5pt\hbox{$\sim$}}}}
\def\udtw{\smash{\rlap{\lower6pt\hbox{$\approx$}}}}

\def\bbbr{{\rm I\!R}} 
\def\bbbn{{\rm I\!N}} 

\def\bbbone{{\mathchoice {\rm 1\mskip-4mu l} {\rm 1\mskip-4mu l}
{\rm 1\mskip-4.5mu l} {\rm 1\mskip-5mu l}}}
\def\bbbc{{\mathchoice {\setbox0=\hbox{$\displaystyle\rm C$}\hbox{\hbox
to0pt{\kern0.4\wd0\vrule height0.9\ht0\hss}\box0}}
{\setbox0=\hbox{$\textstyle\rm C$}\hbox{\hbox
to0pt{\kern0.4\wd0\vrule height0.9\ht0\hss}\box0}}
{\setbox0=\hbox{$\scriptstyle\rm C$}\hbox{\hbox
to0pt{\kern0.4\wd0\vrule height0.9\ht0\hss}\box0}}
{\setbox0=\hbox{$\scriptscriptstyle\rm C$}\hbox{\hbox
to0pt{\kern0.4\wd0\vrule height0.9\ht0\hss}\box0}}}}
\def\bbbe{{\mathchoice {\setbox0=\hbox{\smalletextfont e}\hbox{\raise
0.1\ht0\hbox to0pt{\kern0.4\wd0\vrule width0.3pt
height0.7\ht0\hss}\box0}}
{\setbox0=\hbox{\smalletextfont e}\hbox{\raise
0.1\ht0\hbox to0pt{\kern0.4\wd0\vrule width0.3pt
height0.7\ht0\hss}\box0}}
{\setbox0=\hbox{\smallescriptfont e}\hbox{\raise
0.1\ht0\hbox to0pt{\kern0.5\wd0\vrule width0.2pt
height0.7\ht0\hss}\box0}}
{\setbox0=\hbox{\smallescriptscriptfont e}\hbox{\raise
0.1\ht0\hbox to0pt{\kern0.4\wd0\vrule width0.2pt
height0.7\ht0\hss}\box0}}}}
\def\bbbq{{\mathchoice {\setbox0=\hbox{$\displaystyle\rm Q$}\hbox{\raise
0.15\ht0\hbox to0pt{\kern0.4\wd0\vrule height0.8\ht0\hss}\box0}}
{\setbox0=\hbox{$\textstyle\rm Q$}\hbox{\raise
0.15\ht0\hbox to0pt{\kern0.4\wd0\vrule height0.8\ht0\hss}\box0}}
{\setbox0=\hbox{$\scriptstyle\rm Q$}\hbox{\raise
0.15\ht0\hbox to0pt{\kern0.4\wd0\vrule height0.7\ht0\hss}\box0}}
{\setbox0=\hbox{$\scriptscriptstyle\rm Q$}\hbox{\raise
0.15\ht0\hbox to0pt{\kern0.4\wd0\vrule height0.7\ht0\hss}\box0}}}}
\def\bbbt{{\mathchoice {\setbox0=\hbox{$\displaystyle\rm
T$}\hbox{\hbox to0pt{\kern0.3\wd0\vrule height0.9\ht0\hss}\box0}}
{\setbox0=\hbox{$\textstyle\rm T$}\hbox{\hbox
to0pt{\kern0.3\wd0\vrule height0.9\ht0\hss}\box0}}
{\setbox0=\hbox{$\scriptstyle\rm T$}\hbox{\hbox
to0pt{\kern0.3\wd0\vrule height0.9\ht0\hss}\box0}}
{\setbox0=\hbox{$\scriptscriptstyle\rm T$}\hbox{\hbox
to0pt{\kern0.3\wd0\vrule height0.9\ht0\hss}\box0}}}}
\def\bbbs{{\mathchoice
{\setbox0=\hbox{$\displaystyle     \rm S$}\hbox{\raise0.5\ht0\hbox
to0pt{\kern0.35\wd0\vrule height0.45\ht0\hss}\hbox
to0pt{\kern0.55\wd0\vrule height0.5\ht0\hss}\box0}}
{\setbox0=\hbox{$\textstyle        \rm S$}\hbox{\raise0.5\ht0\hbox
to0pt{\kern0.35\wd0\vrule height0.45\ht0\hss}\hbox
to0pt{\kern0.55\wd0\vrule height0.5\ht0\hss}\box0}}
{\setbox0=\hbox{$\scriptstyle      \rm S$}\hbox{\raise0.5\ht0\hbox
to0pt{\kern0.35\wd0\vrule height0.45\ht0\hss}\raise0.05\ht0\hbox
to0pt{\kern0.5\wd0\vrule height0.45\ht0\hss}\box0}}
{\setbox0=\hbox{$\scriptscriptstyle\rm S$}\hbox{\raise0.5\ht0\hbox
to0pt{\kern0.4\wd0\vrule height0.45\ht0\hss}\raise0.05\ht0\hbox
to0pt{\kern0.55\wd0\vrule height0.45\ht0\hss}\box0}}}}

\def\bbbz{{\mathchoice {\hbox{$\sf\textstyle Z\kern-0.4em Z$}}
{\hbox{$\sf\textstyle Z\kern-0.4em Z$}}
{\hbox{$\sf\scriptstyle Z\kern-0.3em Z$}}
{\hbox{$\sf\scriptscriptstyle Z\kern-0.2em Z$}}}}

\def\diameter{{\ifmmode\oslash\else$\oslash$\fi}}

\def\init{\setcounter{equation}{0}}

\newtheorem{theoreme}{Theorem }[section]
\newtheorem{proposition}[theoreme]{Proposition}
\newtheorem{lemma}[theoreme]{Lemma}
\newtheorem{definition}[theoreme]{Definition}

\def\rr{\bbbr}
\def\cc{\bbbc}
\def\nn{\bbbn}
\def\zz{\bbbz}
\def\one{\bbbone}

\def\q{\hbox{\frak q}}

\def\ie{{i.e.,} }

\def\e{{\rm e}}

\def\i{{\rm i}}
\def\d{{\rm d}}
\def\12{\frac{1}{2}}

\def\proof{{\bf  Proof. }}
\def\slim{\hbox{\rm s-}\lim}
\def\coinf{C_{0}^{\infty}}

\def\bull{$\sqcup \kern -0.645em \sqcap$}
\def\qed{\hbox{\bull}}

\def\EG{{^{\scriptscriptstyle E} \kern -.05cmG}}

\def\supp{{\rm supp\,}}
\def\cH{{\cal H}}
\def\fUbar{\overline{\frak U}}

\def\cU{{\cal U}}
\def\fU{{\frak U}}
\def\cB{{\cal B}}
\def\G{\Gamma}
\def\cF{{\cal F}}
\def\cR{{\frak F}}

\def\ch{{\frak h}}
\def\p{\partial}

\def\x{{\rm x}}

\def\lPi{\mathop{\prod}\limits}

\def\wlim{{\rm w}-\lim}

\def\dG{\d\Gamma}

\def\cW{{\cal W}}
\def\fW{{\frak W}}

\def\cq{{\frak q}}

\def\cD{{\cal D}}
\def\cN{{\cal N}}
\def\cV{{\cal V}}
\def\cM{{\cal M}}

\def\e{{\rm e}}

\def\pfi2{P(\varphi)_{2}}

\newcommand{\beq}{\begin{equation}}
\newcommand{\eeq}{\end{equation}}
\newcommand{\bet}{\begin{theoreme}}
\newcommand{\eet}{\end{theoreme}}
\newcommand{\bel}{\begin{lemma}}
\newcommand{\eel}{\end{lemma}}
\newcommand{\bep}{\begin{proposition}}
\newcommand{\eep}{\end{proposition}}

\newcommand{\bear}[1]{\begin{array}{#1}}
\newcommand{\ear}{\end{array}}

\begin{document}
\def\j{{\rm j}}
\def\bc{{\rm c}}
\def\bp{{\rm p}}
\def\bt{{\rm t}}
\def\q{{\rm q}}
\def\chbar{\overline{\ch}}
\def\cO{{\cal O}}
\def\cA{{\cal A}}
\def\cF{{\cal F}}
\def\stp{ stochastic process}
\def\stps{stochastic processes}
\def\cUbar{\overline{\cal U}}
\def\stpos{stochastically positive}
\def\kms{ KMS system}
\def\Xbar{\overline{X}}
\def\cBbar{{\overline \cB}}
\def\cFbar{{\overline \cF}}
\def\cRbar{{\overline \cR}}

%
%
%
\def\hf{{(\, . \, ,\, . \,)}}
\def\eg{{e.g.\ }}
\def\tq{{\sl q}}
\def\maath{\mathsurround=0pt }
\def\eqalign#1{\null\, \vcenter{\openup1\jot \maath
\ialign{\strut\hfil$\displaystyle{##}$&$\displaystyle{{}##}$\hfil
\crcr#1\crcr}}\,}


\title{Thermal Quantum Fields with Spatially Cut-off  
Interactions in 1+1 Space-time Dimensions\protect\footnotetext{AMS 1991 {\it{Subject 
Classification}}. 81T08, 82B21, 82 B31, 46L55} \protect}          

\author{Christian G\'{e}rard\footnote{ christian.gerard@math.u-psud.fr, Universit\'e Paris Sud XI, F-91405 Orsay, France} \ and Christian D.\ J\"akel\footnote{
christian.jaekel@math.polytechnique.fr, \'Ecole Polytechnique, F-91128 Palaiseau, France}}        
\date{June 2003}
\maketitle
\abstract{We construct  interacting quantum fields in
1+1 space-time dimensions, representing char\--ged or neutral scalar bosons at positive temperature and 
zero chemical potential. 
Our work is based on prior work by Klein and Landau and
H\o egh-Krohn. Generalized path space methods are used to add a spatially cut-off  interaction to 
the free system, which is described in the Araki-Woods representation. It is shown that the 
interacting KMS state is normal w.r.t.\ the Araki-Woods
representation. The observable algebra and the modular conjugation of the interacting system
are
shown to be identical to the ones of the free system
and the interacting Liouvillean is described in terms of
the free Liouvillean and the interaction.}

\tableofcontents

\section{Introduction}
\init\label{introd}
\noindent
Thermal quantum field theory is
supposed to unify both quantum statistical mechanics and elementary
particle physics. The formulation of the general framework should be wide enough to allow a QED
description of ordinary matter.  It should also provide  the necessary tools for  the~QCD 
description of
several experiments currently envisaged with the new Large Hadron Collider (LHC) at CERN. 
While the general theory of thermal quantum fields has made substantial progress in recent years, 
the actual construction of interacting models, which fit into the axiomatic setting, has not yet started (with the 
exception of the very early contributions by H\o egh-Krohn \cite{H-K} and Fr\"ohlich \cite{Fr2}).

Let us briefly recall the formal description of charged scalar fields
in physics. Examples of  scalar 
particle-antiparticle pairs are the mesons $\pi^+$, $\pi^-$, $K^+$,
$K^-$, or $K^0$,  $\overline{K^0}$.  
(In the last case the  `charge' is strangeness). One starts with the classical Lagrangian density 
\[
 {\cal L} = (\p_\nu \varphi) (\p^\nu \varphi^*) - m^2 \varphi \varphi^* - {\lambda \over 4} (\varphi \varphi^*)^2 . 
\]
Here $\varphi (t,x)$ is a complex scalar field over space--time. 
The Lagrangian density ${\cal L}(t,x)$ is invariant under  the global
gauge transformations $\varphi \mapsto {\rm e}^{\i \alpha} \varphi $, $\alpha \in \rr$.
By Noether's theorem  this invariance leads to a conserved current
\[
j_\nu = i ( \varphi^* \partial_\nu \varphi - \varphi \partial_\nu \varphi^*),
\qquad \nu = 0, \dots, 3,
\]
and to a conserved charge
\[ q = \int  {\rm d}^{3} x \, j_0 (t, x). 
\]
The next step, according to the physics literature, is  to setup real
or imaginary time perturbation theory.

The state of art of perturbative thermal field theory  is covered in three
recent books by Kapusta [K], Le Bellac [L-B] and Umezawa [U]. The authors concentrate on theoretical
efforts to understand various hot quantum systems (e.g.,
ultra-relativistic heavy-ion collisions or the phase transitions in the
very early universe) and various physical implications (e.g., spontaneous
symmetry breaking and restoration, deconfinement phase transition).

Constructive thermal field theory allows one  to
circumvent (at least in lower space-time dimensions) the
severe problems (see, e.g., Steinmann \cite{St}) of thermal perturbation
theory, which can otherwise only be removed partially by applying
certain ``resummation schemes". 

A class of models representing scalar neutral bosons with polynomial interactions
in 1+1 space--time dimensions was constructed by H\o egh-Krohn \cite{H-K} 
more than twenty years ago. As he could show, thermal equilibrium states for these models
exist  at all positive temperatures. For neutral particles, the particle density 
(and the energy density) adjust themselves  to the given temperature; 
contrary to the non-relativistic case, a chemical potential adjusting
the particle density can not be introduced, since the mass is no longer a conserved quantity. 
Shortly afterwards,
several related results on the construction and properties of self-interacting 
thermal fields in 1+1 space--time dimensions were
announced by Fr\"ohlich \cite{Fr2}.

Our goal in this and a subsequent paper \cite{GeJ} was twofold: first we wanted to fully understand the 
neutral scalar thermal field with polynomial interaction as
constructed by H\o egh-Krohn \cite{H-K}, with the aim to study thermal scattering
theory, using the framework introduced by Bros and Buchholz in
\cite{BB1}, \cite{BB2}. Secondly we wanted to generalize this
construction to charged fields. This  would allow us to study the 
system at different temperatures and chemical potentials, i.e.,
different charge densities.
A possibility to change the charge density would put this model
closer to non-relativistic models, where the mass is a conserved quantity, giving 
rise to the existence of a chemical potential.

The construction of the full interacting thermal quantum field without cutoffs in \cite{GeJ}
includes several of the original ideas of H\o egh-Krohn \cite{H-K}, but instead
of starting from  the  interacting system in a box
we start from the Araki-Woods representation for the free system in infinite volume.
Using a general method developed by Klein and Landau \cite{KL1}
to treat spatially cutoff perturbations of the free system in infinite volume, we can
eliminate some cumbersome limiting procedures due to the introduction of boxes, when we remove the spatial cutoff.

The present paper is devoted to the construction of neutral and charged
thermal fields with {\em spatially cutoff} interactions in 1+1 space--time
dimensions, using the method of Klein and Landau \cite{KL1}.
Although the excellent paper \cite{KL1} is rather self contained, it did not include the discussion of 
examples. Twenty years ago it might have been evident for the experts
in the field how to apply their method to thermal quantum fields, but we
find it worthwhile to present this application in some detail.

A difference between this paper and \cite{KL1} is the use of
generalized path spaces as in \cite{K}, instead of stochastic processes. This 
compact formulation is convenient for our applications. In addition
we prove several new results concerning the interacting KMS systems
obtained by perturbations of path spaces.

\vskip 1cm

\goodbreak
\subsection{Content of this paper}

Our paper can be divided into several parts. The first part, presented
in Section \ref{sec-1}, discusses the description of neutral  and
charged scalar fields in terms of operator algebras. Its application
to Klein-Gordon fields is discussed in Section \ref{sec2}. As usual the
starting point is a real symplectic space $(X, \sigma)$, which allows
the construction of the Weyl algebra $\fW(X, \sigma)$. The next step
is to introduce on $(X, \sigma)$ a  K\"{a}hler structure, \ie a
compatible Hermitian structure. For charged scalar fields, the
symplectic space $(X, \sigma)$ possesses  also  a canonical `charge' complex
structure $\j$ and a `charge' sesquilinear form $\cq$, such that $\sigma={\rm
Im}\cq$. The maps $X\ni x\mapsto \e^{ \, \j \alpha}x$ for $\alpha\in \rr$
generate the {\em gauge transformations}. Given a regular CCR
representation,  complex quantum fields are defined.

This leads to the notion of a
{\em charged K\"{a}hler structure}, corresponding to the introduction
of another complex structure $\i$ and of  the {\em charge operator} $q$,
relating the two complex structures. Finally the notion of {\em charge
conjugation} is discussed in this abstract framework.  

For Klein-Gordon fields, a  conjugation inducing charge-time reflections is  used to distinguish 
an appropriate abelian sub-algebra 
of the Weyl algebra to which the interaction terms considered later on
will be  affiliated.

Section \ref{stochas} recalls the characterization of a thermal equilibrium state by the KMS property. 
The GNS representation associated to a KMS
state has a number of interesting properties which are briefly recalled.
For instance, the GNS vector is cyclic and separating for the field algebra $\cF$ 
(in our case  the weak closure 
of the Weyl algebra in the GNS representation), and therefore
one can always go over to the weak closure of the relevant operator algebras, and we will do so in the sequel.
Since a KMS state is invariant under time translations, 
a Liouvillean implementing the time evolution is always
available. As has been shown by Araki, the KMS condition allows us to
introduce Euclidean Green's functions.
The notion of {\em stochastically positive KMS systems} due to Klein
and Landau is presented. This notion rests on the introduction of a
distinguished abelian subalgebra $\cU$ of the field algebra
$\cF$. In physics, this algebra is the algebra generated by
the  time-zero
fields.
It is also shown that 
stochastically positive KMS systems are invariant under  a {\em time
reversal} transformation.

In Section \ref{quasi} we recall the notion of a quasi-free KMS system associated to a positive 
selfadjoint operator acting on the one-particle space.
The GNS representation for a quasi-free KMS system has been 
analyzed by Araki and Woods. We briefly recall this framework
and its connection to the Fock representation in a modern notation.
It is shown that the field algebra~$\cF$ is generated by the time-translates
of the abelian algebra $\cU$. 
The observable algebra, consisting of elements of the field algebra which are
invariant under gauge transformations, is introduced.
In Subsection \ref{stocpos} it is shown that the KMS system for the (quasi-)free charged thermal field is indeed 
stochastically positive, if the chemical potential vanishes. However, if the chemical potential is non-zero, then
the charge distinguishes a time direction, and consequently, the system is no longer invariant under time reversal.
Thus it fails to be stochastically positive too, as we show in
Subsection \ref{poskg}. 

Following Klein and Landau, a cyclicity property of the Araki-Woods representation,
which will imply the so-called {\em Markov property} for the free system later on, is shown.
 The Markov property has the consequence that the physical Hilbert
space can naturally be considered as an $L^{2}$-space.

Section \ref{path} recalls the notion of a {\em generalized path
space},
both for the 0-temperature case and the case of positive
temperature. We follow here \cite{K}, \cite{KL1}. Although the
$0$-temperature case is not needed in this paper, it will be useful
later on in
\cite{GeJ}.
A generalized path space consists of a probability space~$(Q, \Sigma, \mu)$, a 
distinguished $\sigma$-algebra $\Sigma_0$, a one-parameter group $t \mapsto U(t)$ and a reflection $R$.
We recall the definition of  {\em OS-positivity} and  the {\em Markov property} 
for both cases.

Section \ref{secst1} is devoted to a discussion of the Osterwalder-Schrader 
{\em reconstruction theorem} in the framework of generalized 
path spaces. This reconstruction theorem associates to a
$\beta$-periodic,  OS-positive
path space a stochastically positive $\beta$-KMS system.

In Section \ref{secst3} we 
recall from \cite{KL1} how to deal with of perturbations, which are given in terms of
{\em Feynman-Kac-Nelson kernels}. The main examples of FKN kernels are
those obtained from a selfadjoint operator $V$ on the physical Hilbert
space $\cH$, where $V$ is  affiliated to $\cU$.

We show that for a class
of perturbations $V$ considered in \cite{KL1}, the perturbed Hilbert space
can be canonically identified with the free Hilbert space in such a
way that the interacting algebras $\cR$, $\cU$ and the modular
conjugation $J$ coincide with the free ones. Moreover, we prove  that the
perturbed Liouvillean $L_{V}$ is equal to
$\overline{\overline{L+V}-JVJ}$, if $L$  is the free Liouvillean. Here
$\overline{H}$ denotes the closure of a linear  operator $H$. 

Finally we show that the  Markov property of a generalized path space
is not destroyed  by the perturbations associated to FKN kernels.

In Section \ref{sec2} we apply the framework of Sections \ref{sec-1}
and \ref{quasi} to charged and neutral Klein-Gordon fields at positive
temperature. The case of the neutral Klein-Gordon field is well known
and reviewed only for completeness. We give more details on the
charged Klein-Gordon field which provides an example of a charge symmetric
K\"{a}hler structure. 
We also compare our setup with the one used in physics textbooks. 
Using the results of Section~\ref{quasi}, we present the
quasi-free KMS system describing a free charged or neutral
Klein-Gordon field at positive temperature. Note that the conjugation
used in the definition of the abelian algebra~$\cU$ corresponds to time
reversal in the neutral case and to the composition of
time-reversal and charge conjugation in the charged case.
We show that the 
KMS system for the charged Klein-Gordon field is not stochastically positive, 
if the chemical potential is unequal to zero. 
The physical reason is that the dynamics of charged particles is only invariant under the combination of 
time reversal and charge conjugation. A non-zero chemical potential 
introduces a disymmetry between particles of positive and negative charge and 
hence breaks time reversal invariance, which itself is a property shared by all stochastically positive KMS systems.

In Section \ref{kgint} we consider Klein-Gordon fields at positive
temperature with spatially cutoff interactions in $1+1$ space-time
dimensions. 
In the neutral  case we will treat the $P(\phi)_{2}$ and the
$\e^{\alpha\phi}\!\:_{2}$ interactions (the later being also
known as the {\em H\o egh-Krohn model}\/). In the charged case we treat the
(gauge invariant) 
$P(\overline{\varphi}\varphi)_{2}$ interaction.

The UV divergences of the interactions are eliminated by Wick
ordering, which is discussed in some details in Subsections \ref{int}
and \ref{wickwick}. As it turns out, the leading order 
in the UV  divergences is independent of the temperature. Thus it is a matter of convenience
whether one uses  thermal Wick ordering or Wick ordering w.r.t.\ the vacuum state.

The $L^{p}$-properties of the interactions needed to apply the
abstract results of Section \ref{secst3} are shown in
Subsections \ref{int.sub1}, \ref{int.sub11} and \ref{chargedp}.

Finally, the main results of this paper, namely the construction and
description of a KMS system representing a Klein-Gordon field at positive
temperature with spatially cutoff interactions, is given in Subsection
\ref{mainres}.

In a forthcoming paper we will consider the translation invariant
$P(\phi)_{2}$ model at positive temperature. 
Following again ideas of H\o egh-Krohn \cite{H-K}, Nelson symmetry will be used
to establish the existence of the model in the thermodynamic limit.

\medskip
\noindent
{\bf Acknowledgments}. The second named author was supported under the FP5 TMR program 
of the European 
Union by the Marie
Curie fellowship HPMF-CT-2000-00881. Both authors benefited from the IHP network HPRN-CT-2002-00277
of the European Union.

\vfill

\section{Real and complex quantum fields}
\init
\label{sec-1}
In this section we present real and complex quantum fields in an abstract
framework. Usually in the physics literature complex quantum fields
are described in the case of Klein-Gordon fields. Although the results
of this section are probably known, we have not found them in the
literature. 
\subsection{Notation}
\label{notat}
Let $X$ be a real vector space. If $X$ is equipped with a
complex structure $\i$, then we will denote by $(X, \i)$ the complex vector
space $X$. If $(X, \i)$ is equipped with a hermitian form $\hf$, then we will denote by $(X, \i , \hf)$ the Hermitian
space $X$. If it is clear from the context which complex or Hermitian
structure is used, $(X, \i)$ or $(X, \i , \hf)$ will simply  be
denoted by $X$. As a rule the complex structure of a Hermitian
space $X$ will be denoted by the letter~$\i$. Sometimes another
`charge' complex structure appears; it will be denoted by the
letter~$\j$.


\subsection{Real fields}
\init
\label{sec0}
We start by recalling the formalism of real quantum fields.

\vskip .3cm
\noindent
{\bf CCR Algebra}

\vskip .2cm
\noindent
Let $(X, \sigma)$ be a real symplectic space. Let $\fW(X, \sigma)$  be the (uniquely determined) $C^*$-algebra generated
by nonzero elements $W(x)$, $x \in X$, satisfying 
\[
\begin{array}{l}W (x_1)W (x_2)=\e^{-\i\sigma(x_1,x_2)/2}
W (x_1+x_2),\\[3mm]
W ^*(x)=W (-x),\ \ \ \ \ W (0)=\one . \\[3mm]
\end{array}
\]
$\fW(X, \sigma)$ is called the {\em Weyl algebra} associated to $(X, \sigma)$.
\vskip .3cm
\noindent
{\bf Regular representations}

\vskip .2cm
\noindent
Let $\cH$ be a Hilbert space. We recall that a representation 
\[
\pi \colon \fW(X, \sigma)\ni W (x) \mapsto W_{\pi}(x)\in {\cal U}(\cH)
\]
is called a {\em regular CCR representation} if
\[
\begin{array}{l}
t\mapsto W_{\pi}(tx)\hbox{ is strongly continuous for any }x\in X.
\end{array}
\]
One can then define {\em field operators} 
\[
\phi_{\pi}(x):= -\i\frac{\d}{\d t}W_\pi(tx)\Big|_{t=0},\: x\in X,
\]
which satisfy in the sense of quadratic forms on
$\cD(\phi_{\pi}(x_{1}))\cap \cD(\phi_{\pi}(x_{2}))$ the commutation
relations 
\beq
[\phi_\pi(x_1),\phi_\pi(x_2)]=\i\sigma(x_1,x_2),\: x_{1}, \:x_{2}\in
X.
\label{e0.3}
\eeq

\vskip .3cm
\goodbreak
\noindent
{\bf K\"{a}hler structures}

\vskip .2cm
\noindent
Let $(X, \sigma)$ be a real symplectic space 
and $\i$ a complex
structure on $X$. The space $(X, \i, \sigma)$ is called a {\em
K\"{a}hler space} if 
\[
\sigma( \i x_{1}, x_{2})= -\sigma(x_{1}, \i x_{2})  \hbox{ and }
\sigma(x, \i x)\hbox{ is positive definite}.
\]
If $(X, \i, \sigma)$ is a K\"{a}hler space, then $(X, \i,
\hf )$ is a Hermitian space for
\[
(x_{1}, x_{2}):= \sigma( x_{1}, \i x_{2})+ \i \sigma(
x_{1}, x_{2}).
\]
The typical example of a K\"{a}hler space is a Hermitian  space $(X,
\i, \hf )$ with its natural complex structure and symplectic form
$\sigma= {\rm Im} \hf$.

\vskip .3cm
\noindent
{\bf Creation and annihilation operators}

\vskip .2cm
\noindent
If $\pi$ is a regular CCR
representation of the Weyl algebra $\fW(X, \sigma)$, and $(X, \sigma)$
is equipped with a K\"{a}hler structure, then the {\em
creation} and {\em annihilation operators} are
defined as follows:
\[
a_\pi^*(x):=\frac{1}{\sqrt2} \bigl(\phi_\pi(x)-\i\phi_\pi(\i x) \bigr),\:
a_\pi(x):=\frac{1}{\sqrt2}\bigl(\phi_\pi(x)+\i\phi_\pi(\i x)\bigr).
\]
Clearly, 
\[
\phi_\pi(x)=\frac{1}{\sqrt2} \bigl(a_\pi^*(x)+a_\pi(x) \bigr),\: x\in X .
\] 
The operators $a_\pi^*(x)$ and $a_\pi(x)$ with domain
$\cD(\phi_\pi(x))\cap\cD(\phi_\pi(\i x))$ are closed and satisfy 
canonical commutation relations in 
the sense of quadratic forms:
\[
[a_\pi(x_1), a_\pi^{*}(x_{2})]= (x_1 , x_2)\one,\:
[a_\pi(x_2),a_\pi(x_1)]=[a_\pi^{*}(x_2),a^*(x_1)]= 0.
\]


\subsection{Complex fields}
\label{sec1}

Let $(X, \j)$ be a complex vector space. Let us assume that $X$ is
equipped with a sesquilinear, symmetric non-degenerate form
$\cq$. If $a\in L(X)$, we say that $a$ is isometric (resp.\
symmetric,  skew-symmetric) if $[a, \j]=0$ and $\cq(ax_{1}, ax_{2})= \cq(x_{1},
x_{2})$ (resp.\  $ \cq(ax_{1}, x_{2})=  \cq(x_{1},
ax_{2})$,  $ \cq(ax_{1}, x_{2})= - \cq(x_{1},
ax_{2})$). 
Clearly $(X, {\rm Im} \cq)$ is a real symplectic space. The
quadratic form~$ \cq$ is called the {\em charge
quadratic form}.

\vskip .3cm
\noindent
{\bf Gauge
transformations}

\vskip .2cm
\noindent
The maps $X\ni x\mapsto \e^{\, \j \alpha}x\in X$ for $\alpha\in \rr$ are called {\em gauge
transformations}.  They are symplectic on $(X, {\rm Im  \cq})$ and
isometric on $(X,  \cq)$.
We have 
\beq
 \cq(x_{1}, x_{2})= {\rm  Im \, } \cq(x_{1}, \j x_{2})+ \i {\rm Im} \cq(x_{1}, x_{2}).
\label{e0.1}
\eeq
\vskip .3cm
\noindent
{\bf Complex fields}

\vskip .2cm
\noindent
Let now $\pi$
be a regular CCR representation of $\fW(X, {\rm Im} \cq )$ on a Hilbert space $\cH$ and let
$\phi_{\pi}(x)$ be  the associated field. 
\goodbreak
Using the complex structure $\j$, we can  define the {\em complex
fields} 
\[
\begin{array}{l}
\varphi_\pi^*(x):=\frac{1}{\sqrt2} \bigl(\phi_\pi(x)-\i\phi_\pi(\j x) \bigr),\\[3mm]
\varphi_\pi(x):=\frac{1}{\sqrt2} \bigl(\phi_\pi(x)+\i\phi_\pi(\j x) \bigr),
\end{array}
\]
with domains $\cD(\phi_{\pi}(x))\cap \cD(\phi_{\pi}(\j x))$. The maps
$X\ni x\mapsto \varphi_{\pi}^{*}(x)\: (\hbox{resp.\
} x\mapsto  \varphi_{\pi}(x))$ are $\j$-linear (resp.\ $\j$-antilinear). 
\begin{lemma}
The operators
$\varphi^{\sharp}_{\pi}(x)$ are closed. In the sense of
quadratic forms on $\cD(\phi_{\pi}(x))\cap \cD(\phi_{\pi}(\j x))$ they satisfy the commutation
relations 
\[
[\varphi_\pi(x_1), \varphi_\pi^{*}(x_{2})]=  \cq(x_1, x_2)\one,\: 
[\varphi_\pi(x_1),\varphi_\pi(x_2)]=[\varphi_\pi^{*}(x_1),\varphi^*(x_2)]= 0.
\]
\end{lemma}
\proof The commutation relations are easily deduced from (\ref{e0.3}). 
Let $u\in \cD(\phi_{\pi}(x))\cap \cD(\phi_{\pi}(\j x))$.
To prove that $\varphi^{\sharp}_{\pi}(x)$ is
closed, we write   
\[
2\|\varphi_{\pi}(x)u\|^{2}= \|\phi_{\pi}(x)u\|^{2}+
\|\phi_{\pi}(\j x)u\|^{2}-  \cq(x, \j x)\|u\|^{2}.
\]
This easily implies that $\varphi_{\pi}(x)$  is closed. The case of
$\varphi^{*}_{\pi}(x)$ is treated similarly. \qed


\subsection{Charge operator}
\label{sec1.2}
\begin{definition}
Let $(X, \j , \cq)$ be as in Subsection \ref{sec1} and $\i$ another complex
structure on $X$. Then $(X, \j, \i,  \cq)$ is called a {\em charged
K\"{a}hler space} if $[\i,\j]=0$ and $(X, \i, {\rm Im} \cq)$ is a
K\"{a}hler space.
\label{0.1}
\end{definition}
Let $(X, \j, \i,  \cq)$ be a charged K\"{a}hler space. Then $\i$ is
antisymmetric for $ \cq$, \ie  $\cq( x_{1}, \i x_{2})= - \cq( \i
x_{1}, x_{2})$, and $\j$ is antisymmetric for $\hf$.

We can introduce the {\em charge operator}\/:
\[
\q 
:= -\i \j.
\]
Note that $[\q, \i]= [\q,\j]=0$, $\q^{2}=1$ and that $\q$ is symmetric and
isometric both for $ \cq$ and~$\hf$.
 Since $\i =\j \q$ we have $\e^{\, \j \alpha}= \e^{\i \alpha \q}$
and the gauge transformations $x \mapsto \e^{ \, \j \alpha}x$, $\alpha\in \rr$, 
form a unitary group on $(X, \i, \hf )$ with infinitesimal
generator $\q$.

The typical example of a charged K\"{a}hler space is a Hermitian  space
$(X, \i, \hf)$  with a distinguished symmetric  operator $\q$
such that $\q^{2}=1$. Let us denote by $X^{\pm}:= {\rm Ker}(\q \mp \one)$ the spaces of positive
(resp.\ negative) charge and by $x^{\pm}$ the orthogonal projection of $x\in X$ 
onto~$X^{\pm}$.  If we set $ \cq(x_{1}, x_{2})= (x_{1}^{+},
x_{2}^{+})-(x_{2}^{-}, x_{1}^{-})$, then $(X, \i \q, \i,  \cq)$ is  a charged
K\"{a}hler space. 
%
%
Note that $X^{+}$ or $X^{-}$ may be equal to
$\{0\}$. 

 Using the fact that $\q$ is symmetric for $\hf$ and
$ \cq$, we see that the spaces $X^{\pm}$ are orthogonal both for
$\hf$ and $ \cq$.
If we set $x^{\pm}= \12(x\pm \q x)$, then the
map
\[
\matrix
{U\colon & X & \to & X^{+}\oplus X^{-} \cr
& x & \mapsto & x^{+}\oplus x^{-}
\cr}
\]
is unitary from $(X, \i, \hf )$ to 
$(X^{+}, \i , \hf )\oplus (X^{-}, \i , \hf )$  and  isometric  from
$(X, \j ,  \cq)$ to~$(X^{+}, \i,
\hf)\oplus (X^{-}, -\i, -\overline{\hf})$.  

If $\pi \colon \fW (X, {\rm Im} \cq)\to {\cal U}(\cH)$
is  a regular CCR representation on a Hilbert space $\cH$, then we can 
introduce, just as in Subsection \ref{sec0}, 
{\em creation} and {\em annihilation operators} 
\[
a_\pi^*(x):=\frac{1}{\sqrt2}\bigl(\phi_\pi(x)-\i\phi_\pi(\i x)\bigr),\:
a_\pi(x):=\frac{1}{\sqrt2}\bigl(\phi_\pi(x)+\i\phi_\pi(\i x) \bigr),
\]
with domains $\cD(\phi_{\pi}(x))\cap \cD(\phi_{\pi}(\i x))$. The maps
$X\ni x\mapsto a_{\pi}^{*}(x)\: (\hbox{resp.\ }a_{\pi}(x))$ are $\i$-linear (resp.\ $\i$-antilinear). 
If $x= x^{+}+ x^{-}$, with $x^{\pm} \in X^{\pm}$, then
\[
\varphi_{\pi}(x)= a_{\pi}(x^{+})+ a^{*}_{\pi}(x^{-}) \hbox{ \ and \ } \varphi_{\pi}^{*}(x)=
a_{\pi}^{*}(x^{+})+ a_{\pi}(x^{-}).
\]
Note that this is consistent with fact that the maps
$X\ni x\mapsto \varphi_{\pi}^{*}(x)\: (\hbox{resp.\
} x\mapsto  \varphi_{\pi}(x))$ are $\j$-linear (resp.\ $\j$-antilinear).

\subsection{Charge conjugation}
\label{sec1.3}

Let $(X, \j,\i,  \cq)$ be a charged K\"{a}hler space.  Assume  that there exists 
some ${\rm c}\in L(X)$  such that
\beq
{\rm c}^{2}=\one, \: {\rm c}\i =\i \bc  , \: \bc  \q =- \q \bc  , \: (x_{1}, \bc  x_{2})=
(\bc  x_{1},x_{2}),\: x_{1}, x_{2}\in X.
\label{e0.4}
\eeq
I.e., $\bc$ is a symmetric involution for $\hf$, which
anticommutes with the charge operator~$\q$. An operator $\bc$ satisfying
(\ref{e0.4}) is called a
{\em charge conjugation}.
Charge conjugations exist in charge-symmetric quantum field theories. A charged
K\"{a}hler space $(X, \j, \i,   \cq, \bc)$ equipped with a charge
conjugation $\bc$
will be called a {\em charge-symmetric K\"{a}hler space}.

It follows from (\ref{e0.4}) that $ \cq(x_{1},
\bc  x_{2})=- \cq(\bc  x_{1}, x_{2})$,
\ie $\bc$ is antisymmetric for $ \cq$. 
Since $\bc  \q =- \q \bc$, we see that  $\bc$ is a unitary
map from $(X^{-}, \i, \hf )$ to
$(X^{+}, \i, \hf)$.


\section{Stochastically positive KMS systems}
\label{stochas}
\init
In this section we recall the notion of a {\em stochastically positive
KMS system} due to Klein and Landau \cite{KL1}. We prove that
stochastically positive KMS systems are invariant under time-reversal.
\subsection{KMS systems}
Let $\cR$ be a $C^{*}$-algebra and $\{\tau_{t}\}_{t\in \rr}$ a 
group of $^{*}$-automorphisms of $\cR$. Let $\omega$ be a
$(\tau, \beta)$-{\em KMS state} on $\cR$, \ie a state such that for each
$A,B\in \cR$ there exists a function $F_{A,B}(z)$ holomorphic in the
strip $\{z\in \cc\mid 0<{\rm Im}z<\beta\}$ and continuous on its closure such
that
\[
F_{A, B}(t)= \omega(A\tau_{t}(B)),\:F_{A, B}(t+\i \beta)=
\omega(\tau_{t}(B)A),\: t\in \rr.
\]
A triple $(\cR, \tau, \omega)$ such that $\omega$ is a $(\tau,
\beta)$-KMS state is called a $\beta$-{\em KMS system}.

Let us now recall some standard facts about KMS systems.
By the GNS construction, one associates to $(\cR, \tau, \omega)$
 a Hilbert space $\cH_{\omega}$, a representation
$\pi_{\omega}$ of $\cR$ on $\cH_{\omega}$, a unit vector~$\Omega_{\omega}$, 
cyclic for $\pi_{\omega}$, and a strongly continuous 
unitary group $\{\e^{-\i tL}\}_{t\in \rr}$ such that
\[
\omega(A)= (\Omega_{\omega},
\pi_{\omega} (A)\Omega_{\omega}),\:\pi_{\omega}(\tau_{t}(A))= \e^{\i t
L}\pi_{\omega}(A)\e^{-\i tL},\:L\Omega_{\omega}=0.
\] 
The KMS condition implies that $\Omega_{\omega}$ is separating for the von Neumann
algebra $\pi_\omega (\cR)''$, \ie $A \Omega _{\omega} = 0 \Rightarrow A=0$ for $A \in \pi_\omega(\cR)''$. 
Consequently, 
the image of $\cR$ under
$\pi_{\omega}$ is isomorphic to~$\cR$; it will therefore not be distinguished from $\cR$. Moreover, we will  
identify an element $A$ of~$\cR$ with its image $\pi_{\omega}(A)$. 

The selfadjoint operator $L$ is called the {\em
Liouvillean} associated to the KMS system $(\cR, \tau, \omega)$.
It is the unique selfadjoint operator whose associated unitary group
generates the dynamics~$\tau$ and such that $L\Omega_{\omega}=0$
(see \eg  \cite[Prop. 2.14]{DJP}).
\begin{proposition}
\label{propliou}
Let $\cR_{1}\subset \cR$ be the set of $A\in \cR$ such that $\tau \colon
t\mapsto \tau_{t}(A)$ is $C^{1}$ for the strong topology on
$\cB(\cH_{\omega})$. Then 
$\cR_{1}\Omega_{\omega}\subset \cD(L)$
is a core for $L$.


\end{proposition}
\proof 
Note first that $A\in \cR_{1}$ iff $A$ is of class $C^{1}(L)$ (see
\cite[Def.\ 6.2.2]{ABG}).
Clearly $\cR_{1}$ is dense in
$\cR$ for the strong operator topology. In fact, if $A\in \cR$, then
the strong integral
$A_{\epsilon}=\epsilon^{-1}\int_{0}^{\epsilon}\tau_{t}(A)\d t$ belongs
to $\cR_{1}$ and converges strongly to $A$ when $\epsilon\to 0$.

Since $\Omega_{\omega}$ is cyclic for $\cR$, this implies that
$\cR_{1}\Omega_{\omega}$ is dense in $\cH_{\omega}$. Moreover, since
$L\Omega_{\omega}=0$, we have
$\e^{\i tL}\cR_{1}\Omega_{\omega}=\cR_{1}\Omega_{\omega}$ and $\cR_{1}\Omega_{\omega}\subset
\cD(L)$. Thus Nelson's theorem  implies that
$\cR_{1}\Omega_{\omega}$ is a core for $L$.


%

\vskip .3cm
\noindent
{\bf Euclidean Green's functions}

\vskip .2cm
\noindent
Let \beq
\label{inbeta}
I_{\beta}^{n+}:=\{(z_{1},\dots, z_{n})\in \cc^n \:| \:{\rm Im}z_{j}<{\rm
Im}z_{j+1}, \: {\rm Im}z_{n}- {\rm Im}z_{1}<\beta\}.
\eeq

It follows from a result of Araki \cite{Ar1, Ar2} that, for
$A_{1}, \dots, A_{n}\in \cR$,  the Green's function
\[
 G(t_{1}, \dots, t_{n};A_{1},\dots, A_{n}):
=\omega \bigl(\prod_{1}^{n}\tau_{t_{i}}(A_{i}) \bigr)
\]
extends to an holomorphic function in  $I_{\beta}^{n+}$, continuous on
$\overline{I_{\beta}^{n+}}$. In particular, one can uniquely define the
{\em Euclidean Green's functions} \[
\EG(s_{1}, \dots, s_{n};A_{1}, \dots, A_{n}):=G(\i s_{1}, \dots, \i s_{n}; A_{1},
\dots, A_{n})
\]
for all $(s_{1}, \dots, s_{n})$ such that $s_{1}\leq \cdots\leq s_{n}$
and $s_{n}-s_{1}\leq \beta$. The correct way to view such an n-tuple
$(s_{1}, \dots, s_{n})$ is as an n-tuple of points on the circle of
length $\beta$, {\em ordered counter-clockwise}.


\subsection{Stochastically positive KMS systems}
In \cite{KL1} Klein and Landau  introduced a class of KMS systems
which they called {\em stochastically positive KMS systems}. To a
stochastically positive
KMS system one can associate a (unique up to equivalence) {\em
generalized path space} $(Q, \Sigma, \Sigma_0, U(t), R, \mu)$ (see Section 5)
which has some special properties, the most important
being the $\beta$-periodicity in $t$ and the {\em Osterwalder-Schrader
{\rm (OS)}-positivity}.

Conversely Klein and Landau have shown in \cite{KL1} that to a generalized path space
satisfying the properties in Definition 5.1
one can
associate a (unique up to unitary equivalence) stochastically positive KMS system.
This is an example of a {\em reconstruction theorem}; similar results are 
well-known in Euclidean QFT.  A
reconstruction theorem allowing to go from Euclidean Green's functions
to a KMS system has recently been proved in a  general context by
Birke and Fr\"ohlich  in \cite{BF}. 

The advantage of the Klein and Landau formalism is that it is relatively easy to
perturb the stochastic process associated to a KMS system, using
functional integral methods.

\begin{definition}
\label{defstocpos}
Let $(\cR, \tau, \omega)$ be a KMS system and 
$\fU\subset \cR$ an abelian $^{*}$-subalgebra. The KMS
system $(\cR, \fU,  \tau , \omega)$ is  called {\em stochastically
positive}  if

\vskip .3cm
\halign{ \indent \indent \indent #  \hfil & \vtop { 
\parindent =0pt 
\hsize=12cm
                            \strut # \strut} \cr 
{\rm (i)}  & the $C^{*}$-algebra generated by $\bigcup_{t\in
\rr}\tau_{t}(\fU)$ is equal to $\cR$;
\cr
{\rm (ii)} &  the Euclidean Green's functions $\EG(s_{1}, \dots, s_{n};
A_{1}, \dots, A_{n})$ are positive for all $A_{1}, \dots, A_{n}\in
\fU^{+}=\{A\in \fU\mid A\geq 0\}$ and for all $(s_{1}, \dots, s_{n})$ such that $s_{1}\leq \cdots\leq s_{n}$
and $s_{n}-s_{1}\leq \beta$. 
\cr}
\end{definition}
It is often more convenient to consider instead of the $C^{*}$-algebras
$\cR$ and $\fU$ their weak closures in the GNS representation, which
we denote by $\cRbar$ and $\fUbar$. We denote by
$\overline{\tau}$ the group $\{\overline{\tau}_{t}\}_{t \in \rr}
$ of $^{*}$-automorphisms of $\cRbar$ defined by
$\overline{\tau}_{t}(A) := \e^{\i tL}A\e^{-i tL}$. The state $\omega$
extends to $\cRbar$ by setting $\overline{\omega}(A):= (\Omega_{\omega},
\pi_{\omega}(A)\Omega_{\omega})$.
The following fact has been shown in \cite[Prop. 3.4]{KL1}.
\begin{proposition}
\label{vonneumann}
Let $(\cR, \fU, \tau, \omega)$ be a stochastically
positive KMS system. Then $(\cRbar, \fUbar, \overline{\tau}, \overline{\omega})$ is also a stochastically positive KMS system (in the $W^*$-sense). I.e.,
\vskip .3cm
\halign{ \indent \indent \indent #  \hfil & \vtop { 
\parindent =0pt 
\hsize=12cm
                            \strut # \strut} \cr 
{\rm (i)}  & the $W^{*}$-algebra generated by $\bigcup_{t\in
\rr}\tau_{t}(\fUbar)$ is equal to $\cRbar$;
\cr
{\rm (ii)} &  the Euclidean Green's functions $\EG(s_{1}, \dots, s_{n};
A_{1}, \dots, A_{n})$ are positive for all $A_{1}, \dots, A_{n}\in
\fUbar^+$ and for all n-tuples $(s_{1}, \dots, s_{n})$ such that $s_{1}\leq \cdots\leq s_{n}$
and $s_{n}-s_{1}\leq \beta$. 
\cr}

\end{proposition}
Now we show that stochastically positive KMS systems are invariant
under {\em time reversal}, a fact that is well known for
$0$-temperature field theories (see for example 
\cite{Si}).
\begin{proposition}
\label{timereversal}
Let $(\cR, \fU, \tau, \omega)$ be a {\em stochastically
positive} KMS system. Then there exists an anti-unitary involution $T$
on $\cH_{\omega}$ such that
\vskip .3cm
\halign{ \indent \indent \indent #  \hfil & \vtop { 
\parindent =0pt 
\hsize=12cm
                            \strut # \strut} \cr 
{\rm (i)}  & $T\cRbar T^{-1}= \cRbar $, $TAT^{-1}= A^{*}$ for $A\in \fUbar$;
\cr
{\rm (ii)} &  $T\Omega_{\omega}= \Omega_{\omega}$, $T \: \overline{\tau}_{t}(A)=
\overline{\tau}_{-t}(A)T $ for $A\in \overline{\cR}, \: t\in \rr$.
\cr}
\end{proposition}
From the properties of $T$ we see that $T$ implements the {\em time
reversal transformation}.

\medskip
\noindent
\proof Let $A_{1},A_{2}\in \fU$. The map $z\mapsto
\omega(A_{1}\tau_{t}(A_{2}))_{|t = iz}$ is holomorphic in $\{0<{\rm Re}z<\beta\}$. 
By stochastic
positivity it is real on $\{{\rm Im}z=0\}$ if $A_{i}=A_{i}^{*}$. 
The Schwarz's reflection principle implies
\[
\omega \bigl(A_{1}\tau_{t}(A_{2})\bigr)_{|t = iz}=
\overline{\omega(A_{1}\tau_{t}(A_{2}))_{|t = i \bar z}} \hbox{ \ for \
}A_{i}\in \fU, \:A_{i}= A_{i}^{*}.
\]
For $z=-\i t$ this yields 
\beq
\label{timerev}
\omega\bigl(A_{1}\tau_{t}(A_{2})\bigr)=
\overline{\omega(A_{1}\tau_{-t}(A_{2}))}=\omega\bigl(\tau_{-t}(A_{2})A_{1}\bigr) \hbox{ \ for \
}A_{i}\in \fU, \:A_{i}= A_{i}^{*}.
\eeq
By $\cc$-linearity this identity extends to all $A_{i}\in \fU$. We can now
define the antilinear operator
\beq
T \colon \sum_{j=1}^{n}\e^{\i t_{j}L}A_{j}\Omega_\omega \mapsto \sum_{j=1}^{n}\e^{-\i
t_{j}L}A^{*}_{j}\Omega_\omega.
\label{reversal}
\eeq
For $u=\sum_{j=1}^{n}\e^{\i t_{j}L}A_{j}\Omega_\omega$ identity (\ref{timerev}) implies 
\[
\begin{array}{rl}
\|u\|^{2}=&\bigl(\sum_{j=1}^{n}\e^{\i t_{j}L}A_{j}\Omega_\omega, \sum_{k=1}^{n}\e^{\i
t_{k}L}A_{k}\Omega_\omega \bigr)\\
=& \sum_{j,k}\bigl(\Omega_\omega, A_{j}^{*}\e^{\i (t_{k}-t_{j})L}A_{k}\Omega_\omega\bigr)=
\sum_{j,k}\omega\bigl(A_{j}^{*}\tau_{t_{k}-t_{j}}(A_{k})\bigr)\\
=&\sum_{j,k}\omega\bigl(\tau_{t_{j}-t_{k}}(A_{k})A_{j}^{*}\bigr)= 
\sum_{j, k}\bigl(\Omega_\omega, A_{k}\e^{\i (t_{k}-t_{j})L}A_{j}^{*}\Omega_\omega\bigr)\\
=&\sum_{j,k}\bigl(\e^{-\i t_{k}L}A_{k}^{*}\Omega_\omega, \e^{-\i t_{j}L}A_{j}^{*}\Omega_\omega \bigr)=\|Tu\|^{2}.
\end{array}
\]
Thus $T$ is a well defined
antilinear operator. Moreover, using property
{\rm (i)} of Definition \ref{defstocpos} and the fact that
$\Omega_{\omega}$ is cyclic for~$\cR$, we conclude that $T$ has a dense domain and a dense range. Hence $T$ extends uniquely to
an anti-unitary operator. Clearly $T$ is an involution. The other
properties of $T$ follow directly from (\ref{reversal}). \qed

\section{Quasi-free KMS states}
\init
\label{quasi}
In this section we recall some well-known facts about quasi-free KMS
states and describe a class of quasi-free KMS states which generate 
stochastically positive KMS systems (see \cite{KL2}, \cite{OGie}).  
\init
\label{sec4}
\subsection{Quasi-free KMS states}
\label{sec4.1}
Let $X_{0}$ be a pre-Hilbert space, $X$ the completion of $X_{0}$. 
Then $(X_{0}, \sigma)$ is a real
symplectic space for $\sigma={\rm Im}\hf$, and we denote by
$\fW(X_{0})$ the Weyl algebra $\fW(X_{0}, \sigma)$. 
Let  ${\rm a}\geq 0$ be  a selfadjoint operator on $X$ such that
$X_{0}\subset \cD({\rm a}^{-\12})$ and $\e^{-\i t{\rm a}}$ preserves
$X_{0}$. Given ${\rm a}\geq 0$ the canonical choice for $X_{0}$ is
$\cD({\rm a}^{-\12})$. 

For  $\beta>0$ one defines
 a state $\omega_{\beta}$ on $\fW(X_{0})$  by the functional
\beq
\label{statedef}
\omega_{\beta}(W(x)):= \e^{-\frac{1}{4}(x, (1+ 2\rho)x)}, \: x\in X_{0},
\eeq
where $\rho:= ( \e^{\beta {\rm a}}-1)^{-1}$. Since $1+2\rho= \frac{1+ \e^{-\beta{\rm a}}}{1-
\e^{-\beta{\rm a}}}$ and ${\rm a}\geq 0$ the form domain of $1+
2\rho$ is equal to $D({\rm a}^{-\12})\supset X_{0}$.

The state $\omega_{\beta}$ is a $(\tau^\circ, \beta)$-KMS state for the
dynamics $\tau^\circ  \colon t \mapsto \tau^\circ_{t}$ defined by
\[
\matrix{ 
\tau^\circ_{t}\colon & \fW(X_{0}) & \to & \fW(X_{0}) \cr
& W(x)& \mapsto & W(\e^{\i t{\rm a}}x).
\cr}
\]
The state $\omega_{\beta}$ is {\em quasi-free} (see [BR]) and the
KMS system $(\fW(X_{0}), \tau^\circ, \omega_{\beta})$ defined above
is called the quasi-free KMS system {\em associated to }${\rm a}$.

The standard example  is the following one:
let ${\rm h}\geq 0$ be a selfadjoint operator representing the {\em one
particle energy}. Assume  that there
exists a selfadjoint operator $\tq$ on $X$ representing the {\em one
particle charge} such that
$\tq^{2}=\one$, $[{\rm h}, \tq]=0$. Then we can
associate a group of {\em gauge transformations} $\{Ê\alpha_{t}
\}_{ t \in [0, 2\pi[}$,
\[
\matrix{ 
\alpha_{t}\colon & \fW(X_{0}) & \to & \fW(X_{0}) \cr
& W(x) & \mapsto & W(\e^{\i t\tq}x),
\cr}
\]
to the charge operator $\tq$.
Let $\mu\in \rr$ such that ${\rm h}- \mu \tq\geq \lambda>0$. Thus the range for the value of the 
chemical potential~$\mu$, which we consider,  excludes Bose-Einstein condensation. It
follows that~${\rm a}:={\rm h}-\mu \tq>0$ and hence~$X_{0}=\cD({\rm
a}^{-\12})=X$. Therefore the unique quasi-free KMS state 
on~$\fW(X)$ at inverse temperature~$\beta$ and chemical potential~$\mu$ is the state~$\omega_{\beta}$  defined
by (\ref{statedef}).

\subsection{Araki-Woods representation}
\label{sec4.3}
Let us consider a quasi-free KMS system associated to a selfadjoint
operator~${\rm a}$ as in Subsection~\ref{sec4.1}.
Let $\Xbar$ be the conjugate Hilbert space to $X$. 
Elements of
$\Xbar$ will be denoted by~$\overline{x}$. Equivalently, we denote by
$X\ni x\mapsto \overline{x}\in \Xbar$ the identity operator, which
is antilinear. 
If ${\rm a}$ is a linear operator on $X$, we denote by
$\overline{{\rm a}}$ the linear operator on $\Xbar$ defined by
$\overline{{\rm a}}\:\overline{x}:= \overline{{\rm a}x}$. If
$\ch$ is a Hilbert space, then
\[
\G(\ch)=\bigoplus_{n=0}^{+\infty}\otimes^{n}_{\rm s}\ch
\]
denotes the bosonic
Fock space over $\ch$.

We set:
\[
\begin{array}{l}
\cH_{\omega}:= \G(X\oplus \Xbar),\\[3mm]
\Omega_{\omega}:= \Omega, \\[3mm]
W_{\omega, {\rm l}}(x):= W_{\rm F}\bigl( (1+ \rho)^{\frac{1}{2}}x\oplus
\overline{\rho}^{\frac{1}{2}}\overline{x}\bigr),\: x\in X_{0},\\[3mm]
W_{\omega, {\rm r}}(x):= W_{\rm F}\bigl( \rho^{\frac{1}{2}}x\oplus
(1+ \overline{\rho})^{\frac{1}{2}}\overline{x}\bigr),\: x\in X_{0},\\[3mm]
\end{array}
\] 
where $W_{F}(.)$ denotes the Fock space Weyl operator on
$\G(X\oplus\Xbar)$ and $\Omega\in \G(X\oplus \Xbar)$ denotes the Fock
vacuum.

The following facts are well known:

\noindent
\vskip .3cm
\halign{ \indent \indent \indent #  \hfil & \vtop { \parindent =0pt \hsize=12cm
                            \strut # \strut} \cr 
{\rm (i)}  & The map $W (x)\mapsto W_{\omega,{\rm l/r}}(x)\in
U(\cH_{\omega})$ defines a regular CCR representations;
\cr
{\rm (ii)}  & $[W_{\omega, {\rm l}}(x), W_{\omega, {\rm r}}(y)]=0$ for $x,y\in
X_{0}$;
\cr
{\rm (iii)}  & $(\Omega_{\omega},W_{\omega, {\rm l}}(x)\Omega_{\omega})=
\omega(W(x))$  for $x\in X_{0}$; 
\cr
{\rm (iv)}  & Let $L:= \dG({\rm a}\oplus-\overline{{\rm a}})$  act 
on $\cH_{\omega}$. Then 
\[
\e^{-\i tL}\Omega_{\omega}= \Omega_{\omega}, \: \e^{\i tL}W_{\omega,
{\rm l}}(x)\e^{-\i tL}= W_{\omega,{\rm l}}(\e^{\i ta}x), \: x\in
X_{0}; 
\]
\vskip -.4cm
\cr
{\rm (v)}  &  The vector $\Omega$ is cyclic for the representations $W_{\omega,
{\rm l}/{\rm r}}(.)$.
\cr}
\vskip .2cm
\noindent
In particular, the Araki-Woods representation is the GNS representation
for the KMS system $\bigl(\fW(X, \sigma), \tau^\circ , \omega\bigr)$ and $L$ is the
associated Liouvillean.

We will only consider the left Araki-Woods representation, thus
will  forget the subscript ${\rm l}$ and write $W_{\omega}(x):=
W_{\omega, {\rm l}}(x)$, $x\in X$. The creation-annihilation
operators associated to $W_{\omega}(.)$ are 
\[
\begin{array}{l}
a^{*}_{\omega}(x)= a^{*}_{F}\bigl( (1+\rho)^{\12}x\oplus 0) + a_{F}(0\oplus
\overline{\rho}^{\12}\overline{x}\bigr),\\
a_{\omega}(x)= a_{F}\bigl( (1+\rho)^{\12}x\oplus 0) + a_{F}^{*}(0\oplus
\overline{\rho}^{\12}\overline{x}\bigr).
\end{array}
\]
\subsection{Field algebras}
\label{sec4.2}
We recall that a {\em conjugation} on a Hilbert space $X$ is an
anti-unitary involution on $X$. Let us assume that $X$ is equipped
with a conjugation $\kappa$. To 
$\kappa$ we associate the real vector space $X_{\kappa}:
=\{x\in X\:| \:\kappa x=x\}$. Let $\omega$ be the quasi-free state
associated to a selfadjoint operator~${\rm a}$, as defined
in Subsection \ref{sec4.1}, and let $\cH_{\omega}$ be the Araki-Woods
space introduced in Subsection~\ref{sec4.3}.

We will denote by $\cW\subset \cB(\cH_{\omega})$ the {\em field algebra},
\ie the von Neumann algebra generated
by the $\{W_{\omega}(x) \mid x\in X\}$ and by
$\cW_\kappa \subset \cB(\cH_{\omega})$ the von Neumann algebra generated
by $\{W_{\omega}(x)\:| \: x\in X_{\kappa}\}$. 
Since the symplectic form $\sigma$ vanishes on $X_{\kappa}$, the algebra
$\cW_\kappa $ is abelian.

\begin{lemma}\label{4.1}
Assume that ${\rm a}= {\rm h}-\mu \tq$, where ${\rm h}$ and $ \tq $ are 
selfadjoint operators such that $[{\rm h}, \tq ]=0$, $\tq^{2}=1$, ${\rm h}\geq m> 0$ and $|\mu|< m$. Let $\kappa$ be a
conjugation on $X$ such that $[{\rm h} ,\kappa]=0$. Then $\cW$ is
the von Neumann algebra generated by $\{\e^{\i tL}A\e^{-\i tL} \mid t\in \rr, \: A\in
\cW_\kappa \}$.
\end{lemma}
\proof Clearly $\{\e^{\i tL}A\e^{-\i tL} \mid t\in \rr, \: A\in
\cW_\kappa \}\subset \cW$, so it suffices to prove the converse
inclusion. Using the CCR, the facts that $(1+ \rho)^{\12}$ and
$\rho^{\12}$ are bounded, and the fact that  the map
\[
 X\oplus \overline{X}\ni x_{1}\oplus \overline{x_{2}}\mapsto
W_{F}( x_{1}\oplus \overline{x_{2}})\in \cB(\cH_{\omega}) 
\]
is continuous for the strong topology, it suffices to verify that
\beq
E={\rm Vect}_{\rr}\{\e^{\i t( {\rm h}-\mu \tq)}x,\: t\in \rr, \; x\in
X_{\kappa}\}\hbox{ is dense in }X.
\label{e4.1}
\eeq
Clearly $\overline{E}$ contains $X_{\kappa}$, and by
differentiating with respect to $t$, we see that $\overline{E}$
contains also $\{\i({\rm h}- \mu \tq)x \mid x\in X_{\kappa}\cap
\cD({\rm h})\}$. We now claim that for each $x\in X$ there exists
$x_{1}\in X_{\kappa}$ and $x_{2}\in X_{\kappa}\cap \cD({\rm h})$
such that
\[
x= x_{1}+ \i ({\rm h}-\mu \tq)x_{2}.
\]
This will imply (\ref{e4.1}). In fact, the $\rr$-linear map $r=\12 \mu
\tq{\rm h}^{-1}(1-\kappa)$ has norm less than~$|\mu|m^{-1}<1$, so for
$x\in X$ we
can find $y\in X$ such that $y- ry=x$. If $x_{1}= \12(y+ \kappa y)$ 
and~$x_{2}= \12 (\i {\rm h})^{-1}( y-\kappa y)$, then both are elements of $
X_{\kappa}$, since $[{\rm h}, \kappa]=0$. Now
\[
x_{1}+ \i ({\rm h}- \mu \tq)x_{2}= y -\frac{\i}{2}\mu {\tq} {\rm h}^{-1}(
y- \kappa y)= y -ry=x  \square .
\]

\subsection{Observable algebras}
\label{sec4.3b}
The gauge transformations $\alpha_{t}$ on $\fW(X_{0}, \sigma)$ can be
unitarily implemented in the Araki-Woods representation:
\[
\alpha_{t} \bigl(W_{\omega}(x)\bigr)= \e^{\i tQ}W_{\omega}(x)\e^{-\i tQ},
\]
where $Q:= \dG( \tq\oplus -\overline{\tq})$.

We denote by $\cA$ the {\em observable algebra}
\[
 \cA:= \bigl\{ A\in \cW \mid \e^{\i tQ}A\e^{-\i tQ}= A, \: t\in [0, 2\pi[ \bigr\} 
\]
and by $\cA_{\kappa}$ the {\em abelian observable algebra} $\cA_{\kappa}:=
\cA \cap \cW_\kappa $.

\begin{lemma}\label{4.2}
Assume that ${\rm h}\geq m> 0$ and  $|\mu|< m$. Let $\kappa$ be a
conjugation on $X$ such that $[{\rm h}, \kappa]=0$. Then $\cA$ is
the von Neumann algebra generated by $\{\e^{\i tL}A\e^{-\i tL} \mid t\in \rr, \: A\in
\cA_{\kappa}\}$.
\end{lemma}
\proof Clearly $\e^{\i tL}A\e^{-\i tL}\in \cA$, if $A\in
\cA_{\kappa}$, since $[L, Q]=0$. Conversely, let $A\in \cA$. By
Lemma~\ref{4.1} there exists a net $\{ A_{i} \}_{i\in I}$ in the  algebra generated by $ \{\e^{\i tL}A\e^{-\i tL}, \: t\in \rr, \: A\in
\cA_{\kappa}\}$ such that
$A= \slim A_{i}$. For $R\in \cB(\cH_\omega)$, let $R^{\rm av}:=
(2\pi)^{-1}\int_{0}^{2\pi}\e^{\i t Q}R\e^{-\i tQ}\d t$ be the average
of $R$ with respect to the gauge group. Then by dominated convergence
$\slim A_{i}^{\rm av}= A^{\rm av}=A$. Since $[L, Q]=0$, we have
$(\e^{\i tL}R\e^{-\i tL})^{\rm av}= \e^{\i tL}R^{\rm av}\e^{-\i tL}$,
which implies the lemma \qed .
\begin{lemma}\label{4.3}
We have $\overline{\cA \Omega_{\omega}}=\{u\in \cH_\omega \mid Qu=0\}$.
\end{lemma}
\proof Since $Q\Omega_{\omega}=0$ we have $\overline{\cA\Omega_{\omega}}\subset {\rm
Ker}Q$. Let now $u\in {\rm Ker}Q$. If $\{ A_{i} \in \cW\}_{i \in I}$ is a net such that
$\lim A_{i}\Omega_{\omega}= u$, then
\[
 u=\frac{1}{2\pi}\int_{0}^{2\pi}\e^{\i tQ}u \,  \d t=
\lim\frac{1}{2\pi}\int_{0}^{2\pi}\e^{\i tQ}A_{i}\e^{-\i tQ}\Omega_{\omega} \, \d t=
\lim_{n \to \infty} A_{i}^{\rm av}\Omega_{\omega},
\]
which proves the lemma since $A_{i}^{\rm av}\in \cA$  \qed .

\subsection{Stochastic positivity}
\label{stocpos}
In this subsection we give a criterion for the stochastic positivity
of a quasi-free KMS system.

The following lemma is due to Klein and Landau \cite{KL2}.
\begin{lemma}
Let ${\rm a}\geq 0$ be a selfadjoint operator on a Hilbert space $X$. Let
$\rr\ni s\to r(s)\in \cB(X)$ be the $\beta$-periodic operator-valued
function defined by
\[
r(s)= \frac{\e^{- s{\rm a}}+ \e^{(s-\beta){\rm a}}}{1-\e^{-\beta {\rm a}}}, \: 0\leq
s< \beta.
\]
Then, for $x_{i}\in X$ and $s_{i}\in \rr$, one has 
\[
\sum_{i,j}\bigl( x_{i}, r(s_{j}-s_{i})x_{j} \bigr)\geq 0.
\]
\label{p1.0}
\end{lemma}
\proof Using the spectral decomposition of ${\rm a}$, we can assume that
$x_{i}\in \cc$ and ${\rm a}\geq 0$ is a positive real number. Hence it is sufficient
to verify that $r(s)$ is a distribution of positive type. But this
follows from Bochner's theorem and the fact that  the Fourier
transform of~$r$ is $\sum_{n\in \zz}r_{n}\delta(. -2\pi/n)$, where
$r_{n}= \frac{2{\rm a}}{ {\rm a}^{2}+
(2\pi n/\beta)^{2}}\geq 0$ 
\rm 
\qed .

\begin{theoreme}
\label{p1.1}
Let $X$ be a Hilbert space equipped with a conjugation $\kappa$ and
${\rm a}\geq m>0$ a selfadjoint operator on $X$ such that $[{\rm a},
\kappa]=0$. Let $X_{\kappa}\subset X$ be the real vector space
associated to $\kappa$.

Let $\bigl(\cW, \tau^\circ, \omega\bigr)$ be the quasi-free KMS system
associated to ${\rm a}$ and let $\cW_\kappa\subset \cW$ be the abelian  von Neumann algebra generated by $\{W_\omega(x) \mid
x\in X_{\kappa}\}$. Then the KMS system $\bigl(\cW, \cW_\kappa, \tau^\circ,\omega \bigr)$ is stochastically positive.
\end{theoreme}
\proof We start by computing the Euclidean Green's functions. 
Using the CCR we get, for $x_{j}\in X$ and $1\leq j\leq n$,
\[
\prod_{1}^{n}W(x_{j})= \prod_{1\leq i\leq j\leq n}\e^{-\frac{\i}{2}\sigma(x_{i},
x_{j})}W \bigl(\sum_{1}^{n}x_{j} \bigr).
\]
We denote by  
\[
G\bigl(t_{1}, \dots, t_{n}; W(x_{1}), \dots,
W(x_{n}) \bigr)=\omega \bigl(\prod_{j=1}^{n}W(\e^{\i t_{j}{\rm a}}x_{j}) \bigr)
\]
the Green's functions for the Weyl operators $W(x_{j})$, $1\leq j\leq n$. Now
\[
\begin{array}{rl}
& G\bigl(t_{1}, \dots, t_{n}; W(x_{1}), \dots, W(x_{n})\bigr)\\[2mm]
= &\lPi_{1\leq i< j\leq
n}\e^{-\i {\rm Im}( x_{i}, \e^{\i
(t_{j}-t_{i}){\rm a}}x_{j})}\e^{-\frac{1}{4}(\sum_{1}^{n}\e^{\i
t_{j}{\rm a}}x_{j}, (1+ 2 \rho)\sum_{1}^{n}\e^{\i
t_{j}{\rm a}}x_{j})}\\[2mm]
=&\prod_{1}^{n}\e^{-\frac{1}{4}( x_{i}, (1+ 2 \rho) x_{i})} \lPi_{1\leq i< j\leq
n}\e^{-\frac{1}{2}R(t_{j}- t_{i})( x_{i}, x_{j})},
\end{array}
\]
where
\[
R( t)( x, y)= \bigl( x, (1-\e^{-\beta {\rm a}})^{-1}\e^{\i t{\rm a}}y \bigr)+ \bigl(y, \e^{-\beta
{\rm a}}(1-\e^{-\beta {\rm a}})^{-1}\e^{\i t{\rm a}}x \bigr).
\]
For $x,y\in X$ the function $t\mapsto R(t)( x,y)$ has an holomorphic
extension to $0<{\rm Im }z<\beta$ such that  
the function $(t_{1}, \dots, t_{n})\mapsto G\bigl(t_{1}, \dots, t_{n}; W(x_{1}), \dots, W(x_{n})\bigr)$ is holomorphic in the set~$I_{\beta}^{n+}$
defined in (\ref{inbeta}) and continuous
on $\overline{I_{\beta}^{n+}}$ with
holomorphic extension 
\[
(\zeta_{1}, \dots, \zeta_{n})\mapsto \prod_{1}^{n}\e^{-\frac{1}{4}( x_{i}, (1+2 \rho)x_{i})}
\prod_{1\leq i< j\leq
n}\e^{-\frac{1}{2}R(\zeta_{j}- \zeta_{i})( x_{i}, x_{j})}.
\]
Hence the euclidean Green's functions  
\[
\EG \bigl(  s_{1}, \dots, s_{n}; W(x_{1}), \dots, W(x_{n})\bigr)=
\prod_{1}^{n}\e^{- \frac{1}{2}C(0)( x_{i}, x_{i})}\prod_{1\leq i<j\leq
n}\e^{-C(s_{j}-s_{i})( x_{i}, x_{j})},
\] 
where
\[
 C(s)(x,y):= \frac{1}{2} \bigl( x, (1-\e^{-\beta {\rm a}})^{-1}\e^{-s {\rm a}}y \bigr)+
\frac{1}{2} \bigl( y, (1-\e^{-\beta {\rm a}})^{-1}\e^{(s-\beta){\rm a}}x \bigr).
\]
Using  the fact that $\kappa {\rm a}= {\rm a}\kappa$ we get 
\[
C(s)( x, y)=\frac{1}{2} \Bigl(x, \frac{\e^{-s {\rm a}}+ \e^{(s-\beta){\rm a}}}{1-\e^{-\beta {\rm a}}}y \Bigr),
\hbox{ for }x, y\in X_{\kappa}.
\]
Thus, for $
x_{j}\in X_{\kappa}$ and $1\leq j\leq n$,

\beq
\EG \bigl( s_{1}, \dots, s_{n}; W(x_{1}), \dots, W(x_{n})\bigr) =\prod_{1\leq
i, j\leq n}\e^{-\frac{1}{2}C(|s_{i}-s_{j}|)( x_{i}, x_{j})}. \label{pe1.0}
\eeq
We will now prove the stochastic positivity. We will use the
Araki-Woods representation described in Subsection \ref{sec4.3}.
The operators of the form $F\bigl(\phi_{\omega}(x_{1}), \dots,
\phi_{\omega}(x_{n}) \bigr)$ for $x_{i}\in X_{\kappa}$ and $F\in
\coinf(\rr^{n})$ (resp.\ $F\in \coinf(\rr^{n})$ and $F\geq 0$) are
strongly dense in $\cW_\kappa$  (resp.\ in $\cW_\kappa^{+}$).
We have to show that if $(s_{1}, \dots, s_{n})$ is a $n$-tuple such that $s_{1}\leq \cdots\leq s_{n}$
and $s_{n}-s_{1}\leq \beta$, 
and
$A_{i}\in \cW_\kappa^{+}$, then
\beq
\EG ( s_{1}, \dots, s_{n}; A_{1},\dots, A_{n}) \geq 0.
\label{pe1.1}
\eeq
By \cite[Thm.\ 2.2]{KL1} and a density argument it suffices to prove
(\ref{pe1.1}) for $A_{i}$ of the form given above.

Let now $m\in \nn$, $m\geq 1$, $k_{i}\in \nn$ with $k_{i}\geq 1$ 
for $1\leq i\leq n$ and $\sum_{1}^{n}k_{i}=m$, $l_{i}:=
\sum_{j\leq i-1}k_{j}$.
For ${\frak t}= (t_{1}, \dots, t_{m})\in \rr^{m}$, $x_{1}, \dots,
x_{m}\in X_{\kappa}$, and $F_{i}\in \coinf(\rr^{k_{i}})$ with
$F_{i}\geq 0$ we set ${\frak  t}_{i}= (t_{l_{i}}, \dots, t_{l_{i+1}})\in
\rr^{k_{i}}$ and take 
\[
 \begin{array}{rl}
A_{i}=&F_{i} \bigl( \phi_{\omega}(x_{l_{i}}),\dots,
\phi_{\omega}(x_{l_{i+1}}) \bigr)\\
=&(2\pi)^{-k_{i}}\int\hat{F}_{i}(t_{l_{i}}, \dots,
t_{l_{i+1}})W_{\omega} \bigl(\sum_{l_{i}}^{l_{i+1}}t_{j}x_{j} \bigr) \:  \d t_{l_{i}}\dots
\d t_{l_{i+1}}.
\end{array}
\]
Now set $f_{i}({\frak  t}_{i})= \sum_{l_{i}}^{l_{i+1}}t_{j}x_{j}$. It follows that 
\[
\begin{array}{rl}
&\EG (  s_{1}, \dots, 
s_{n}; A_{1}, \dots, A_{n})\\[2mm]=&(2\pi)^{-m}\int \prod_{1}^{n}  \d {\frak 
t}_{i} \: \hat{F}_{i}({\frak t_{i}})G \bigl( \i s_{1},
\dots, \i s_{n}; W(f_{1}({\frak t_{1}})),\dots , W(f_{n}({\frak
t_{n}}))\bigr). 
\end{array}
\]
We recall that by (\ref{pe1.0}) 
\[
\EG \bigl( s_{1},
\dots,  s_{n}; W (f_{1}({\frak t_{1}})),\dots, W ( f_{n}({\frak t_{n}}))\bigr)= \prod_{1\leq
i, j\leq n}\e^{-\frac{1}{2}C(|s_{i}-s_{j}|)( f_{i}({\frak t}_{i}),
f_{j}({\frak t}_{j}))}=:
\e^{-Q({\frak t}_{1}, \dots, {\frak t}_{m})},
\]
where $Q({\frak t}_{1}, \dots, {\frak t}_{m})$ is a quadratic form.
Applying Lemma \ref{p1.0}, we see that $Q$ is a positive quadratic form,
and hence the inverse Fourier transform
${\cal F}^{-1} \bigl(\e^{-Q(\dots)} \bigr)$ is a positive function. This implies
that
\[
\EG( s_{1}, \dots, 
s_{n}; A_{1}, \dots, A_{n}) = (F_{1}\otimes\cdots \otimes F_{n})*{\cal F}^{-1}\bigl(\e^{-Q} \bigr)(0)
\]
is positive as the value at $0$ of the convolution of two positive
functions \qed .

\subsection{Markov property}
In this subsection we show a result which implies that the generalized
path space  associated to the quasi-free KMS system $(\cW, \cW_\kappa, \tau^\circ, \omega)$
considered in Subsection \ref{sec4.1} has the {\em Markov property } (see
Subsection \ref{secst2}).
\begin{lemma}
\label{markov}
Let $X$ be a Hilbert space equipped with a conjugation $\kappa$ and
${\rm a}\geq m>0$ a selfadjoint operator on $X$ such that $[{\rm a},
\kappa]=0$. Let $X_{\kappa}\subset X$ be the real vector space
associated to $\kappa$.

Let $(\cW(X), \tau^\circ, \omega )$ be the quasi-free KMS system
associated to ${\rm a}$ and let $\cW_\kappa\subset \cW$ be the abelian 
von Neumann algebra generated by $\{W_\omega (x)\mid
x\in X_{\kappa}\}$. Let $(\cH_{\omega}, L,
\Omega_{\omega})$ be the
Araki-Woods objects defined in Subsection \ref{sec4.3}.  Then the space
$\{A\e^{-\frac{\beta}{2}L}B\Omega, \: A, B\in \cW_\kappa\}$ is dense in
$\cH_{\omega}$. 
\end{lemma}
\proof The function
\[
\begin{array}{rl}
\e^{\i tL} W_{\omega, l} (y) \Omega_\omega
&= W_{\omega, l} (\e^{\i t{\rm a}}y) \Omega_\omega
\\
&=
W_{F}  \bigl((1+\rho)^{\12} \e^{\i t{\rm a}} y \oplus
(\overline{\rho})^{\12}\e^{-\i t\overline{\rm a}}\overline{y}  \bigr)\\
&=
\e^{\i a_F^*  \bigl((1+\rho)^{\12} \e^{\i t{\rm a}} y \oplus (\overline{\rho})^{\12}\e^{-it\overline{\rm a}}\overline{y}  \bigr)}
\e^{- {1 \over 2} ( y, (1+ 2 \rho) y) } \Omega_\omega 
\end{array}
\]
is analytic in $\{0<{\rm Im z}<\frac{\beta}{2}\}$ and continuous
on $\{0\leq {\rm Im z}\leq \frac{\beta}{2}\}$, and
\[
\begin{array}{rl}
\e^{-\beta L/2} W_{\omega, l} (y) \Omega_\omega
&= \e^{i a_F^*  \bigl((1+\rho)^{\12} \e^{- \beta{\rm a}/2} y \oplus (\overline{\rho})^{\12}\e^{\beta \overline{\rm a}/2}\overline{y}  \bigr)}
\e^{- {1 \over 2} ( y, (1+ 2 \rho) y) } \Omega_\omega \\
&=W_{\omega, r} (y) \Omega_\omega.
\end{array}
\]
Hence, for $A=W_{\omega, {\rm l}}(x)$ and $B= W_{\omega, {\rm r}}(y)$, one
has
\beq
\label{uti}
 A\e^{-\frac{\beta}{2}L}B\Omega= W_{\omega, {\rm l}}(x)W_{\omega, {\rm
r}} (y) \Omega= W_{F}  \bigl( (1+\rho)^{\12}x\oplus
\overline{\rho}\overline{x}  \bigr)W_{F}  \bigl( \rho^{\12}y\oplus
(1+\overline{\rho})^{\12} \: \overline{y}  \bigr)\Omega.
\eeq
Let ${\cal M}$ be the von Neumann algebra  generated by
$\{W_{\omega, {\rm l}}(x), W_{\omega, {\rm r}}(y)\mid x, y\in
X_{\kappa}\}$. By (\ref{uti})  the von
Neumann algebra generated by $\bigl\{W_{F}  \bigl((1+\rho)^{\12}x+
\rho^{\12}y\oplus \overline{\rho}^{\12}\overline{x}+
(1+\overline{\rho})^{\12}\overline{y} \bigr) \mid x,y\in X_{\kappa}\bigr\}$ is 
equal to ${\cal M}$. Since $[a, \kappa]=0$,
the operator
\[
\left(\begin{array}{cc}(1+\rho)^{\12}&\rho^{\12}\\ 
\rho^{\12} &(1+\rho)^{\12}
\end{array}\right):X\oplus X\to X\oplus X
\]
sends $X_{\kappa}\oplus X_{\kappa}$ into itself. It 
is invertible with inverse
\[
\left(\begin{array}{cc}(1+\rho)^{\12}&-\rho^{\12}\\ 
-\rho^{\12} &(1+\rho)^{\12}
\end{array}\right).
\]
Thus ${\cal M}$ is equal to the von Neumann algebra
generated by $\{W_{F}( x\oplus \overline{y}), \: x,y\in X_{\kappa}\}$.
It is well known that if $\ch$ is a Hilbert space and ${\rm c}$ is a
conjugation on $\ch$, then the vacuum vector~$\Omega$ is cyclic in the
Fock space $\G(\ch)$ for the algebra generated by $\{W_{F}(h)\:| \: {\rm
c}h=h\}$ (see e.g.\ \cite[Sect.\ 5.2]{DG} and references therein). We apply
this result to $\ch= X\oplus \overline{X}$, ${\rm c}=
\kappa\oplus\overline{\kappa}$ and obtain the lemma. \qed

\section{Generalized path spaces}
\init\label{path}
In \cite{KL1} a canonical isomorphism is constructed between
a stochastically positive $\beta$-KMS system $(\cW, \cW_\kappa,  \tau^\circ , \omega)$ and
a $\beta$-periodic   {\em stochastic process} $(Q, \Sigma, \mu,X_{t})$
indexed by the circle~$S_{\beta}$ of length ${\beta}$, with values in the
compact Hausdorff space $K= \hbox{\rm Sp} \: (\cW_\kappa)$, the spectrum 
of~$\cW_\kappa$. 

We recall that a {\em
stochastic process} $(Q, \Sigma, \mu,X_{t})$ indexed by an interval
$I\subset \rr$ with values in a topological space $K$ consists of

\vskip .3cm
\halign{ \indent \indent \indent #  \hfil & \vtop { \parindent =0pt \hsize=12cm
                            \strut # \strut} \cr 
{\rm (i)}  & a probability space $(Q, \Sigma, \mu)$;
\cr
{\rm (ii)} & a family $\{X_{t}\}_{t\in I}$ of measurable functions $X_{t}\colon Q\to
K$.
\cr}

\vskip .2cm
\noindent
Typically it is required that the map $I\in t\mapsto X_{t}$
is continuous in measure.

\vskip .3cm
The stochastic process $(Q, \Sigma, \mu,X_{t})$ associated to a
stochastically positive $\beta$-KMS system in \cite{KL1} satisfies four important
properties: {\em stationarity}, {\em symmetry}, $\beta$-{\em
periodicity} and {\em Osterwalder-Schrader positivity} (see
\cite[Sect.\ 4]{KL1}).

It turns out that the only really important feature of such a stochastic
process is the underlying {\em generalized path space}, which consists
of the sub $\sigma$-algebra $\Sigma_{0}$ generated by the functions
$F(X_{0})$ for $F\in C(K)$, the automorphism group $U(t)$ of
$L^{\infty}(Q, \Sigma, \mu)$ generated by the time translations
$U(t)\colon F(X_{t_{1}}, \dots, X_{t_{t}})\mapsto F(X_{t_{1}+t}, \dots,
X_{t_{n}+t})$ for $F\in C(K^{n})$ and the automorphism $R$
of $L^{\infty}(Q, \Sigma, \mu)$ generated by $R\colon F(X_{t_{1}}, \dots,
X_{t_{t}})\mapsto F(X_{-t_{1}}, \dots,
X_{-t_{n}})$. 

In particular the detailed knowledge of the random
variables $X_{t}$ and of the topological space $K$ is not necessary.

(Note that  time translations on  the path space will correspond to
imaginary time translations on the physical Hilbert space).

The analog of the constructions of \cite{KL1} for
$\beta=\infty$ done by Klein in \cite{K} is formulated in terms of
generalized path spaces. Using this essentially equivalent formulation
turns out to be more convenient in applications.
We now proceed to a more precise description of this structure, taken
from \cite{KL1} and~\cite{K}.

If $\Xi_{i}$, for $i$ in an index set $I$, is a family of subsets of a set
$Q$, we denote by $\bigvee_{i\in I} \Xi_{i}$ the $\sigma$-algebra
generated by $\bigcup_{i\in J} U_{i}$, where $U_{i}\in \Xi_{i}$ and $J$  are countable
subsets of $I$.
\begin{definition}
 A {\em generalized path
space} $(Q, \Sigma, \Sigma_{0}, U(t), R, \mu)$ consists of 

\vskip .3cm
\halign{ \indent \indent \indent #  \hfil & \vtop { \parindent =0pt \hsize=12cm
                            \strut # \strut} \cr 
{\rm (i)} & a probability space $(Q, \Sigma, \mu)$;
\cr
{\rm (ii)} & a distinguished sub $\sigma$-algebra $\Sigma_{0}$;
\cr
{\rm (iii)} & a one-parameter group $\rr\ni t\mapsto U(t)$ of measure
preserving $^{*}$-automorphisms
of $L^{\infty}(Q, \Sigma, \mu)$, which is strongly continuous in
measure;
\cr
{\rm  (iv)} & a measure preserving $^{*}$-automorphism $R$ of $L^{\infty}(Q, \Sigma,
\mu)$ such that $RU(t)= U(-t)R$, $R^{2}=\one$, $RE_{0}= E_{0}R$, where
$E_{0}$ is the conditional expectation w.r.t. the $\sigma$-algebra $\Sigma_{0}$.
\cr}
Moreover one requires that
\halign{ \indent \indent \indent #  \hfil & \vtop { \parindent =0pt \hsize=12cm
                            \strut # \strut} \cr 
{\rm (v)} & $\Sigma=\bigvee_{t\in \rr}U(t)\Sigma_{0}$.
\cr}
\end{definition}

It follows from (iii) and (iv) that $U(t)$ extends to a strongly
continuous group of isometries of $L^{p}(Q, \Sigma, \mu)$, 
and $R$ extends to an isometry of $L^{p}(Q, \Sigma, \mu)$, for $1\leq
p<\infty$.

We say that the path space $(Q, \Sigma, \Sigma_{0}, U(t), R, \mu)$ is
$\beta$-{\em periodic} for $\beta>0$ if $U(\beta)=\one$. On a
$\beta$-periodic path space we can consider the one-parameter group
$U(t)$ as indexed by the circle $S_{\beta}=[-\beta/2, \beta/2]$.

For $I\subset \rr$ we denote by $E_{I}$ the conditional expectation
with respect to the $\sigma$-algebra $\Sigma_{I}:=\bigvee_{t\in
I}U(t)\Sigma_{0}$.

\begin{definition}

{\rm ($0$-temperature case):} A path space $(Q, \Sigma, \Sigma_{0}, U(t), R, \mu)$ is {\em
OS-positive} if \hfill\linebreak 
$E_{[0, +\infty[}RE_{[0, +\infty[}\geq 0$ as an operator on $L^{2}(Q,
\Sigma, \mu)$.

{\rm (Positive temperature case:)} A $\beta$-periodic path space $(Q, \Sigma, \Sigma_{0}, U(t), R, \mu)$ is {\em
OS-positive} if $E_{[0,\beta/2]}RE_{[0, \beta/2]}\geq 0$ as an operator on $L^{2}(Q,
\Sigma, \mu)$.
\end{definition}
In order to For simplify the
notation we set  $E_{0}= E_{\{0\}}$, 
$\Sigma_{+}=\Sigma_{[0, +\infty[},\:E_{+}=E_{[0, +\infty[}$,
$\Sigma_{-}=\Sigma_{]-\infty, 0]}$ and $E_{-}= E_{]-\infty, 0]}$.
If the path space $(Q, \Sigma, \Sigma_{0}, U(t), R, \mu)$ is
$\beta$-periodic, we set $\Sigma_{+}=\Sigma_{[0, \beta/2]}, \:E_{+}= E_{[0,\beta/2 ]}$, 
$\Sigma_{-}=\Sigma_{[-\beta/2, 0]}$ and $E_{-}=E_{[-\beta/2, 0]}$.
\begin{definition}
A path space $(Q, \Sigma, \Sigma_{0}, U(t), R, \mu)$ is a {\em
Markov path space} if it has the
\vskip .3cm
\halign{ \indent \indent \indent #  \hfil & \vtop {
\parindent =0pt \hsize=12cm
                            \strut #
\strut} \cr 
{\rm (i)} & {\em reflection property:} $RE_{0}= E_{0}$ (resp.\ $RE_{\{Ê0, \beta/2\} }= E_{\{Ê0, \beta/2\} }$); 
\cr
{\rm (ii)} & {\em Markov property:} $E_{+} E_{-}= E_{+} E_{0} E_{-}$ (resp.\ $E_+ E_-= E_+ E_{\{Ê0, \beta/2\} }E_-$).
\cr}
\vskip .2cm
\noindent
It follows that $E_{+}RE_{+}= E_{-}E_{+}= E_{+}E_{-}= E_{0}$ (resp.\ $E_{+}RE_{+}= E_{-}E_{+}= E_{+}E_{-}= E_{\{Ê0, \beta/2\} }$) .
\end{definition}
A Markov path space is OS-positive because $E_{0} $ (resp.\ $E_{\{Ê0, \beta/2\} }$) is positive as an orthonormal projection.
An OS-positive path space satisfies the reflection property (see \cite[Prop.\
1.6]{K}).

Let $(\cF, \cU, \tau, \omega)$ be a stochastically positive $\beta$-KMS system.
Let $K:={\rm Sp} (\cU)$ be the spectrum of the abelian $C^{*}$-algebra
$\cU$, which equipped with the weak topology is a compact Hausdorff
space. Let $Q:=K^{[-\beta/2, \beta/2]}$ be equipped with the product
topology and  let $\Sigma$ be the Baire $\sigma$-algebra on $Q$.
\begin{theoreme}{\rm \cite{KL1}.}
Let $(\cF, \cU, \tau, \omega)$ be a stochastically positive
$\beta$-KMS system. Then there exists a Baire probability measure
$\mu$ on $Q$, a sub $\sigma$-algebra $\Sigma_{0}\subset \Sigma$, a
measure preserving group $U(t)$ of $^{*}$-automorphisms of
$L^{\infty}(Q, \Sigma, \mu)$ and a measure preserving automorphism~$R$
of $L^{\infty}(Q, \Sigma, \mu)$ such that $(Q, \Sigma, \Sigma_{0},
U(t), R, \mu)$ is an OS-positive $\beta$-periodic generalized path
space.
\end{theoreme}
A more precise relationship between the $\beta$-KMS system and the
generalized path space will be given in Theorem \ref{stp1.4}.

\section{Reconstruction theorems}
\label{secst1}
\init
In this section we recall  reconstruction theorems of Klein \cite{K}
and Klein and
Landau \cite{KL1} which associate a stochastically positive $\beta$-KMS system
to an OS-positive generalized path space $(Q, \Sigma, \Sigma_{0}, U(t), R, \mu)$.

To simplify notation, we allow the
parameter $\beta$ to take values in $]0, +\infty]$. The case~$\beta=+\infty$ corresponds to the $0$-temperature
case. If $\beta<\infty$, then the OS-positive path spaces will be
assumed to be $\beta$-periodic. 

\subsection{Physical Hilbert space}
Set $\cH_{OS}:= L^{2}(Q, \Sigma_{+}, \mu)$ and
\[
(F, G):= \int_{Q}R(\overline{F})G\d \mu,\: F, G\in \cH_{OS}.
 \]
By OS-positivity 
\[
 0\leq (F, F)\leq \|F\|^{2}_{\cH_{OS}}.
\]
If we set ${\cal N}:={\rm Ker}E_{+}RE_{+}$, then $(\cdot, \cdot)$ is a
positive definite sesquilinear form on $\cH_{OS}/{\cal N}$. 

The {\em physical Hilbert space}, denoted by $\cH_{\rm phys}$ (or
simply by $\cH$) is
\[
\cH :=\hbox{ completion of }\cH_{OS}/{\cal N} \hbox{ for
}(\cdot, \cdot).
\]
If we denote by ${\cal V}\colon \cH_{OS}\to \cH_{OS}/{\cal N}$ the canonical
projection, then ${\cal V}$ extends uniquely to a contraction with
dense range: $\cH_{OS}\to \cH$. In fact  
\[
({\cal V}F, {\cal V}F)= (F, F) \leq
\|F\|^{2}_{\cH_{OS}}.
\]
In the physical Hilbert space $\cH$ we find a {\em
distinguished vector}\/:
\[
\Omega:= \cV (1).
\]
\subsection{Selfadjoint operator}
{\bf The $0$-temperature case}
\begin{proposition}{\rm \cite[Thm.\ 1.7]{K}}.
Let $(Q, \Sigma, \Sigma_{0}, U(t), R, \mu)$ be an OS-positive generalized path
space. For $t\geq 0$ the time evolution $U(t)$ maps $ {\cal N}\to {\cal N}$. Hence the 
linear operator
\[
P(t)\colon \cH_{\rm OS}/{\cal N}\ni{\cal V}(F)\mapsto {\cal V}(U(t)F)\in
\cH_{\rm OS}/{\cal N}
\]
is well defined  for $t\geq 0$. 

The family $\{P(t)\}_{t\geq 0}$ uniquely extends to a
strongly continuous selfadjoint semigroup of contractions
$\{\e^{-tH}\}_{t\geq 0}$ on $\cH$, where $H$ is a positive
selfadjoint operator such that $H\Omega=0$.
\end{proposition}

\noindent
{\bf The positive temperature case}

\vskip .3cm
\noindent
We first recall the definition of a local symmetric semigroup
(\cite{KL3}, \cite{Fr}):
\begin{definition}
\label{st1.2ter}
Let $\cH$ be a Hilbert space and $T>0$. A {\em local symmetric
semigroup} \ \ $(P(t), {\cal D}_{t}, T)$ is a family 
$\{P(t), \cD_{t}\}_{t\in [0, T]}$ of linear operators $P(t)$ and vector subspaces $\cD_{t}$ 
of~$\cH$ such that

\vskip .3cm
\halign{ \indent \indent \indent #  \hfil & \vtop { \parindent =0pt \hsize=12cm
                            \strut # \strut} \cr 
{\rm (i)} & $D_{0}=\cH$, $\cD_{t}\supset\cD_{s}$ if $0\leq t\leq s\leq T$ and $\cD=
\cup_{0<t\leq T}D_{t}$ is dense in $\cH$;
\cr
{\rm (ii)} & $P(t)\colon \cD_{t}\to \cH$ is a symmetric linear operator with
$P(0)=\one$, $P(s)\cD_{t}\subset \cD_{t-s}$ for $0\leq s\leq t\leq T$ and
$P(t)P(s)= P(t+s)$ on $\cD_{t+s}$ for $t,s ,t+s\in [0, T]$.
\cr
{\rm (iii)} & $t\mapsto P(t)$ is weakly continuous, \ie for $u\in \cD_{s}$
and $0\leq t\leq s$ the map 
$t\mapsto (u, P(t)u)$ is continuous.
\cr}
\end{definition}
The following theorem was shown in \cite{KL3} and \cite{Fr}.
\begin{theoreme}
Let $(P(t), {\cal D}_{t}, T)$ be a local symmetric semigroup on $\cH$. Then there
exists a unique selfadjoint operator $L$ on $\cH$ such that

\vskip .3cm
\halign{ \indent \indent \indent #  \hfil & \vtop { \parindent =0pt \hsize=12cm
                            \strut # \strut} \cr 
{\rm (i)} & $\cD_{t}\subset \cD(\e^{-tL})$, $\e^{-tL}_{|\cD_{t}}=P(t)$ for
$0\leq t\leq T$;
\cr
{\rm (ii)} & $\cD_{]0,T']}:= \cup_{0<t\leq T'}\cup_{0<s<t}P(s)\cD_{t}$
is a core for $L$ for   $0<T'\leq T$.
\cr} 
\label{st1.2b}
\end{theoreme}

\begin{proposition}{\rm \cite[Lemma 8.3]{KL1}}.
Let $(Q, \Sigma, \Sigma_{0}, U(t), R, \mu)$ be a $\beta$-periodic OS-positive path
space.
Set ${\cal M}_{t}:= L^{2}(Q, \Sigma_{[0, \beta/2-t]}, \mu)$ for $0\leq
t\leq \beta/2$. Then

\vskip .3cm
\halign{ \indent \indent \indent #  \hfil & \vtop { \parindent =0pt \hsize=12cm
                            \strut # \strut} \cr 
{\rm (i)} & $U(s)\colon {\cal M}_{t}\cap
{\cal N}\to {\cal M}_{t-s}\cap {\cal N}$ for $0\leq s\leq  t\leq \beta/2$. If $D_{t}:=
{\cal V}({\cal M}_{t})$, then the linear operator
\[
\matrix{
P(s)\colon & \cD_{t} & \to &\cD_{t-s},\cr
&\cV (F) & \mapsto & \cV( U(s)F)
\cr}
\]
is well defined;
\cr
{\rm (ii)} & $(P(t), \cD_{t}, \beta/2)$ is a local symmetric semigroup.
\cr}
\label{st1.3}
\vskip .2cm
\noindent
By Theorem \ref{st1.2b} there exists a unique selfadjoint operator
$L$ such that $P(t)_{|\cD_{t}}= \e^{-tL}$. Moreover $L\Omega=0$.
 \end{proposition}

\subsection{Algebras of operators}
\label{itit}
\noindent
 {\bf  Abelian $C^{*}$-algebra  ${\cal U}$} 
\vskip .2cm
\noindent
Let $f\in L^{\infty}(Q, \Sigma_{0}, \mu)$. Since $\Sigma_{0}\subset
\Sigma_{+}$, $f$ acts as a multiplication operator on
$\cH_{OS}$, which we will still denoted by $f$.
\begin{proposition}{\rm \cite[Lemma 2.2]{KL1}}.
For $f\in L^{\infty}(Q, \Sigma_{0}, \mu)$ the multiplication operator $f$ preserves ${\cal N}$. 
Hence
\[
\tilde{f}{\cal V}(F):= {\cal V}(fF) 
\]
defines a unique element of $\cB(\cH)$ with
$\|\tilde{f}\|= \|f\|_{\infty}$.
Let ${\cal U}\subset \cB(\cH)$ be defined by
\[
{\cal U}:=\{\tilde{f}\mid f\in L^{\infty}(Q,\Sigma_{0}, \mu)\}.
\]
Then ${\cal U}$ is a von Neumann algebra isomorphic to $L^{\infty}(Q,
\Sigma_{0}, \mu)$ and $\Omega$ is a separating vector for~${\cal U}$.
\label{st1.2}
\end{proposition}
We will denote by ${\cal U}^{+}$  the
set of positive elements in ${\cal U}$.

\medskip
\noindent
{\bf Full algebra $\cF$ and automorphism group}
\begin{definition}
Let $\cF\subset \cB(\cH)$ denote the von Neumann algebra generated by 
$\{ \e^{\i tH}A\e^{-\i tH}  \mid   A \in \cU, t \in \rr \} $ for $\beta= \infty$ (resp.\ $\{ 
\e^{\i tL}A\e^{-\i tL} \mid   A \in \cU, t \in \rr \} $ for $\beta< \infty$). We denote 
by~$\{\tau_{t}\}_{t\in \rr}$ the strongly continuous group of
automorphisms of $\cF$ defined by $\tau_{t}(B)= \e^{\i tH}B\e^{-\i
tH}$ for $B\in \cF$, $t\in \rr$ and $\beta= \infty$ (resp.\ 
$\tau_{t}(B)= \e^{\i tL}B\e^{-\i
tL}$ for $B\in \cF$, $t\in \rr$ and $\beta< \infty$).
\end{definition}

\subsection{$\beta$-KMS system associated to a $\beta$-periodic path
space}
In case $\beta<\infty$ one can associate to a $\beta$-periodic OS
positive path space a stochastically positive $\beta$-KMS system (see
\cite{KL1}). (The analog object in case $\beta=\infty$ is called a
{\em positive semigroup structure} \cite{K}). 
Let, for $n\in \nn$ and $\beta>0$, 
\[
J_{\beta}^{n+}:= \{(t_{1}, \dots,
t_{n})\in \rr^{n}\mid t_{i}\geq 0, \: t_{1}+ \cdots+ t_{n}\leq
\beta/2\}.\]
\begin{theoreme}{\rm \cite{KL1}}.
\label{stp1.4}
Let $L$ be the selfadjoint operator associated to the local symmetric
semigroup $(P(t), \cD_{t}, \beta/2)$. It follows that

\vskip .3cm
\halign{ \indent \indent \indent #  \hfil & \vtop { \parindent =0pt \hsize=12cm
                            \strut # \strut} \cr 
{\rm (i)} & $\Omega\in \cD(L)$ and $L\Omega=0$;
\cr
{\rm (ii)} & if $n\in \nn$, $(t_{1}, \dots, t_{n})\in
J_{\beta}^{n+}$  and  $A_{1}, \dots, A_{n}\in {\cal U}$, then
$A_{n} \bigl( \prod_{n-1}^{1}\e^{-t_{j}L}A_{j} \bigr) \Omega\in \cD(\e^{-t_{n}L})$.
The vector span of  these vectors is dense in $\cH$;
\cr
{\rm (iii)} & if $f_{1}, \dots, f_{n}\in L^\infty ( Q, \Sigma_0, \mu)$ and $0\leq s_{1}\leq
\cdots\leq s_{n}\leq \beta/2$, then
\[
 \cV \bigl(\prod_{1}^{n}U(s_j) f_{j} \bigr)=
\e^{-s_{1}L}\tilde{f}_{1} \bigl(\prod_{2}^{n}\e^{-(s_{j}-
s_{j-1})L}\tilde{f}_{j} \bigr) \Omega,
\]
where $\tilde{f}_{j}$ is defined in Proposition \ref{st1.2}.
\cr
{\rm (iv)} & if $n\in \nn$, $(t_{1}, \dots, t_{n})\in J_{\beta}^{n+}$ and
$A_{1}, \dots, A_{n}$, $B_{1}, \dots, B_{n}\in \cU^{+}$, then
\[
\Bigl(
A_{n} \bigl( \prod_{n-1}^{1}\e^{-t_{j}L}A_{j} \bigr) \Omega \, , \,
B_{n} \bigl( \prod_{n-1}^{1}\e^{-t_{j}L}B_{j} \bigr) \Omega \Bigr)\geq 0;
\]
\cr
{\rm (v)} & $\|\e^{-\beta/2 L}A\Omega\|=\|A^{*}\Omega\|$ for all $A\in
\cU$.
\cr}
\end{theoreme}
 \begin{theoreme}{\rm \cite{KL1}}.
Let  $\omega_\Omega$ be the state on $\cF$ defined by $\omega_\Omega (B)= (\Omega,
B\Omega)$. Then $(\cF, \cU, \tau, \omega_\Omega)$ is a stochastically
positive $\beta$-KMS system.
\label{st1.5}
\end{theoreme}
Finally let $J$ be the modular conjugation associated to the KMS
system $(\cF, \tau , \omega_\Omega)$. 
\begin{proposition}{\rm \cite{KL1}}.
 The modular conjugation $J$ is the unique extension of 
\beq
J\cV(F)=
\cV(R_{\beta/4}\overline{F}),
\label{est1.2}
\eeq 
where
 \[
R_{\beta/4}:=U(\beta/4)RU(-\beta/4)= RU(-\beta/2)= U(\beta/2)R
\]
 is the reflection at $t=\beta/4$ in
$\cH_{OS}$.
\end{proposition}
\subsection{Markov property for $\beta$-periodic path spaces}
\label{secst2}
We recall a characterization of the Markov property for a
$\beta$-periodic path space in terms of the associated stochastically positive
$\beta$-KMS system due to Klein and Landau \cite{KL1}.
\begin{theoreme}
\label{st2.2}
A $\beta$-periodic OS-positive path space $(Q, \Sigma, \Sigma_{0}, U(t), R, \mu)$ 
satisfies the Mar\--kov
property iff the vectors $A\e^{-\frac{\beta}{2}L}B\Omega$ for $A, B\in
\cU$ are dense in $\cH$. In this case   \[
\cH =L^{2}( Q, \Sigma_{\{0, \beta/2\}}, \mu) .\]
\end{theoreme}
\proof 
The first statement of the theorem is shown in \cite[Thm.\ 11.2]{KL1}.
The second statement is obvious: it follows from the Markov property
that $E_{[0, \beta/2]}RE_{[0, \beta/2]}=E_{\{0, \beta/2\}}$ is  a projection, hence
$\cH_{OS}/{\cal N}$ is canonically identified with $E_{\{0,
\beta/2\}}\cH_{OS}= L^{2}( Q, \Sigma_{\{0, \beta/2\}}, \mu)$. 

\begin{theoreme}
Let $(\cW, \cW_\kappa, \tau^\circ, \omega_\beta )$ be the quasi-free KMS system
associated to a selfadjoint operator ${\rm a}\ge0$ and a conjugation
$\kappa$ with $[a, \kappa]=0$.  
Then the OS-positive generalized path space $(Q, \Sigma, \Sigma_{0},
U(t), R, \mu)$ associated to  $(\cW(X), \cW_\kappa (X), \tau^\circ, \omega_\beta  )$
satisfies the Markov property.
\end{theoreme}
\proof Stochastic positivity of the quasi-free KMS system  $(\cW, \cW_\kappa, \tau^\circ, 
\omega_\beta  )$ 
was shown in Theorem 4.5.
The Markov property follows from Lemma 4.6 and Theorem 6.10
\qed .

\section{Perturbations of generalized path spaces}
\init\label{secst3}
In this section we recall some results concerning
perturbations of OS-positive path spaces. 

\subsection{FKN kernels}
\label{fkn}
Let $(Q, \Sigma, \Sigma_{0}, U(t), R, \mu)$ be an OS-positive path
space.
\begin{definition}
\label{fkn1}
A {\em Feynman-Kac-Nelson} (FKN) {\em kernel} is a family $\{F_{[a,
b]}\}$ of real measurable functions on $(Q, \Sigma, \mu)$ such that,
for $0\leq
b-a\leq \beta$,

\vskip .3cm
\halign{ \indent \indent \indent #  \hfil & \vtop { \parindent =0pt \hsize=12cm
                            \strut # \strut} \cr 
{\rm (i)} & $F_{[a,b]}>0$ $\mu$-a.e.;
\cr
{\rm (ii)} & $F_{[a,b]}\in L^{1}(Q, \Sigma, \mu)$ and $F_{[a,b]}$ is continuous in
$L^{1}(Q, \Sigma, \mu)$ as a function of $b$;
\cr
{\rm (iii)} & $  F_{[a,b]}F_{[b,c]}= F_{[a,c]}$ for $a\leq b\leq c$,
$c-a\leq \beta$;
\cr
{\rm (iv)} & $U(s)F_{[a,b]}= F_{[a+s, b+s]}$ for $s\in \rr$;
\cr
{\rm (v)} & $RF_{[a,b]}=F_{[-b, -a]}$.
\cr}
\end{definition}

The main examples of FKN kernels are those associated to 
a selfadjoint operator $V$ affiliated to $\cU$. In
\cite{KL1} and \cite{K} perturbations associated to more general FKN kernels are
considered. However, the present case is sufficient for our applications.

Let $V$ be a selfadjoint operator
affiliated to $\cU$. Since by Proposition 6.5 the algebra
$\cU$ is
isomorphic to $L^{\infty}(Q, \Sigma_{0}, \mu)$, we can uniquely associate to $V$
a real function on $Q$, measurable with respect to $\Sigma_{0}$, which we will still denote by $V$.
\begin{proposition}
\label{st3.1b}
Let $(Q, \Sigma, \Sigma_{0}, U(t), R, \mu)$ be a $\beta$-periodic OS-positive 
path space and let
$V$ be a selfadjoint operator affiliated to $\cU$ such
that $V\in L^{1}(Q, \Sigma_{0}, \mu)$, and $\e^{-TV}\in L^{1}(Q,
\Sigma_{0}, \mu)$ for some $T>0$ if $\beta=\infty$ or $\e^{-\beta
V}\in L^{1}(Q, \Sigma_{0}, \mu)$ if $\beta<\infty$. Then 

\vskip .3cm
\halign{ \indent \indent \indent #  \hfil & \vtop { \parindent =0pt \hsize=12cm
                            \strut # \strut} \cr 
{\rm (i)} & the family of
functions  $F_{[a,
b]}:=\e^{-\int_{a}^{b}U(t)V\d t}$ for $0\leq b-a\leq{\rm inf}(T,
\beta)/2$ is a FKN kernel; 
\cr
{\rm (ii)} & $F_{[0, s]}\in L^{2}(Q, \Sigma_{[0, s]},\mu)$ for
$0\leq s\leq {\rm inf}(T, \beta)/2$ and the map $s \mapsto F_{[0, s]}$ is continuous in $L^{2}(Q, \Sigma_{[0,
\beta/2]}, \mu)$.
\cr}
\end{proposition}
\proof 
All properties required in Definition
\ref{fkn1} except from property {\rm (ii)} follow directly from the
definition of $U(t)$ and the properties of the path space $(Q, \Sigma,
\Sigma_{0}, U(t), R, \mu)$. Let us now verify {\rm (ii)}.
 Writing $V=
V_{+}-V_{-}$, where $V_{\pm}$ is the positive/negative part of $V$, we
have $F_{[a,b]}\leq \exp \bigl( \int_{a}^{b}U(t)V_{-}\d t \bigr)$, and hence
$F_{[0, s]}^{2}\leq \exp \bigl(2\int_{0}^{\beta/2}U(t)V_{-}\d t \bigr)$. 
Since $\mu$ is a probability measure, we have
$V_{-}, \:\e^{\beta V_{-}}\in L^{1}(Q, \Sigma_{0},\mu)$. 
We recall the following bound from \cite[Thm.~6.2 (i)]{KL0}:
\beq
\|\e^{-\int_{a}^{b} U(t)V\d t}\|_{L^{p}(Q, \Sigma, \mu)}\leq
\|\e^{-(b-a)V}\|_{L^{p}(Q,\Sigma, \mu)},\: 1\leq p<\infty.
\label{bound}
\eeq
This yields
\[
\|F_{[0, s]}^{2}\|_{L^{1}(Q, \Sigma, \mu)}\leq
\|\e^{2\int_{0}^{\beta/2}U(t)V_{-}\d t}\|_{L^{1}(Q, \Sigma, \mu)}\leq
\|\e^{\beta V_{-}}\|_{L^{1}(Q, \Sigma,\mu)}<\infty.
\]
Hence $F_{[0,s]}\in L^{2}(Q, \Sigma_{[0, \beta/2]}, \mu)$ for $0\leq
s\leq {\rm inf}(T, \beta)/2$. The continuity w.r.t.\ to $s$ follows from the
dominated convergence theorem. This completes the proof of {\rm (ii)}.

The proof of property {\rm (ii)} from Definition \ref{fkn1} for $0\leq
a$ follows from {\rm (ii)} and
the fact that $L^{2}(Q, \Sigma, \mu)\subset
L^{1}(Q, \Sigma, \mu)$. The case $b\leq 0$ is reduced to the case
$a\geq 0$ using property {\rm (v)}. Finally the case $a<0<b$ follows
from the identity $F_{[a,b]}= F_{[a, 0]}F_{[0, b]}$ \qed .

\subsection{Selfadjoint operator associated to a FKN kernel}
In this subsection we recall a result of Klein and Landau \cite{KL1},
allowing us to construct a selfadjoint operator starting from a FKN
kernel associated to a selfadjoint operator $V$, which is affiliated to $\cU$.
To keep the exposition compact, we will use the convention for the
parameter~$\beta$ explained at the beginning of Section \ref{secst1}.

Let $(Q, \Sigma, \Sigma_{0}, U(t), R, \mu)$ be an  OS
positive path space and $V$ a selfadjoint operator affiliated to $\cU$
such that $V\in L^{1}(Q, \Sigma, \mu)$ and $\e^{-TV}\in L^{1}(Q,
\Sigma_{0}, \mu)$ for some $T>0$. Let~$F_{[a, b]}$ be the
associated FKN kernel. 

Let, for $0< t<T/2$,  ${\cal M}_{t}$ be the linear span of $\bigcup_{0\leq
s\leq T/2-t}F_{[0,
s]}L^{\infty}(Q, \Sigma_{[0, T/2 -t]}, \mu)$. Set 
\[
\matrix{ U_{V}(s)\colon &
{\cal M}_{t} & \to & L^{2}(Q, \Sigma_{+}, \mu)\cr
& \psi & \mapsto & F_{[0,s]}U(s)\psi,\: 
\cr} \quad 0\leq s\leq t.
\]

\begin{lemma}
\label{pt2} \
\vskip .3cm
\halign{ \indent \indent \indent #  \hfil & \vtop { \parindent =0pt \hsize=12cm
                            \strut # \strut} \cr 
{\rm (i)} & For $\psi \in {\cal M}_{t}$ the map
\[
[0, t]\ni s\mapsto U_{V}(s)\psi\in L^{2}(Q, \Sigma_{+}, \mu)
\]
is  continuous on $[0, t]$.
\cr
{\rm (ii)} & $U_{V}(s)\colon {\cal M}_{t}\cap {\cal N}\to {\cal N}$ for $0\leq
s\leq t<T/2$.
\cr}
\end{lemma}
\proof Using the definition of ${\cal M}_{t}$ and the properties of
the FKN kernel $F_{[a,b]}$  it suffices to show that for  
$\psi\in L^{\infty}(Q, \Sigma_{[0, T/2 -t]}, \mu)$ the map $s\to U_{V}(s)\psi$ is
 continuous at $s=s'$, $0<s'\leq t<T/2$. 
For
$0\leq s, s'\leq  t<T/2$ we have  \[
U_{V}(s')\psi- U_{V}(s)\psi= F_{[0,s']}\bigl(U(s')\psi-
U(s)\psi\bigr)+ \bigl(F_{[0,s']}- F_{[0,s]}\bigr)U(s)\psi.
\] Hence
\[
\begin{array}{rl}
\|U_{V}(s')\psi - U_{V}(s)\psi\|_{2}^{2} \leq &\int_{Q}F_{[0,
s']}^{2}|U(s')\psi- U(s)\psi|^{2}\d \mu+ \int_{Q}(F_{[0,s]}- F_{[0,
s']})^{2}|U(s)\psi|^{2}\d \mu\\
\leq &\int_{\{|U(s')\psi- U(s)\psi|(q)>\epsilon\}}F_{[0,
s']}^{2}|U(s')\psi- U(s)\psi|^{2}\d \mu\\
&+ \int_{\{|U(s')\psi-
U(s)\psi|(q)\leq\epsilon\}}F_{[0,
s']}^{2}|U(s')\psi- U(s)\psi|^{2}\d \mu \\
&+  \|F_{[0,s']}- F_{[0,
s]}\|^{2}_{2}\|\psi\|^{2}_{\infty}.
\end{array}
\]
The last term on the r.h.s.\ tends to $0$ if $s\to s'$ as a consequence of Proposition
\ref{st3.1b}. The second term on the r.h.s.\ is less than
$\epsilon^{2}\|F_{[0,s']}\|^{2}_{2}$. To estimate the first term, we
write the function $f:= F_{[0,s']}^{2}$ as $f\one_{\{|f(q)|\leq M\}}+
f\one_{\{|f(q)|>M\}}$. It follows that
\[
\begin{array}{rl}
&\int_{\{|U(s')\psi- U(s)\psi|(q)>\epsilon\}}  f
|U(s')\psi - U(s)\psi|^{2}\d \mu \\[2mm]
\leq &4M\|\psi\|^{2}_{\infty} \int\one_{\{|U(s')\psi-
U(s)\psi|(q)>\epsilon\}}\d\mu + 4\|f\one_{\{|f(q)|>M\}}\|_{1}\|\psi\|^{2}_{\infty}.
\end{array}
\]
Since $f\in L^{1}(Q, \Sigma_{+}, \mu)$, the second term tends to $0$
as $M\to \infty$. Since $U(t)$ is strongly continuous in measure,
the first term tends to $0$ as $s\to s'$. Picking first $\epsilon\ll
1$, then $M\gg 1$ and finally $|s-s'|\ll 1$ we obtain {\rm (i)}.

Let us now prove {\rm (ii)}. Let $0\leq s\leq t<T/2$. 
Note that $U_{V}(s)$ sends
${\cal M}_{s}$ into $L^{2}(Q, \Sigma_{+}, \mu)$. Let us fix $\psi\in
{\cal M}_{t}$. First we consider the case $s<t$.
For $0<r\leq s$ and $s+r\leq t$ we have 

\[
\begin{array}{rl}
(U_{V}(s)\psi, U_{V}(s)\psi)&=\int_{Q}F_{[0,
s]}U(s)\overline{\psi}RF_{[0,s]}U(s)\psi\d\mu\\
&=\int_{Q}F_{[-r,s-r]}U(s-r)\overline{\psi}U(-r)RF_{[0,s]}U(s)\psi \, \d\mu\\
&=\int_{Q}F_{[-r,s-r]}U(s-r)\overline{\psi}RF_{[r,s+r]}U(s+r)\psi  \, \d\mu\\
&=\int_{Q}F_{[0,s-r]}U(s-r)\overline{\psi}RF_{[0,
s+r]}U(s+r)\psi  \, \d\mu\\
&=(U_{V}(s-r)\psi, U_{V}(s+r)\psi).
\end{array}
\]
Since $\hf$ is positive, the Cauchy-Schwartz inequality implies
\[
\bigl(U_{V}(s)\psi, U_{V}(s)\psi\bigr)\leq \bigl(U_{V}(s-r)\psi,
U_{V}(s-r)\psi\bigr)^{\12}\bigl(U_{V}(s+r)\psi, U_{V}(s+r)\psi\bigr)^{\12}.
\]
Thus, by induction,
\[
\bigl(U_{V}(s)\psi, U_{V}(s)\psi \bigr)\leq
\|U_{V}(s-nr)\psi\|\prod_{j=0}^{n-1}\bigl(U_{V}(s-(j-1)r)\psi,
U_{V}(s-(j-1)r)\psi\bigr)^{\12}.
\]
If we pick $0<r<s$, $s=nr$, such that $s+r\leq t$, then
$(\psi, \psi)=0$ implies $\bigl(U(s)\psi, U(s)\psi\bigr)=0$. Finally, {\rm (ii)} for
$s=t$ follows from {\rm (ii)} for $s<t$ and {\rm (i)}  \qed .

\begin{theoreme}
\label{pt1}
Let $0<t<T/2$, $\cD_{t}= {\cal V}({\cal M}_{t})$ and $0\leq s\leq t$. Then
\[
\matrix{P_{V}(s)\colon &
\cD_{t} & \to & \cH \cr
&\cV(\psi) & \mapsto & \cV(F_{[0,s]}U(s)\psi) 
\cr} 
\]
is a well defined linear operator, and $(\cD_{t}, P_{V}(t), T/2)$ is
a local symmetric semigroup on~$\cH$. We denote by $H_{V}$
the associated selfadjoint operator.
\end{theoreme}
\proof The fact that $P_{V}(s)$ is well defined follows from Lemma 7.3 (ii).
Property {\rm (ii)} of Definition \ref{st1.2ter} follows
from the properties of the FKN kernel $F_{[a,b]}$. Monotonicity of the
family $\{\cD_{t}\}$ w.r.t.\ inclusions is immediate. That $\cD=
\cup_{0<t\leq T}D_{t}$ is dense  in $\cH$ follows from the fact that $\cD$
contains ${\cal V}\bigl(L^{\infty}(Q, \Sigma_{+}, \mu) \bigr)$. Finally property
{\rm (iii)} follows from the continuity property stated in Lemma \ref{pt2} 
\qed .

\medskip

\begin{theoreme}{\rm \cite[Thm.\ 16.4]{KL1}}.
\label{pt3}
Let $V$ be a selfadjoint operator affiliated to $\cU$ such
that $V\in L^{1}(Q, \Sigma_{0}, \mu)$ and $\e^{-TV}\in L^{1}(Q,
\Sigma_{0}, \mu)$ for some $T>0$. Assume in addition that either $V\in
L^{2+\epsilon}(Q, \Sigma_{0}, \mu)$ for $\epsilon>0$ or that $V\in
L^{2}(Q, \Sigma_{0}, \mu)$ and $V\geq 0$. Let, for $\beta = \infty$, $H$ (resp.~$L$ 
for $\beta < \infty$)
denote the selfadjoint generator of the unperturbed
semi-group $t \mapsto P(t)$.
Then $H+V$ (resp.\ $L+V$) is essentially
selfadjoint and the operator $H_{V}$ (for both cases) constructed in Theorem \ref{pt1}
is equal to $\overline{H+V}$ (resp.\ $\overline{L + V}$).
\end{theoreme}

\subsection{Perturbations in the positive temperature case}
\label{otot}
The following theorem is shown in \cite{KL1}:
\begin{theoreme}{\rm \cite{KL1}}.
\label{st3.2}
Let $(Q, \Sigma, \Sigma_{0}, U(t), R, \mu)$ be a $\beta$-periodic 
OS-positive path space, $V$ a selfadjoint operator on $\cH$
affiliated to $\cU$, which  satisfies the hypotheses of Proposition
\ref{st3.1b}. Let
$F=\{F_{[a,b]}\}$ be the associated
$\beta$-periodic FKN kernel. Then the path space $(Q, \Sigma,
\Sigma_{0}, U(t), R, \mu_{V})$,
 where 
\[\d\mu_{V}:= {F_{[-\beta/2,\beta/2]}\d\mu \over \int_Q   F_{[-\beta/2,
\beta/2]} \d \mu \, ,}
\] is a $\beta$-periodic OS-positive path space.
\end{theoreme}
By the reconstruction theorem  recalled in Section \ref{secst2}, one can
associate to the perturbed path space  $(Q, \Sigma, \Sigma_{0}, U(t),
R, \mu_{V})$  a
physical Hilbert space $\cH_{V}$, a distinguished vector $\Omega_{V}$,
an abelian von Neumann algebra $\cU_{V}$, a selfadjoint operator
$L_{V}$ and a von Neumann algebra $\cF_{V}$. If $\omega_{V}$ and
$\tau_{V}$
are the state and $W^{*}$-dynamics associated to $\Omega_{V}$ and $ L_{V}$,
then 
$(\cF_{V},\cU_{V}, \tau_{V}, \omega_{V})$ is a stochastically positive $\beta$-KMS
system.

Our next aim is to construct
canonical identifications
between the perturbed objects and perturbations of the original objects associated to
the path space $(Q, \Sigma, \Sigma_{0}, U(t), R, \mu)$.

\medskip
\noindent
{\bf Identification of the physical Hilbert spaces}
 
\medskip
\noindent
We first show that 
there is a canonical unitary operator between $\cH_{V}$
and $\cH$.
\begin{proposition}
\label{st3.3}
Assume that $V, \:\e^{-\beta V}\in L^{1}(Q, \Sigma_{0}, \mu)$. Set 
\[
\matrix{
\hat{I}\colon & L^{\infty}( Q, \Sigma_{+}, \mu)/\cN_{V} & \to & \cH_{OS}/\cN \cr
&&\cr
& \cV_{V}(\psi) & \mapsto & {\cV(F_{[0,
\beta/2]}\psi) \over \bigl(\int_Q   F_{[-\beta/2,
\beta/2]} \d \mu\bigr)^{\12}}.
\cr}
\]
Then $\hat{I}$ is a well defined  isometry from $\cH_{OS,
V}/\cN_{V}$ into $\cH_{OS}/\cN$ with dense range and domain. Hence
$\hat{I}$ uniquely extends to a unitary map $\hat{I} \colon \cH_{ V}\to \cH$.
\end{proposition}
\proof Note that $\mu_{V}$ is absolutely continuous w.r.t.\
$\mu$. Thus $L^{\infty}(Q, \Sigma, \mu_{V})= L^{\infty}(Q, \Sigma, \mu)$.
If $\psi\in L^{\infty}(Q, \Sigma, \mu)\cap \cN_{V}$, then $\int_Q  \, 
R\overline{\psi}\psi \d \mu_{V}= \int_Q \d \mu R\overline{F_{[0,
\beta/2]}\psi}F_{[0, \beta/2]}\psi =0$. Hence~$F_{[0,
\beta/2]}\psi\in \cN$. Consequently $\hat{I}$ is well defined. $\hat{I}$ is
clearly isometric since
\[
(\cV_{V}\psi, \cV_{V}\psi)_{V}= \frac{\int_Q R\overline{\psi}\psi \d \mu_{V} }{ \int_Q F_{[-\beta/2,
\beta/2]} \d \mu } =\frac{  \int_Q R\overline{F_{[0, \beta/2]}\psi }F_{[0,
\beta/2]}\psi \d \mu }{ \int_Q F_{[-\beta/2,
\beta/2]} \d \mu } =(\hat{I}\cV_{V}\psi, \hat{I}\cV_{V}\psi).
\]
$\hat{I}$ is densely defined since $L^{\infty}(Q, \Sigma_{+}, \mu)$ is
dense in $\cH_{OS, V}$. Since $\cV_{V}$ is a contraction,
$L^{\infty}(Q, \Sigma_{+}, \mu)/\cN_{V}$ is dense in $\cH_{OS, V}/\cN_{V}$
and hence in $\cH_{ V}$. Finally, we note that ${\rm Ran}\hat{I}$
contains $\cV\bigl(F_{[0, \beta/2]} L^{\infty}(Q, \Sigma_{+}, \mu)\bigr)$.
Since $F_{[0, \beta/2]}>0$ a.e., $F_{[0, \beta/2]} L^{\infty}(Q,
\Sigma, \mu)$ is dense in~$\cH_{OS}$ and hence its image under $\cV$  is
dense in $\cH$ \qed .

\medskip
\medskip
\noindent
{\bf Identification of the abelian algebra}

\begin{proposition}
\label{st3.4b}
For $f\in L^{\infty}(Q,\Sigma_{0}, \mu)$ one has 
\[
 \hat{I}\tilde{f}\psi= \tilde{f}\hat{I}\psi,\: \psi\in \cH_{V},
\]
and, consequently,  $\hat{I}\cU_{V}=\cU \hat{I}$.
\end{proposition}
\proof This follows immediately from the definitions of $\tilde{f}$ in
Proposition \ref{st1.2} and $\hat{I}$ in 
Proposition 7.7  \qed .

\medskip
\noindent
{\bf Identification of the $C^{*}$-dynamics}

\medskip
\noindent
Applying Theorem \ref{pt1} we obtain a selfadjoint operator $H_{V}$
from the FKN kernel associated to $V$. It will be 
called the {\em pseudo-Liouvillean} generated by
$V$.

\begin{proposition} One has
\label{st3.5}

\vskip .3cm
\halign{ \indent \indent \indent #  \hfil & \vtop { \parindent =0pt \hsize=12cm
                            \strut # \strut} \cr 
{\rm (i)} & 
$\hat{I}\Omega_{V}=\|\e^{-\beta H_{V}/2}\Omega\|^{-1}\e^{-\beta H_{V}/2}\Omega$;
\cr
{\rm (ii)} & for $0\leq s_{1}\leq\cdots\leq s_{n}\leq \beta/2$ and
$A_{1}, \dots, A_{n}\in \cU$ 

\[
\begin{array}{rl}
&\hat{I}\e^{-s_{1}H_{V}}A_{1}
\bigl(\prod_{2}^{n}\e^{(s_{j-1}-s_{j})H_{V}}A_{j}
\bigr)\Omega_{V}\\[2mm]
=&{\e^{-s_{1}H_{V}}A_{1}\bigl(\prod_{2}^{n}\e^{(s_{j-1}-s_{j})H_{V}}A_{j} \bigr)\e^{(s_{n}-\beta/2)H_{V}}\Omega
\over \|\e^{-\beta H_{V}/2}\Omega\| 
} \, \, ;
\end{array}
\]
\cr
{\rm (iii)} & for $t_{1}, \dots, t_{n}\in \rr$,
$A_{1}, \dots, A_{n}\in \cU$  and $\psi\in \cH_{ V}$
\[
\hat{I} \bigl( \prod_{1}^{n}\e^{\i t_{j}L_{V}}A_{j}\e^{-\i
t_{j}L_{V}}\bigr) \psi= \bigl(\prod_{1}^{n}\e^{\i t_{j}H_{V}}A_{j}\e^{-\i
t_{j}H_{V}} \bigr)\hat{I}\psi \, ;
\]
\cr
{\rm (iv)} & $\hat{I}J_{V}= J\hat{I}$.
\cr}
\end{proposition}
Note that in {\rm (ii)} and {\rm (iii)} we identify $\cU$ with
$L^{\infty}(Q, \Sigma_{0}, \mu)$.

\medskip
\noindent
{\bf Identification of the observable algebras}

\medskip
\noindent
We recall that the observable algebra and the dynamics associated to the
perturbed path space  $(Q, \Sigma, \Sigma_{0}, U(t), R, \mu_{V})$ are
the von Neumann algebra $\cF_{V}$ generated by $\{\e^{\i
tL_{V}}A\e^{-\i tL_{V}} \mid A\in \cU_{V}, \: t\in
\rr\}$ and the automorphism group $\tau_V \colon t \mapsto \tau_V (t)$,  $t\in \rr$,
where
\[
\tau_{V}(t)(B)= \e^{\i tL_{V}}B\e^{-\i tL_{V}}, \: B\in
\cF_{V} .
\]
\begin{proposition} \
\label{st3.6}

\vskip .3cm
\halign{ \indent \indent \indent #  \hfil & \vtop { \parindent =0pt \hsize=12cm
                            \strut # \strut} \cr 
{\rm (i)} & $\hat{I}\tau_{V}(t)(B)\hat{I}^{-1}= \e^{\i
tH_{V}}\hat{I}B\hat{I}^{-1}\e^{-\i tH_{V}}$ for $B\in
\cF_{V}$ and $t\in \rr$;
\cr
{\rm (ii)} & Assume  that either $V\in
L^{2+\epsilon}(Q, \Sigma_{0}, \mu)$ for $\epsilon>0$ or that $V\in
L^{2}(Q, \Sigma_{0}, \mu)$ and $V\geq 0$. It follows that
$\hat{I}\cF_{V}\hat{I}^{-1}= \cF$.
\cr}
\end{proposition}
\proof {\rm (i)} follows from Proposition \ref{st3.5} {\rm (iii)}. To
prove {\rm (ii)} we recall from Theorem \ref{pt3} that, 
under the assumptions of the proposition,  $L+V$
is essentially selfadjoint on $\cD(L)\cap \cD(V)$ and $H_{V}=
\overline{L+V}$. Hence, by Trotter's formula,
\[
\e^{\i tH_{V}}= \slim_{n\to\infty}(\e^{\i tL/n}\e^{\i
tV/n})^{n}.
\] 
Thus
\[
\e^{\i tH_{V}}A\e^{-\i tH_{V}}=\wlim_{n\to +\infty}(\e^{\i tL/n}\e^{\i
tV/n})^{n}A(\e^{-\i tV/n}\e^{-\i
tL/n})^{n}.
\]
Since $\e^{\i
sV}\in \cU\subset \cF$, $A\in \cF$ implies that $\e^{\i sV}A\e^{-\i sV}\in \cF$. 
Moreover, $\e^{\i sL}A\e^{-i sL}\in \cF$ by
definition. So $\e^{\i tH_{V}}A\e^{-\i tH_{V}}\in \cF$,
if $A\in\cU$, and hence 
\[
\hat{I}\cF_{V}\hat{I}^{-1}\subset \cF.
\] 
According to Tomita's theorem (see, e.g., [BR])
$\cF'=J\cF J$ and 
$\cF_{V}'=J_{V}\cF_{V}J_{V}$. Using Proposition \ref{st3.5} {\rm
(iv)}. Thus
\[
(\hat{I}\cF_{V}\hat{I}^{-1})'= \hat{I}\cF_{V}'\hat{I}^{-1}= \hat{I}J_{V}\cF_{V}J_{V}\hat{I}^{-1}=
J\hat{I}\cF_{V}\hat{I}^{-1}J\subset J\cF J= \cF'.
\]
Taking commutants we obtain
\[
\cF=\cF''\subset (\hat{I}\cF_{V}\hat{I}^{-1})''=
\hat{I}\cF_{V}\hat{I}^{-1}.
\]
Hence $\cF= \hat{I}\cF_{V}\hat{I}^{-1}$\qed .

The results in this section are summarized in the following theorem.
\begin{theoreme}
Let $(\cF, \cU, \tau , \omega)$ be a stochastically positive $\beta$-KMS system. Let
$\cH, \Omega, L$ be the associated GNS Hilbert spaces, GNS vector and
Liouvillean. 
Let $V$ be a selfadjoint operator on $\cH$, affiliated to
$\cU$, such that 
\[
\matrix {
V,\: \e^{-\beta V}\in L^{1}(Q, \Sigma_{0}, \mu)  \hbox{ and}
& \hbox{  either }
& V\in L^{2+\epsilon}(Q, \Sigma_{0}, \mu), \:\epsilon>0, 
\cr 
& \hbox{ or } & V\in
L^{2}(Q, \Sigma_{0}, \mu) \hbox{ and } V\geq 0.}
\] 
Then

\vskip .3cm
\halign{ \indent \indent \indent #  \hfil & \vtop { \parindent =0pt \hsize=12cm
                            \strut # \strut} \cr 
{\rm (i)} & $L+V$ is essentially selfadjoint on $\cD(L)\cap
\cD(V)$; 
\cr
{\rm (ii)} & $\Omega\in
\cD(\e^{-\frac{\beta}{2}H_{V}})$, where $H_{V}=\overline{L+V}$; 
\cr
{\rm (iii)} & $(\cF, \cU, \tau_{V}, \omega_{V})$ is a stochastically positive
$\beta$-KMS system for \\ $\tau_{V, t}(A)= \e^{\i tH_{V}}A\e^{-\i
tH_{V}}$,  $\omega_{V}(A)=
\|\e^{-\frac{\beta}{2}H_{V}}\Omega\|^{-2}(\e^{-\frac{\beta}{2}H_{V}}\Omega,
A\e^{-\frac{\beta}{2}H_{V}}\Omega)$, $A\in \cF$.
\cr}

\end{theoreme}

\vskip .3cm
\noindent
{\bf Perturbed Liouvillean}

\vskip .2cm
\noindent
In the next theorem, we identify the Liouvillean for the perturbed
system.
\begin{theoreme}
\label{liouvillean}
Assume that $V$ is a selfadjoint operator affiliated to $\cU$ such
that 
\beq
\e^{-\beta V}\in L^{1}(Q, \Sigma_0, \mu)  \eeq
and
\beq
\label{hyp}
\begin{array}{l}
V\in L^{p}(Q, \Sigma_0, \mu),\:
\e^{-\frac{\beta}{2}V}\in L^{q}(Q, \Sigma_0, \mu) \, \hbox{ for }\, p^{-1}+ q^{-1}=
\12,\:2< p, \, \, q< \infty,\\ 
\hbox{or }V\in L^{2}(Q, \Sigma_{0}, \mu)  \hbox{ and }V\geq 0. 
\end{array}
\eeq
Let $L_{V}$ be the Liouvillean associated to the $\beta$-KMS system
$(\cF, \tau_{V }, \omega_{V})$. Then $H_{V}-JVJ$ is essentially
selfadjoint on $\cD(H_{V})\cap \cD(JVJ)$ and $L_{V}=
\overline{H_{V}-JVJ}$.
\end{theoreme}
\begin{lemma}
\label{l1}
For $A\in\cU$  one has
$JA\Omega_{V}=\|\e^{-\frac{\beta}{2}H_{V}}\Omega\|^{-1}\e^{-\frac{\beta}{2}H_{V}}A^{*}\Omega$.
\end{lemma}
\proof 
Let us set $c= \|\e^{-\frac{\beta}{2}H_{V}}\Omega\|^{-1}$. Then $A\Omega_{V}=
c\cV(A F_{[0,
\beta/2]})$. Moreover,  
$JA\Omega_{V}=
c\cV( U (\beta/2) A^{*} F_{[0,
\beta/2]})$, since $F_{[0,\beta/2]}$ is invariant under $R_{\beta/4}$.
Since $A^{*}$ belongs to the space $\cM_{\beta/2}=L^{\infty}(
Q, \Sigma_{0}, \mu)$ defined in
Section 7.2, $\cV(A^{*})= A\Omega\in
\cD(\e^{-\frac{\beta}{2}H_{V}})$ and
\[
c\e^{-\frac{\beta}{2}H_{V}}A^{*}\Omega= c\cV(U (\beta /2) A^{*} F_{[0,
\beta/2]})= JA\Omega_{V} \, \, \qed .
\] 
\begin{lemma}\label{l2}
Let $f_{1}$ be a real function in $L^{2}(Q, \Sigma_0, \mu)$ such that
$f_{1} F_{[0, \beta/2]}\in L^{2}(Q, \Sigma_{[0, \beta/2]},\mu)$.
Then $\Omega_{V} $ and $\Omega$ are vectors in $ \cD(f_{1})$. The vector $f_{1}\Omega$ is in $
\cD\bigl(\e^{-\frac{\beta}{2}H_{V}}\bigr)$ and satisfies $Jf_{1}\Omega_{V}=
\|\e^{-\frac{\beta}{2}H_{V}}\Omega\|^{-1}\e^{-\frac{\beta}{2}H_{V}}f_{1}\Omega$.
\end{lemma}
\proof 
Since $f_{1}\in L^{2}(Q, \Sigma_0, \mu)$, we have $\Omega\in \cD(f_{1})$.
Now $f_{1} F_{[0, \beta/2]}\in L^{2}(Q, \Sigma_{[0,
\beta/2]},\mu)$, thus~$\Omega_{V}\in \cD(f_{1})$.
Let  $f_{n}= f_{1}\one_{\{|f_{1}|\leq n\}}$. By dominated convergence
$f_{n} F_{[0, \beta/2]}\to f_{1} F_{[0, \beta/2]}$ in
$L^{2}(Q, \Sigma_{[0, \beta/2]}, \mu)$, \ie
\[
f_{1}\Omega_{V}= \cV(f_{1} F_{[0, \beta/2]})=\lim_{n\to
\infty}\cV(f_{n} F_{[0, \beta/2]}) =\lim_{n\to\infty}f_{n}\Omega_{V}.
\]
Applying Lemma \ref{l1} to $A=f_{n}$ we obtain, for $u\in
\cD(\e^{-\frac{\beta}{2}H_{V}})$,
\[
\begin{array}{rl}
&(\e^{-\frac{\beta}{2}H_{V}}u, f_{1}\Omega)= \lim_{n\to
\infty}(\e^{-\frac{\beta}{2}H_{V}}u, f_{n}\Omega)\\[2mm]
= &\lim_{n\to
\infty}(u, \e^{-\frac{\beta}{2}H_{V}}f_{n}\Omega)=\lim_{n\to
\infty}(u, Jf_{n}\Omega_{V})= (u, Jf_{1}\Omega_{V}).
\end{array}
\]
This shows that $f_{1}\Omega\in \cD(\e^{-\frac{\beta}{2}H_{V}})$ and
$\e^{-\frac{\beta}{2}H_{V}}f_{1}\Omega= Jf_{1}\Omega_{V}$ \qed .

\begin{lemma}
\label{l3}
Assume that $V$ is a selfadjoint operator, affiliated to $\cU$, which satisfies
(\ref{hyp}). Then 
\[
\Omega_{V}\in \cD(H_{V})\cap \cD(V) \hbox{ and } (H_{V}- JVJ)\Omega_{V}= (H_{V}-
JV)\Omega_{V}=0.
\]
\end{lemma}
\proof 
We first verify that $V$ satisfies the hypotheses of Lemma \ref{l2},
\ie that \beq
\label{check}
V \e^{-\int_{0}^{\beta/2}U(t)V \d t}\in L^{2}(Q,
\Sigma_{[0, \beta/2]},\mu).
\eeq Let $2\leq p, q\leq \infty$ be as in
(\ref{hyp}). If $p=2$, then $V\geq 0$ a.e., thus
(\ref{check}) is clearly satisfied. If~$q<\infty$, then, 
applying H\"{o}lder's inequality, it suffices to
prove that 
\[
V \in L^{p}(Q, \Sigma, \mu)
\hbox{ and }
\e^{-\int_{0}^{\beta/2} U(t) V \d t} \in L^{q}(Q,\Sigma, \mu).\]
Applying  (\ref{bound}) we find
\[
\|\e^{-\int_{0}^{\beta/2} U(t) V \d t}\|_{L^{q}(Q, \Sigma, \mu)}\leq \|\e^{-\frac{\beta}{2}V}\|_{q}<\infty.
\]
Let $u\in \cD\bigl(\e^{-\frac{\beta}{2}H_{V}}\bigr)\cap \cD(H_{V})\cap
\cD\bigl(H_{V}\e^{-\frac{\beta}{2}H_{V}}\bigr)$ and set $c:=
\|\e^{-\frac{\beta}{2}H_{V}}\Omega\|^{-1}$. Then
\[
(H_{V}u, \Omega_{V})=c(\e^{-\frac{\beta}{2}H_{V}}H_{V}u,
\Omega)=c(\e^{-\frac{\beta}{2}H_{V}}u,
H_{V}\Omega)=c(\e^{-\frac{\beta}{2}H_{V}}u, V\Omega),
\]
since $\Omega\in \cD(V)\cap \cD(L)$ and $H_{V}\Omega= L\Omega+
V\Omega=V\Omega$. Applying  Lemma \ref{l2} to $f_{1}=V$  we obtain
\[
c(\e^{-\frac{\beta}{2}H_{V}}u, V\Omega)= c(u,
\e^{-\frac{\beta}{2}H_{V}}V\Omega)= (u, JV\Omega_{V}).
\]
This implies, together with  
$J\Omega_{V}=\Omega_{V}$, that $\Omega_{V}\in \cD(H_{V})$ and
$H_{V}\Omega_{V}=JV\Omega_{V}=JVJ\Omega_{V}$ \qed .

\medskip

\noindent
{\bf Proof of Theorem \ref{liouvillean}.}
Let $\cF_{1}$ be the set of $A\in \cF$ such that $t\mapsto
\tau_{V,t}(A)$ is $C^{1}$ for the strong topology and let $A\in
\cF_{1}$. Since $H_{V}$
implements the dynamics $\tau_{V,t}$, we see that~$A\in
C^{1}(H_{V})$. By \cite{ABG}, this implies that $A\colon \cD(H_{V})\to
\cD(H_{V})$. Since $\Omega_{V}\in \cD(H_{V})$, the vector
$A\Omega_{V}\in \cD(H_{V})$. Since $JVJ$ is affiliated to $\cF'$, Lemma \ref{l3} 
implies  
\[
\matrix{
L_{V}A\Omega_{V} & = \i^{-1}\frac{\d}{\d t}\tau_{V,t}(A)\Omega_{V}\:_{|t=0}=
H_{V}A\Omega_{V}- AH_{V}\Omega_{V} \hfill
\cr
&= H_{V}A\Omega_{V}- AJVJ\Omega_{V}=
H_{V}A\Omega_{V}- JVJA\Omega_{V}. \hfill }
\]
This yields $L_{V}u= H_{V}u-JVJu$ for $u\in \cF_{1}\Omega_{V}$. By
Proposition \ref{propliou}, we know that $\cF_{1}\Omega_{V}$ is a
core for $L_{V}$. This implies that $L_{V}$ is the closure of
$H_{V}-JVJ$ on $\cF_{1}\Omega_{V}$ and hence also the closure of
$H_{V}-JVJ$ on $\cD(H_{V})\cap \cD(JVJ)$ \qed .

\subsection{Markov property for perturbed of path spaces}
In this subsection we show that the Markov property of a path space is
preserved  by the perturbations described in Subsection \ref{fkn}.
\begin{proposition}
\label{st3.7}
Let $(Q, \Sigma, \Sigma_{0}, U(t), R, \mu)$ be a generalized path
space satisfying the
Mar\--kov property and let $\{F_{[a, b]}\}$ be a FKN kernel. Then $(Q, \Sigma, \Sigma_{0}, U(t), R, \mu_F)$ 
satisfies the Markov property.
\end{proposition}
\proof 
Let $(Q, \Sigma, \mu)$ be a probability space, $F\in L^{1}(Q, \Sigma,
\mu)$ with $F>0$ $\mu$-a.e.\  and set $\d\mu_{F}= (\int
F\d\mu)^{-1}F\d\mu$.

If $B\subset\Sigma$ is a $\sigma$-algebra and $f$ is
$\Sigma$-measurable, then we denote by $E_{B}(f)$, (resp.\ $E_{B}^{F}(f)$) the
conditional expectation of $f$ w.r.t.\ $B$ for the measure $\mu$ (resp.\
$\mu_{F}$). Then (see \cite[Sect. 2.4]{Loeve})  
\beq
E_{B}(fg)= E_{B}(f)g, \: E_{B}^{F}(fg)= E_{B}^{F}(f)g\: \: \mu\hbox{-a.e.\
if } g\hbox{ is } B\hbox{-measurable} 
\label{ste3.4}
\eeq
and
\beq
\label{ste3.3}
E_{B}^{F}(f)= \frac{E_{B}(Ff)}{E_{B}(F)} \, \, \,  \mu\hbox{-a.e.}
\eeq
To simplify the notation, let us set $E_{0}= E_{ \{Ê0\} }$ if $ \beta= + \infty$ and $E_{0}=E_{\{0, \beta/2\}}$ if $\beta < \infty$. Set
$F_{+}= F_{[0,\beta/2]}$ and
$F_{-}=F_{[-\beta/2, 0]}$, so that $F=F_{-}F_{+}$. Set $E_{+}^{(F)}=
E_{[0, \beta/2]}^{(F)}$ and $E_{-}^{(F)}= E_{[-\beta/2, 0]}^{(F)}$.  Finally set
$E_{0}^{(F)}= E_{ \{Ê0\} }^{(F)}$ if $ \beta= + \infty$ and $E_{0}^{(F)}=E_{\{0, \beta/2\}}^{(F)}$ if $\beta < \infty$.

Let now $f$ be $\Sigma$-measurable. Then
\[
E_{+}^{F}(f)=
\frac{E_{+}(Ff)}{E_{+}(F)}=\frac{E_{+}(F_{-}F_{+}f)}{E_{+}(F_{-}F_{+})}=\frac{E_{+}(F_{-}f)}{E_{+}(F_{-})}, 
\]
using (\ref{ste3.3}), (\ref{ste3.4}) and the fact that $F_{+}$ is
$\Sigma_{[0, \beta/2]}$-measurable. Next 
\[
\frac{E_{+}(F_{-}f)}{E_{+}(F_{-})}=\frac{E_{+}(F_{-}f)}{E_{+}E_{-}(F_{-})}=\frac{E_{+}(F_{-}f)}{E_{0}(F_{-})},
\]
by the Markov property for $(Q, \Sigma, \mu)$ and the fact that
$F_{-}$ is $\Sigma_{[-\beta/2, 0]}$-measurable. Since $E_{0}(F_{-})$
is $\Sigma_{[-\beta/2, 0]}$-measurable, we have, by (\ref{ste3.3}) and
(\ref{ste3.4}),
\[
E_{-}^{F}E_{+}^{F}(f)=
\frac{E_{-}(FE_{+}(F_{-}f))}{E_{0}(F_{-})E_{-}(F)}=
\frac{E_{-}(F_{-}F_{+}E_{+}(F_{-}f))}{E_{0}(F_{-})E_{-}(F_{-}F_{+})}=
\frac{E_{-}(F_{+}E_{+}(F_{-}f))}{E_{0}(F_{-})E_{-}(F_{+})},
\]
since $F_{-}$ is $\Sigma_{[-\beta/2, 0]}$-measurable. 

Now
\[
\frac{E_{-}(F_{+}E_{+}(F_{-}f))}{E_{0}(F_{-})E_{-}(F_{+})}=
\frac{E_{0}(Ff)}{E_{0}(F_{+})E_{0}(F_{-})},
\]
by the Markov property for $(Q, \Sigma, \mu)$ and the fact that
$F_{+}$ is $\Sigma_{[0, \beta/2]}$-measurable. Finally 
\[
\begin{array}{rl}
E_{0}(F_{-})E_{0}(F_{+})= &E_{+}E_{-}(F_{-})E_{0}(F_{+})=
E_{+}(F_{-}E_{0}(F_{+}))\\=& E_{+}(F_{-}E_{-}(F_{+}))=
E_{+}E_{-}(F_{-}F_{+})= E_{0}(F).
\end{array}
\]
This yields $E_{-}^{F}E_{+}^{F}(f)= E_{0}^{F}(f)$ $\mu$-a.e.\ and
completes the proof  \qed .
\section{Free Klein-Gordon fields at positive temperature}

\init
\label{sec2}
In this section we recall some results about the complex Klein-Gordon
field and show that it provides an example of a charge symmetric K\"{a}hler
structure. 

The classical Klein-Gordon equation describing a charged particle of mass $m$ is 
\[
\p_{t}^{2}\Phi- \p_{x}^{2}\Phi + m^{2}\Phi=0, \: (t,x)\in
\rr^{d+1},\:  
\]
where $\Phi\colon \rr^{d+1} \to \cc $ is a complex valued function.
For later use we recall the discrete symmetries of the Klein-Gordon equation,
namely the {\em parity} $\bp$, {\em time reversal} $\theta$ and {\em charge
conjugation} $\bc$: 
\[
\bp \Phi(t,x):= \Phi(t, -x), \: \theta \Phi(t,x)= \overline{\Phi}(-t,x)\hbox{
and }\bc  \Phi(t,x)= \overline{\Phi}(t,x).
\]
In particular, real solutions of the Klein-Gordon equation without
external field describe neutral scalar particles. 
In the sequel only time-reversal and charge conjugation will play a
role.
\subsection{The complex Klein-Gordon field}
\label{ckg}

Let us now describe the abstract Klein-Gordon equation that we will
consider in the sequel.

\vskip .3cm
\noindent
{\bf Abstract Klein-Gordon equation}

\vskip .2cm
\noindent
Let $\ch$ be a Hilbert space. We denote by $\i$ the complex structure
on $\ch$ and by $\hf_{\ch}$ the
scalar product on $\ch$. We assume that $\ch$ is equipped with a
conjugation denoted by $\Phi \to \overline{\Phi}$.

Let \beq
\epsilon\geq m >0
\label{e2.01}
\eeq be a real selfadjoint operator on $\ch$, \ie such
that $\overline{\epsilon \Phi}= \epsilon \overline{\Phi}$.

For $0\leq s\leq 1$ we denote by $\ch_{s}$ the Hilbert space
$\cD(\epsilon^{s})$ with complex structure $\i$ and scalar product $v, u \mapsto (v,
\epsilon^{2s}u)_{\ch}$ and by $\ch_{-s}$ the completion of $(\ch, \i)$ for the 
norm $(v, \epsilon^{-2s }v)_{\ch}$. The space $\ch_{-s}$ can be identified
with the anti-dual of $\ch_{s}$ using the sesquilinear form 
$\langle v, u\rangle=(v, u)_{\ch}$ for $v\in \ch_{-s}$ and $u\in \ch_{s}$.

We
consider the abstract Klein-Gordon equation
\[
\hbox{(KG)}\:\: (\p_{t}^{2}\Phi) (t)+ \epsilon^2\Phi (t)=0,
\]
where $\Phi(t)$ is a function of $t\in \rr$ with values  in $\ch$.
This (complex) KG equation describes a classical field of scalar charged
particles. 

The complex structure on $\ch$ yields a complex
structure on the space of solutions of (KG), associated to the $U(1)$
gauge group.  Following  the convention
of Subsection \ref{notat} this `charge' complex structure will be
denoted by $\j$. It is defined by
\[
(\j \Phi)(t):= \i \Phi(t)  \hbox{ for }\Phi\hbox{ a solution of (KG)
and }t\in \rr.
\]

The following  quantity does not depend on $t$: 
\[
q(\Psi, \Phi):= \i \bigl(\Psi(t), (\p_{t}\Phi)(t) \bigr)_{\ch}-
\i \bigl((\p_{t}\Psi)(t), \Phi (t)\bigr)_{\ch} .
\]
Hence it defines a symmetric sesquilinear form on the space of solutions
of (KG).
The following transformations preserve the solutions of (KG):
\vskip .2cm

-- {\em gauge transformations} $\Phi(t)\mapsto \e^{\, \i
\alpha}\Phi(t)=(\e^{\, \j \alpha}\Phi)(t)$,
$\alpha\in [0, 2\pi]$;

-- {\em time-reversal} $\theta \colon \Phi(t)\mapsto \overline{\Phi}(-t)$;

-- {\em charge conjugation}  $\bc  \colon \Phi(t)\mapsto \overline{\Phi (t)}$.

\vskip .3cm
\noindent
{\bf Energy space}

\vskip .2cm
\noindent
It is convenient to identify a solution of (KG) with its Cauchy data at $t=0$, 
\[
f:= (\Phi(0),(\p_{t}\Phi)(0))\in \ch\times \ch.
\]
To do so one introduces the
{\em energy space } ${\cal E}:= \ch_{1}\oplus \ch$ equipped
with the norm
\[(f,f)_{\cal E}= (f_{1}, \epsilon^2f_{1})_{\ch}+ (f_{2}, f_{2})_{\ch},
\]
where we set $f= (f_{1}, f_{2})$.  Note that the complex structure
$\j$ becomes $\i\oplus \i$ on ${\cal E}$. Setting $f_{t}= \bigl(\Phi (t),
(\p_{t}\Phi) (t) \bigr)$ one can rewrite the Klein-Gordon equation as the
first order system:
\[
\j (\p_{t}f)_{t}= Lf_{t} \hbox{ for }L=\left(\begin{array}{cc}
0&\i \\ -\i \epsilon^2&0
\end{array}\right).
\] 
It is convenient to diagonalize $L$ using the unitary map
\[
\matrix{
U_{0}\colon &{\cal E} & \to & \ch\oplus \ch \cr
& f & \mapsto & u= (u_{1}, u_{2}),
\cr}
\]
where
\[
U_{0}:=\frac{1}{\sqrt{2}}\left(\begin{array}{cc}
\epsilon&\i\\ \epsilon &-\i
\end{array}\right) \hbox{ and } U_{0}^{-1}=\frac{1}{\sqrt{2}}\left(\begin{array}{cc}
\epsilon^{-1}&\epsilon^{-1}\\ -\i &\i
\end{array}\right).
\]
It follows that 
\[
U_{0}LU_{0}^{*}= \left(\begin{array}{cc}
\epsilon&0\\ 0&-\epsilon
\end{array}\right).
\]
In particular, $L$ is selfadjoint on ${\cal E}$ with domain
$U^{-1}(\ch_{1}\times\ch_{1})$ and the evolution $\rr\ni t\mapsto
\e^{-\j tL}$
is a strongly continuous unitary group. Therefore the space of
solutions of (KG) can be identified with ${\cal E}$. On ${\cal E}$ the
symmetric form $q$ is 
\[
q(g, f)= \i (g_{1}, f_{2})_{\ch}- \i (g_{2}, f_{1})_{\ch}.
\]

\vskip .2cm
\noindent
{\bf Charged K\"{a}hler space structure}

\vskip .2cm
\noindent
On ${\cal E}$ we put the `energy' complex structure $\i:= \j\frac{L}{|L|}$.

\begin{proposition}
The space $({\cal E},\j, \i , q)$ is a charged K\"{a}hler space.
\end{proposition}
\proof  Clearly $[\i, \j]=0$. We have to
prove that
\[
(g,f):= {\rm Im}q(g, \i f)+ \i {\rm  Im}q(g, f)
\]
is a positive definite symmetric sesquilinear form on $({\cal E},
\i)$. If $U_{0}f= (u_{1}, u_{2})$ and $U_{0}g=(v_{1}, v_{2})$, then
\[
\begin{array}{l}
q(g, f)= -(v_{2}, \epsilon^{-1}u_{2})_{\ch}+ (v_{1},
\epsilon^{-1}u_{1})_{\ch},\\[3mm]
q(g, \i f)= -(v_{2}, -\i \epsilon^{-1}u_{2})_{\ch}+ (v_{1},
\i \epsilon^{-1}u_{1})_{\ch}= \i(v_{1},
\epsilon^{-1}u_{1})_{\ch}+\i(v_{2}, \epsilon^{-1}u_{2})_{\ch},\\[3mm]
\end{array}
\]
and consequently 
\beq
(g,f)= (v_{1}, \epsilon^{-1}u_{1})_{\ch}+ \overline{(v_{2},\epsilon^{-1}u_{2})_{\ch}}.
\label{e1.1b}
\eeq
\qed

\begin{definition}
We denote by $\bigl({\cal E}_{\rm q}, \i, \hf \bigr)$ the completion of 
$({\cal E}, \i)$ for the scalar product~$\hf$. 
\end{definition}
\begin{proposition}
The space ${\cal E}_{\rm q}$ is equal to the space
$\ch_{\12}\oplus\ch_{-\12}$ equipped with the complex structure
\[
\i=\left(\begin{array}{cc}
0&-\epsilon^{-1}\\\epsilon &0
\end{array}\right)
\]
and the scalar product $(g,f)= {\rm Re}(g_{1}, \epsilon f_{1})_{\ch}+ {\rm
Re}(g_{2}, \epsilon^{-1}f_{2})_{\ch}+ \i \bigl({\rm Re}(g_{1}, f_{2})_{\ch}-
{\rm Re}(g_{2}, f_{1})_{\ch} \bigr)$.
\end{proposition}

\vskip .2cm
\noindent
{\bf Standard form of the complex  Klein-Gordon field}

\vskip .2cm
\noindent
It is convenient to introduce the map 
\[
\begin{array}{l}
U_{\rm q}(f_{1}, f_{2}):= \frac{1}{\sqrt{2}} \bigl( \epsilon^{\12} f_{1}+ \i
\epsilon^{-\12}f_{2},
\epsilon^{\12} \overline{f}_{1} + \i \epsilon^{-\12}\overline{f}_{2}\bigr)=:(u_{1}, u_{2} ).
\end{array}
\]
Using (\ref{e1.1b}) we obtain that $U_{\rm q}$
extends to a unitary map
\[
U_{\rm q}\colon \bigl({\cal E}_{\rm q}, \i, (\cdot, \cdot)\bigr)\to (\ch, \i)\oplus
(\ch, \i).
\]
Let us describe the various objects after conjugation by $U_{\rm q}$. We will
denote by the same letter an object acting on ${\cal E}_{\rm q}$  and
its conjugation by $U_{\rm q}$ acting on
$\ch\oplus\ch$. 

\vskip .3cm
\halign{ \indent \indent \indent #  \hfil & \vtop { \parindent = 0pt \hsize=12cm
                            \strut # \strut} 
\cr 
--    & 
{\em symmetric form}: after conjugation by $U_{\rm q}$ the symmetric
form $q(g,f)$ becomes
\[   
q \bigl((v_{1}, v_{2}),(u_{1}, u_{2}) \bigr)= (v_{1},
u_{1})-(u_{2}, v_{2}).
\]
\cr 
--    & 
{\em `charge'  complex structure}: after conjugation by $U_{\rm q}$ the complex
structure $\j$ becomes
\[
 \j = \left(\begin{array}{cc}
\i &0\\ 0 &-\i
\end{array}\right).
\]
\cr 
--    & 
{\em Hamiltonian}: the infinitesimal generator of $\rr\ni t\mapsto
\e^{-\j
tL}$ on $\bigl({\cal E}_{\rm q}, \i , \hf \bigr)$ is the {\em
Hamiltonian}, denoted by ${\rm h}$. After conjugation by $U_{\rm q}$,
\[
{\rm h} =\left(\begin{array}{cc}
\epsilon &0\\ 0 &\epsilon
\end{array}\right).
\]
In particular ${\rm h}$ is positive.
\cr 
--    & 
{\em Gauge transformations:} the infinitesimal generator of $[0, 2\pi]\ni
\alpha\mapsto \e^{-\j \alpha}$ on $\bigl({\cal E}_{\rm q}, \i , \hf \bigr)$ is the {\em charge operator} 
${\rm q}$. After conjugation by
$U_{\rm q}$,
\[
{\rm q}= \left(\begin{array}{cc}
1&0\\ 0 &-1
\end{array}\right).
\]
We have  ${\rm q}= -\i \j$. Hence 
${\rm q}$ is a charge operator in the sense of Subsection
\ref{sec1.2}. 
\cr 
--    & 
{\em Time reversal:}  we have  $\theta (f_{1}, f_{2})= (\overline{f}_{1},
-\overline{f}_{2})$, and after conjugation by $U_{\rm q}$,  
\[
\theta (u_{1}, u_{2})= (\overline{u}_{1}, \overline{u}_{2}).
\]
\cr 
--    & 
{\em charge conjugation:} we have  $\bc (f_{1}, f_{2})=
(\overline{f}_{1}, \overline{f}_{2})$, and after conjugation by
$U_{\rm q}$,  
\[
\bc (u_{1}, u_{2})= (u_{2}, u_{1}).
\]
We see that $({\cal E}_{\rm q}, \j ,\i,q, \bc)$ is a charge-symmetric K\"{a}hler space.
\cr}

\vskip 1cm

From now on we will set $X:= \ch\oplus \ch$ with
elements $x= (x^{+}, x^{-})$ and equip $X$ with the complex
structures 
\[
\i = \left(\begin{array}{cc}
\i &0\\ 0 &\i
\end{array}\right) \hbox{ and } \j = \left(\begin{array}{cc}
\i &0\\ 0 &-\i
\end{array}\right),
\]
with the symmetric form and the scalar product 
\[
q( y, x)= ( y^{+}, x^{+})- (x^{-}, y^{-} ) \hbox{ and } (y, x):= ( y^{+}, x^{+})+ (y^{-}, x^{-}),
\]
the Hamiltonian and the charge operator  
\[
{\rm h}= \left(\begin{array}{cc}
\epsilon &0\\ 0 &\epsilon
\end{array}\right) \hbox{ and } {\rm q}=\left(\begin{array}{cc}
\one &0\\ 0 &-\one
\end{array}\right),
\]
and the time-reversal and the charge conjugation  
\[
\theta (x^{+}, x^{-})= (\overline{x^{+}}, \overline{x^{-}})  \hbox{ and } \bc (x^{+},
x^{-})= (x^{-}, x^{+}). 
\]
From the discussion above we obtain the following theorem.
\begin{theoreme}
The map $U_{\rm q}\colon ({\cal E}_{\rm q}, \j ,\i,q,
\bc)\to (X, \j, \i, q, \bc)$ is unitary between $\bigl({\cal
E}_{\rm q}, \i, \hf\bigr)$ and $\bigl(X, \i, \hf \bigr)$, and isometric between $({\cal
E}_{\rm q}, \j, q)$ and $(X, \j, q)$. It satisfies
\[
U_{\rm q}aU_{\rm q}^{-1}= a  \hbox{ for }a={\rm h},\: {\rm q},\:{\rm
t},\: {\rm c}. 
\] 
\end{theoreme}

For later use we set $\kappa:= \theta \bc$ and $X_{\kappa}:=\{x\in X| \kappa x=x\}=\{(x^{+},
\overline{x}^{+}),\: x^{+}\in \ch\}$. 
Note that in terms of solutions of  (KG) we have
$\kappa\Phi(t,x)= \Phi(-t, x)$ and an element of $X_{\kappa}$
corresponds to a solution of (KG) with Cauchy data $(u,0)$, where $u\in
\ch_{\12}$. 

We see that $\kappa$ is a conjugation on $(X, \i, \hf)$ and hence
${\rm Im\hf}$ vanishes on $X_{\kappa}$. Since~$[\kappa ,
\j]=0$, the vector space $X_{\kappa}$ is a complex vector space  for the
complex structure~$\j$.

For comparison with the physics literature, let us consider the case
$\ch=L^{2}(\rr^{d}, \d x)$ and $\epsilon= (-\Delta_{x}+ m^{2})^{\12}$. Then $\ch_{-\12}$
is the Sobolev space $H^{-\12}(\rr^{d})$. In the physics
literature one defines for $u\in \coinf(\rr^{d})$ the time-zero field
$\phi_{\rm p}(u)$ to be the Hermitian field associated with the
solution of (KG) with Cauchy data $\bigl(\frac{1}{2\pi}\epsilon^{-1}u, 0\bigr)$.

After the unitary transformation $U_{\rm q}$,
$\bigl(\frac{1}{2\pi}\epsilon^{-1}u, 0 \bigr)$ becomes the element
\[
\frac{1}{\sqrt{2}2\pi} \bigl( \epsilon^{-\12}u,
\epsilon^{-\12}\overline{u} \bigr)\in L^{2}(\rr^{d} )\oplus
L^{2}(\rr^{d}),
\] \ie 
\[
 \phi_{\rm p}(u)= \frac{1}{\sqrt{2}2\pi}\phi \bigl( \epsilon^{-\12}u,
\epsilon^{-\12}\overline{u}\bigr).
\] 
In the physics litterature one also considers the  {\em
complex time-zero  field} $\varphi_{\rm p}(u)$ defined as $\phi_{\rm p}(u)+ \i \phi_{\rm p}(\i u)$, \ie
\[
\varphi_{\rm p}(u)= \frac{1}{2\pi}\varphi \bigl(\epsilon^{-\12}u,
\epsilon^{-\12}\overline{u} \bigr).
\]

\subsection{The real Klein-Gordon field}
\init
\label{sec3}
We now quickly discuss the real Klein-Gordon field.
\vskip .3cm
\noindent
{\bf Abstract real Klein-Gordon equation}

\vskip .2cm
\noindent
Let $\ch_{\rr}$  be a real Hilbert space.  
Let $\epsilon \geq m>0$ be
a  selfadjoint operator on $\ch_{\rr}$. We
consider the Klein-Gordon equation:
\[
 \p_{t}^{2}\Phi(t) + \epsilon^2\Phi(t) =0,
\]
where $\Phi$ is a function of $t\in \rr$ with values in   $\ch_{\rr}$.
The real Klein-Gordon equation describes a classical field of scalar neutral
particles.

Let us denote by  $\ch:=\cc\ch_{\rr}$ the complexification of
$\ch_{\rr}$ with its canonical scalar product~$(\cdot, \cdot)_{\ch}$.
The space $\ch$ is  equipped with the canonical conjugation $\ch\ni
\Phi\mapsto \overline{\Phi}$, $\Phi\in \ch$.

On the space of 
real solutions of the Klein-Gordon equation, the charge conjugation $\bc$ acts as
identity and the time-reversal $\theta $ takes the form $\theta \colon\Phi(t)\mapsto
\Phi (-t)$.
We will still denote by $\epsilon$ the complexification of
$\epsilon$ acting on $\ch$. We can now apply the results of Subsection \ref{ckg} to the Hilbert space
$\ch$.

The real energy space is ${\cal E}_{\rr}:= {\cal
E}\cap \ch_{\rr}\times \ch_{\rr}$. The image of ${\cal E}_{\rr}$ under
the transformation~$U$ is
\[
U{\cal E}_{\rr}=: {\cal S}_{\rr}= \bigl\{(u_{1}, u_{2})\in \ch\oplus\ch|
u_{2}= \overline{u_{1}} \bigr\}.
\]

Note that $\e^{-\j tL}$ preserves ${\cal E}_{\rr}$. More general, 
if $F\colon \rr\to \cc$ is a bounded measurable function
such that $\overline {F}(\lambda)= F(-\lambda)$ then $F(L)$ preserves
${\cal E}_{\rr}$. Therefore $\i$ preserves~${\cal E}_{\rr}$ and hence
defines a  complex structure on ${\cal E}_{\rr}$. 
The space $({\cal E}_{\rr}, \i , q)$ is a K\"{a}hler space. 
\begin{definition}
We denote by $\bigl({\cal E}_{{\rm q}, \rr}, \i, \hf \bigr)$ the closure of
$({\cal E}_{\rr}, \i)$ for the scalar product~$\hf$. 
\end{definition}
\begin{proposition}
The space ${\cal E}_{{\rm q}, \rr}$ is equal to $\ch_{\12, \rr}\oplus
\ch_{-\12, \rr}$ equipped with the complex structure
\[
\i=\left(\begin{array}{cc}
0&-\epsilon^{-1}\\\epsilon &0
\end{array}\right)
\]
and the scalar product $(g,f)= (g_{1}, \epsilon f_{1})_{\ch}+ 
(g_{2}, \epsilon^{-1}f_{2})_{\ch}+ \i \bigl((g_{1}, f_{2})_{\ch}-
(g_{2}, f_{1})_{\ch} \bigr)$.
\end{proposition}
\vskip .2cm
\noindent
{\bf Standard form of the real Klein-Gordon field}

\vskip .2cm
\noindent
We set
\[
\matrix{
U_{\rr}\colon & {\cal E}_{\rr} & \to & \ch \cr
& f & \mapsto &(\epsilon^{\12} f_{1}+ \i\epsilon^{-\12} f_{2}).
\cr}
\]
Then $U_{\rr}$ extends to a unitary map between $\bigl({\cal E}_{{\rm q},
\rr}, \i , \hf \bigr)$ and $\ch$.
 Let us
describe the various objects after conjugation by $U_{\rr}$: 

\vskip .3cm
\halign{ \indent \indent \indent #  \hfil & \vtop { \parindent = 0pt \hsize=12cm
                            \strut # \strut} 
\cr 
$ $-    & 
{\em Hamiltonian}: The infinitesimal generator of $\rr\ni t\mapsto \e^{-\j tL}$ 
on $\bigl({\cal E}_{{\rm q}, \rr}, \i , (\cdot, \cdot)\bigr)$ is the {\em
Hamiltonian} denoted by ${\rm h}$. After conjugation by $U_{\rr}$,
\[{\rm h}=\epsilon.
\]
In particular, ${\rm h}$ is positive.
\cr 
$ $-    & 
{\em Time reversal:}  
We have $\theta (f_{1}, f_{2})= (f_{1},-f_{2})$.
After conjugation by $U_{\rr}$, one finds  $\theta u_{1}=\overline{u}_{1}$.
\cr}

\vskip .3cm
\noindent
From the discussion above we obtain the following theorem.
\begin{theoreme}
There exist a map $U_{\rr}$ between $({\cal E}_{{\rm q},\rr}, \i,q,
\theta)$ and $(\ch, \j, q, \theta)$ which is unitary between $\bigl({\cal
E}_{\rm q,\rr}, \i, \hf \bigr)$ and $\bigl(\ch, \j, \hf \bigr)$,  and satisfies
\[
U_{{\rm q},\rr}aU_{{\rm q},\rr}^{-1}= a  \hbox{ for }a={\rm h},\:{\rm
t}. 
\] 
\end{theoreme}

\vskip .3cm
\noindent
For later use we set $\kappa:= \theta $ and $\ch_{\kappa}:=\{h\in \ch\mid h=\overline{h}\}$. 

\subsection{Free Klein-Gordon fields at positive temperature}
\label{poskg}
We can now apply the results of Section \ref{quasi} to the real and
complex Klein-Gordon fields.

In the complex case we set $X=\ch\oplus\ch$, ${\rm h}=\epsilon\oplus
\epsilon$, ${\rm q}=\one\oplus -\one$ and introduce for $|\mu|<m$ the state
$\omega_{\beta, \mu}$ on
$\fW(X)$ defined by the functional 
\[
\omega_{\beta,\mu}(W(x)):= \e^{-\frac{1}{4}(x, (1+ 2\rho)x)}, \: x\in X,
\]
where $\rho= ( \e^{\beta {\rm a}}-1)^{-1}$ and ${\rm
a}={\rm h}-\mu {\rm q}$. As recalled in Section \ref{quasi},
$\omega_{\beta,\mu}$ is a $(\tau, \beta)$-KMS state for the dynamics
$\tau_{t} \bigl(W(x)\bigr)=W(\e^{\i t{\rm a}}x)$, which is invariant under the
gauge transformations $\alpha_{t} \bigl(W(x)\bigr)= W(\e^{\i t{\rm q}}x)$.
For $\mu=0$ the state $\omega_{\beta, \mu}$ will be denoted by
$\omega_{\beta}$. 

In the real case we set $X=\ch$, ${\rm h}=\epsilon$ and consider 
the state on $\fW(X)$
defined by the functional
\[
\omega_{\beta} \bigl(W(x)\bigr):= \e^{-\frac{1}{4}(x, (1+ 2\rho)x)}, \: x\in X,
\]
where $\rho= (\e^{-\beta {\rm
\epsilon}}-1)^{-1}$. It is a $(\tau, \beta)$-KMS state for the dynamics
$\tau_{t}\bigl(W(x)\bigr)=W(\e^{\i t\epsilon}x)$.

In both cases we denote by $\cF$ and $\cU$ the algebras defined in
Subsection \ref{sec4.2}; note that $\cU$ is defined w.r.t.\ the appropriate conjugation
$\kappa$. 

Applying Theorem~\ref{p1.1} we obtain that
the KMS system $(\cF, \cU, \tau, \omega_{\beta})$ is
stochastically positive both for real and complex Klein-Gordon fields.
Moreover, by Lemma \ref{markov} and Theorem~\ref{st2.2}, the stochastic process
associated to  $(\cF, \cU,\tau, \omega_{\beta})$ satisfies the Markov property.

In the next lemma we show that for $\mu\neq 0$, the KMS system $(\cF,
\cU, \tau, \omega_{\beta,\mu})$ is {\em not} stochastically
positive. The same is true, if we restrict the KMS state
$\omega_{\beta, \mu}$ to gauge invariant observables (see Subsection
\ref{sec4.3b}). 

The physical reason for this fact is that  a system of
charged particles is only invariant under the combination of time
reversal and charge conjugation. A nonzero chemical
potential introduces a disymmetry between particles of positive and
negative charge and hences breaks time reversal invariance, which is a
necessary property shared by all stochastically positive KMS systems, as 
we have seen in
Proposition \ref{timereversal}.
\begin{lemma}
For $\mu\neq 0$ the KMS systems $(\cF,
\cU, \tau, \omega_{\beta,\mu})$and $(\cA, \cA_{\kappa}, \tau,
\omega_{\beta, \mu})$ are not stochastically positive.
\end{lemma}
\proof Using the results of Subsection \ref{sec1.2} we have:
\[
\varphi_{\omega}(x)= a_{\omega} (x^{+})+ a_{\omega}^{*}(x^{-}),\:
\varphi^{*}_{\omega}(x)= a_{\omega}^{*}(x^{+})+ a_{\omega}(x^{-}),\\
\]
which, by an easy computation using the results recalled in Subsection
\ref{sec4.3},  implies
\[
\begin{array}{rl}
 \varphi^{*}_{\omega}(x)\varphi_{\omega}(x)\Omega_{\beta, \mu}
 & = a^{*}_{F} \bigl(
(1+\rho)^{\12}x^{+}\oplus \overline{\rho}^{\12}\overline{x^{-}}\bigr)a^{*}_{F} \bigl(
(1+\rho)^{\12}x^{-}\oplus \overline{\rho}^{\12}\overline{x^{+}}\bigr)\Omega_{\beta, \mu}\\
& \qquad + \bigl((x^{-}, (1+\rho)x^{-})+ (x^{+}, \rho x^{+})\bigr)\Omega_{\beta, \mu}.
\end{array}
\]
Set $H=\dG({\rm h}\oplus -{\rm h})$ and $Q= \dG({\rm q}\oplus
-{\rm q})$, so that $L =H-\mu Q$. Then 
\[ \eqalign{ \e^{-sL}\varphi^{*}_{\omega}(x)\varphi_{\omega}(x)\Omega_{\beta, \mu} &=
\e^{-s H}\varphi^{*}_{\omega}(x)\varphi_{\omega}(x)\Omega_{\beta, \mu} \cr
&= a^{*}_{F} \bigl(
(1+\rho)^{\12}\e^{-s{\rm h}}x^{+}\oplus
\overline{\rho}^{\12}\e^{s\overline{\rm h}}\overline{x^{-}} \bigr)a^{*}_{F} \bigl(
(1+\rho)^{\12}\e^{-s{\rm h}}x^{-}\oplus
\overline{\rho}^{\12}\e^{s\overline{\rm h}}\overline{x^{+}} \bigr)\Omega_{\beta, \mu} \cr
& \quad+ \bigl((x^{-}, (1+\rho)x^{-})+ (x^{+}, \rho x^{+}) \bigr)\Omega_{\beta, \mu}.
\cr}\]
Thus, for $x, y\in X$,
\[ \eqalign{ &
\bigl(\varphi^{*}(y)\varphi(y)\Omega_{\beta, \mu}, \e^{-
sL}\varphi^{*}(x)\varphi(x)\Omega_{\beta, \mu} \bigr)\cr
=&\bigl((1+\rho)^{\12}y^{+}\oplus \overline{\rho}^{\12}\overline{y^{-}},(1+\rho)^{\12}\e^{-s{\rm h}}x^{+}\oplus
\overline{\rho}^{\12}\e^{s\overline{\rm h}}\overline{x^{-}} \bigr)\cr
&\times \bigl((1+\rho)^{\12}y^{-}\oplus
\overline{\rho}^{\12}\overline{y^{+}},(1+\rho)^{\12}\e^{-s{\rm h}}x^{-}\oplus
\overline{\rho}^{\12}\e^{s\overline{\rm h}}\overline{x^{+}} \bigr)\cr
&+\bigl((x^{-}, (1+\rho)x^{-})+ (x^{+}, \rho x^{+})\bigr) \bigl((y^{-},
(1+\rho)y^{-})+ (y^{+}, \rho y^{+})\bigr).\cr}
\]
Let us now restrict ourselves to $x,y\in X_{\kappa}$, \ie $x=(u,
\overline{u})$, $y=(v,\overline{v})$, $u,v\in \ch$. We obtain
$x^{+}=u$, $x^{-}=\overline{u}$, $y^{+}=v$ and $y^{-}=\overline{v}$. If
we set $\rho^{\pm}=(
\e^{\beta(\epsilon\mp \mu)}-1)^{-1}$, then
\[
\begin{array}{rl}
\bigl(\varphi^{*}(y)\varphi(y) \Omega_{\beta, \mu} \, ,  \, \tau_{t}(\varphi^{*}(x)\varphi(x)) \Omega_{\beta, \mu} \bigr)_{| t= is} 
=& \bigl(v, (\e^{-s\epsilon}(1+ \rho^{+})+\e^{s\epsilon}\rho^{-})u \bigr)
\\
& \: \: \: \quad \times \bigl(u, (\e^{-s\epsilon}(1+ \rho^{-})+ \e^{s\epsilon}\rho^{+})v \bigr)\\
&+ \bigl(u, (1+ \rho^{+}+ \rho^{-})u \bigr) \bigl( v, (1+ \rho^{+}+ \rho^{-})v \bigr)).
\end{array}
\]
This quantity is not real if $s\neq 0$ and $\mu \neq 0$. Since 
$\varphi_{\omega}^{*}(x)\varphi_{\omega}(x)$ is a positive operator
affiliated to~$\cA_\kappa$  this shows that the KMS systems  $(\cF,
\cU, \tau, \omega_{\beta, \mu})$ and $(\cA, \cA_\kappa, \tau, \omega_{\beta, \mu})$ are  not
stochastically positive \qed .

\section{Scalar quantum fields at positive temperature with spatially
cutoff interactions}
\label{kgint}
\init
In this section we present the main results of this paper, namely the
construction of scalar quantum fields at positive temperature in one space dimension with spatially
cutoff interactions. For the real
scalar quantum field the two kinds of interactions that we will consider  
are the spatially cutoff $P(\phi)_{2}$ and  
$\e^{\alpha\phi}\,\!_{2}$ models (the later one is known as the H\o egh-Krohn
model). The first model is specified by the formal interaction 
$\int g(\x)P(\phi(x))\d\x$, where $P(\lambda)$ is a real 
polynomial, which is bounded from below. The second model is specified by $\int g(\x)\e^{\alpha \phi(\x)}\d\x$ for
$|\alpha|<\sqrt{2\pi}$. In both cases $g$ is a positive function in
$L^{1}(\rr)\cap L^{2}(\rr)$.

For the complex scalar field we will consider the spatially cutoff
$P(\varphi^*\varphi)_{2}$ interaction, specified by the formal
interaction term $\int g(\x)P(\varphi^{*}(\x)\varphi(\x))\d\x$.

\subsection{Some preparations}
\init\label{int}
In this subsection we prove some auxiliary results, which we will need to prove
some properties of the interaction terms later on.
We first recall a result of Klein and Landau \cite{KL1}.
\begin{lemma}
\label{int.1}
Let $(\cF, \cU, \tau, \omega)$ be a stochastically positive KMS system and let
$\cH_{1}$ be the closure of $\cU\Omega$. Let $\cU_{1}:=
\cU_{|\cH_{1}}$. Then $\Omega$ is a cyclic and separating vector for
$\cU_{1}$, and $\cU_{1}$ and $\cU$ are isomorphic as
$C^{*}$-algebras.
\end{lemma}
\begin{lemma}
\label{int.2}
Let $(\cF, \cU, \tau, \omega)$ be the stochastically positive KMS system
introduced in Section~\ref{stocpos}. 
Let $X_{\rho}$ be the vector space $X$ equipped with the scalar
product 
$(x,x)_{\rho}= (x,(1+2\rho)x)$ and set
\[
\matrix{
j\colon&X_{\rho}& \to & X\oplus\overline{X} \cr
&x & \mapsto & (1+\rho)^{\12}x\oplus \overline{\rho}^{\12}\overline{\kappa x}. \cr}
\]
Then 

\vskip .3cm
\halign{ \indent  \indent \indent #  \hfil & \vtop { \parindent =0pt \hsize=12cm
                            \strut # \strut} \cr 
{\rm (i)} & $\G(j)$ is an isometry from $\G(X_{\rho})$ into $\G(X\oplus
\overline{X})$ such that 
\[
\G(j)\e^{\i \phi(x)}= W_{\omega}(x)\G(j),\: x\in X_{\kappa};
\]
\cr
{\rm (ii)} & $\cH_{1}= \G(j)\G(X_{\rho})\equiv L^{2}(Q, \Sigma_0, \mu)$.
\cr}
\end{lemma}
\proof 
The map $x\to \overline{\kappa x}$ is $\cc$-linear from $X$
to $\Xbar$, hence $j$ is $\cc$-linear. From the results recalled in Subsection
\ref{sec4.3} and  the functional properties of $\G(j)$ we obtain that
$\G(j)\e^{\i \phi(x)}= W_{F}(jx)\G(j)$. Now $W_{F}(jx)= W_{\omega}(x)$
for $x\in X_{\kappa}$, and this proves {\rm (i)}.

Let us now prove {\rm (ii)}. The fact that $\cH_{1}$ is isomorphic to
$L^{2}(Q, \Sigma_0, \mu)$ follows from the definition of $\cU$ in
Subsection \ref{itit}. To prove the second equality, we
note that 
$\kappa$ extends to a conjugation on $X_{\rho}$, 
since $[\kappa, \rho]=0$. By a well-known
result on Fock spaces, which we already recalled in the proof of Lemma \ref{markov}, the vacuum vector
$\Omega\in \G(X_{\rho})$ is cyclic for $\{W(x) \mid
x\in X_{\rho}, \kappa x=x\}$.

Let now $u\in \G(X_{\rho})$. Because of the result recalled above  we find
\[
u=\lim_{n\to \infty}u_{n},\:
u_{n}=\sum_{1}^{N}\lambda_{j}W(x_{j})\Omega,\: x_{j}\in X_{\rho}, \;
\kappa x_{j}= x_{j}.
\]
It follows that
\[
\G(j)u=\lim_{n\to \infty}v_{n},\:
v_{n}=\sum_{1}^{N}\lambda_{j}W_{\omega}(x_{j})\Omega.
\]
Since $v_{n}\in \cU\Omega$ we have $\G(j)u\in \cH_{1}$ and hence $\G(j)\G(X_{\rho})\subset
\cH_{1}$. Let us now prove the converse inclusion: let $v\in \cH_{1}$
with
\[
v=\lim_{n\to \infty}v_{n},\:
v_{n}=\sum_{1}^{N}\lambda_{j}W_{\omega}(x_{j})\Omega,\: x_{j}\in
X, \:
\kappa x_{j}= x_{j}.
\]
Then
\[
v_{n}=\G(j)u_{n} \hbox { for } u_{n}=\sum_{1}^{N}\lambda_{j}W(x_{j})\Omega.
\]
Since $\G(j)$ is isometric, $u_{n} \to u\in \G(X_{\rho})$
and $v=\G(j)u$. This shows that $\cH_{1}\subset \G(j)\G(X_{\rho})$ 
\qed .
\subsection{Wick ordering}
\label{wickwick}
We recall some well known facts concerning the Wick ordering of
Gaussian random variables. Let $(Q, \Sigma_0, \mu)$ be a probability
space, $F$ a real vector space equipped with a positive quadratic form $f\mapsto
c(f,f)$, called a {\em covariance}. Let  $F\ni f\mapsto \phi(f)$ be a $\rr$-linear map from
$F$
to the space of real measurable functions on $Q$.
 
The {\em Wick ordering} $ \, :\phi(f)^{n}: \, $ with respect to the
covariance $c$ is defined using a generating series:
\beq
\label{wickdef}
:\!\e^{\alpha\phi(f)}\!:_{c} \: :
=\sum_{0}^{\infty}\frac{\alpha^{n}}{n!}:\!\phi(f)^{n}\!:_{c}=
\e^{\alpha\phi(f)}\e^{-\frac{\alpha^{2}}{2}c(f,f)}.
\eeq
Thus
\beq
\label{wick}
:\!\phi(f)^{n}\!:_{c}=
\sum_{m=0}^{[n/2]}\frac{n!}{m!(n-2m!)}\phi(f)^{n-2m}\Bigl(-\12 c(f,f) \Bigr)^{m}.
\eeq
If now $c_{1}$, $c_{2}$ are two covariances on $F$, then
\beq
\label{wickreordering2}
:\!\e^{\alpha \phi(f)}\!:_{c_{2}}= :\!\e^{\alpha\phi(f)}\!:
_{c_{1}}\e^{-\frac{\alpha^{2}}{2}(c_{2}-c_{1})(f,f)}. 
\eeq
This implies the following {\em Wick reordering identities} (see e.g.\ \cite{GJ}):
\beq
\label{wickreordering}
:\!\phi(f)^{n}\!:_{c_{2}}=
\sum_{m=0}^{[n/2]}\frac{n!}{m!(n-2m!)}:\!\phi(f)^{n-2m}\!:
_{c_{1}}\Bigl(-\12 (c_{2}-c_{1})(f,f)\Bigr)^{m}.
\eeq
\subsection{The spatially cutoff $P(\phi)_{2}$ interaction}
\label{int.sub1}

We recall from Section \ref{sec3}  that the real Klein-Gordon field in
one space dimension 
is described by the Weyl algebra $\fW(\ch)$, where $\ch=L^{2}(\rr, \d k)$.
Let $\chi\in \coinf(\rr)$ be a real cutoff function with
$\int_{\rr}\chi(\x)\d \x=1$.
For $\x\in \rr$ and $\Lambda\in [1, +\infty[$ an ultraviolet cutoff
parameter, we define~$f_{\Lambda, \x}\in
\ch$ by 
\[
f_{\Lambda, \x}(k):=\frac{1}{(4\pi)^{\12}}\e^{-\i
k.\x}\hat{\chi}\Bigl(\frac{k}{\Lambda}\Bigr)\epsilon(k)^{-\12}.
\]
We set 
\[
\phi_{\Lambda}(\x):= \sqrt{2}\phi_{\omega}(f_{\Lambda,\x})=
a_{\omega}^{*}(f_{\Lambda,\x})+ a_{\omega}(f_{\Lambda,\x}),\: \x\in \rr.
\]
Note that $f_{\Lambda, \x}\in \ch_{\kappa}$, so
$\phi_{\Lambda}(\x)$ is affiliated to $\cU$; i.e., $\phi_{\Lambda}(\x)$ can be considered as
a measurable function on $(Q, \Sigma_0, \mu)$.

In order to define the spatially cut-off $P(\phi)_{2}$ interaction 
we fix a real polynomial of 
degree~$2n$, which is bounded from below, namely
\beq
P(\lambda)= 
\sum_{j=0}^{2n}a_{j}\lambda^{j}  \hbox{ with }a_{2n}>0,
\label{polynome}
\eeq
and a real function $g\in L^{1}_{\rr}(\rr, \d x)\cap L^{2}(\rr, \d x)$ 
with $g\geq 0$.

We set
\[
V_{\Lambda}=\int g(\x):\!P(\phi_{\Lambda}(\x))\!:_{0}\d \x, 
\]
where $:\: :_{0}$ denotes the Wick ordering with respect to the
covariance at temperature $0$ given by $c_{0}(f,f)=\12(f,f)_{\ch}$. 

For technical reasons we will
also need to consider similar UV cutoff interactions with the Wick
ordering done with respect to the covariance at inverse temperature
$\beta$ given
by $c_{\beta}( f,f)=\12 (f,f)_{\rho}=\12 (f, (1+2\rho)f)$, $f \in \ch$. We set
\[
V_{\Lambda, \beta}=\int g(\x):\!P(\phi_{\Lambda}(\x))\!:_{\beta}\d \x,
\]
where $:\: :_{\beta}$ denotes  Wick ordering with respect to
$c_{\beta}$. Note that $V_{\Lambda}$ and $V_{\Lambda,\beta}$ are
affiliated to~$\cU$. 
We first collect some properties of these auxiliary
interactions.
\begin{lemma}
\label{int.3}
The family $\{ V_{\Lambda,\beta} \}$ is Cauchy in all spaces $L^{p}(Q,
\Sigma_0, \mu)$ for $1\leq p<\infty$
and converges when $\Lambda\to \infty$ to a function
$V_{\beta}\in L^{p}(Q, \Sigma_0, \mu)$, $1\leq p<\infty$,
which satisfies $\e^{-tV_{\beta}}\in L^{1}(Q, \Sigma_0, \mu)$ for all $t>0$.
We set
\[
V_{\beta}=:\int g(\x):\!P(\phi(\x))\!:_{\beta}\d \x.
\]
\end{lemma}
\proof We use the identification of $L^{2}(Q, \Sigma_0, \mu)$ with
$\G(\ch_{\rho})$ presented in Lemma \ref{int.2}. Then Wick ordering
with respect to $c_{\beta}$ coincides with Wick ordering with
respect to the Fock vacuum on $\G(\ch_{\rho})$. By exactly the same
arguments as those used in the $0$-temperature case (see e.g.\
\cite{SHK} or \cite[Sect. 6]{DG} for a recent survey) we obtain that, for $0\leq p\leq 2n$, 
the cuttoff interaction $V_{\Lambda, \beta}$ is a linear combination of Wick monomials of the form 
\[
\sum\limits_{r=0}^{p}\left(\begin{array}{c}p\\r\end{array}\right) \int
 w_{p,\Lambda}(k_{1},\ldots, k_{r}, k_{r+1}, \ldots, k_{p}) 
 a^{*}(k_{1})\cdots a^{*}(k_{r})
a(-k_{r+1})\cdots a(-k_{p})\d k_{1}\cdots \d k_{p}, 
\]
where 
\[
w_{p,\Lambda}(k_{1}, \cdots, k_{p})= 
\hat{g} \bigl(\sum_{1}^{p}k_{i} \bigr)\prod_{1}^{p}\hat{\chi}\Bigl(\frac{k_{i}}{\Lambda}\Bigr)\epsilon(k_{i})^{-\12}.
\]
Recalling that $1+2\rho= \frac{1+\e^{-\beta\epsilon}}{1-\e^{-\beta
\epsilon}}$ we see that \[
w_{p, \Lambda}\in
\otimes^{p}\ch_{\rho}=L^{2} \Bigl(\rr^{p},
\prod_{1}^{p}\frac{1+\e^{-\beta\epsilon(k_{i})}}{1-\e^{-\beta
\epsilon(k_{i})}}\d k_{1}\dots, \d k_{p} \Bigr).
\]
The sequence $\{ w_{p, \Lambda} \}$ is Cauchy in this space. Consequently $w_{p, \Lambda}\to w_{p,
\infty}$ when $\Lambda\to \infty$, where
\[
w_{p,\infty}(k_{1}, \cdots, k_{p})= 
\hat{g} \bigl(\sum_{1}^{p}k_{i} \bigr)\prod_{1}^{p}\epsilon(k_{i})^{-\12}.
\] 
We can now apply these Wick monomials to the Fock vacuum and conclude that $V_{\Lambda,
\beta}\Omega$ converges to a vector $V_{\beta}\Omega$ in $\G(\ch_{\rho})$,
or equivalently that $V_{\Lambda, \beta}$ converges to $V_{\beta}$ in 
$L^{2}( Q, \Sigma_0, \mu)$. Since $V_{\lambda,\beta}\Omega$ is a finite
particle vector, it follows from a standard argument (see e.g.\
\cite[Thm.\ 1.22]{Si1} or \cite[Lemma 5.12]{DG}) that $V_{\Lambda,
\beta}\to V_{\beta}\in L^{p}(Q, \Sigma_0, \mu)$ for all
$1\leq p<\infty$.

We will now prove that $\e^{-tV_{\beta}}\in L^{1}(Q, \Sigma_0, \mu)$. We argue as
in the $0$-temperature case: we first verify that $\|w_{p,
\Lambda}-w_{p, \infty}\|\leq C\Lambda^{-\epsilon_{0}}$ for some
$\epsilon_{0}>0$ and therefore $\|V_{\Lambda,
\beta}-V_{\beta}\|_{L^{2}(Q, \Sigma_0, \mu)}\leq C\Lambda^{-\epsilon_{0}}$.
Applying again \cite[Lemma 5.12]{DG} we find 
\beq\label{int.e0}
\|V_{\Lambda,
\beta}-V_{\beta}\|_{L^{p}(Q, \Sigma_0, \mu)}\leq
C(p-1)^{n}\Lambda^{-\epsilon_{0}},\: p>1.
\eeq Using the Wick
ordering identities (\ref{wick})  we obtain as identities between
functions on~$K$ (see, e.g., \cite[Lemma 6.6]{DG}):
\[
: P(\phi_{\Lambda}(\x)) :_{\beta}\geq
-C \bigl(\|\phi_{\Lambda}(\x)\Omega\|^{2n} +1\bigr).
\]
Now $\|\phi_{\Lambda}(\x)\Omega\|=
C\|\epsilon^{-1}\hat{\chi}\Bigl(\frac{\cdot}{\Lambda} \Bigr)\|_{\ch_{\rho}}\leq
C(\ln (\Lambda))^{\12}$. This yields 
\beq
\label{int.e01}V_{\Lambda, \beta}\geq -C\ln
(\Lambda)^{n}.
\eeq Applying  now \cite[Lemma V.5]{Si1} we deduce from
(\ref{int.e0}) and (\ref{int.e01}) that $\e^{-tV_{\beta}}\in L^{1}(Q, \Sigma_0, \mu)$ for all $t>0$  \qed .

\begin{proposition}
\label{int.4}
The family $\{ V_{\Lambda} \}$ is Cauchy in all spaces $L^{p}(Q,
\Sigma_0, \mu)$ for $1\leq p<\infty$ 
and converges when $\Lambda\to \infty$ to a function
$V\in L^{p}(Q, \Sigma_0, \mu)$, $1\leq p<\infty$, which satisfies
$\e^{-tV}\in L^{1}(Q, \Sigma_0, \mu)$ for all $t>0$.
We set
\[
V=:\int g(\x):\!P(\phi(\x))\!:_{0}\d \x.
\]
\end{proposition}
\proof 
With the help of the Wick reordering identity (\ref{wickreordering}) we find, for
$f\in \ch_{\kappa}$,
\[
\begin{array}{rl}
:\!P(\phi_{\omega}(f))\!:_{0} & = \sum_{j=0}^{2n}a_{j}
:\!\phi_{\omega}(f)^{n}\!:_{0}\\
&=\sum_{j=0}^{2n}\sum_{m=0}^{[j/2]}a_{j}\frac{j!}{m!(j-2m!)}:
\!\phi(f)^{j-2m}\!:
_{\beta}\bigl(-\12 (c_{0}-c_{\beta})(f,f) \bigr)^{m}.
\end{array}
\]
For $f= f_{\Lambda, x}$ 
\[
\begin{array}{rl}
r_{\Lambda}:=&(c_{\beta}-c_{0})(f_{\Lambda, x},f_{\Lambda, x})= ( f_{\Lambda,
0}, \rho f_{\Lambda, 0})\\
=&\int\e^{-\beta\epsilon(k)}\hat{\chi}\bigl(\frac{k}{\Lambda} \bigr)\d k=
r_{\infty}+ O(\Lambda^{-\infty}),
\end{array}
\]
where $r_{\infty}=\int\e^{-\beta\epsilon(k)}\d k$.

On the other hand,
\[
\int_{Q}|\phi_{\omega}(f_{\Lambda,\x})|^{p}\d\mu\in O \bigl(| c_{\beta}(f_{\Lambda,
\x}, f_{\Lambda,\x})|^{p})\in O(\ln(\Lambda)^{p} \bigr).
\]
Therefore 
\[
:\!P(\phi_{\Lambda}(\x))\!:_{0}= :\!\tilde{P}(\phi_{\Lambda}(\x))\!:_{\beta} + 
O \bigl(\ln(\Lambda)^{2n}\Lambda^{-\infty} \bigr) \hbox{ uniformly for
}\x\in \supp g, 
\]
where
\[
 \tilde{P}(\lambda)=\sum_{j=0}^{2n}\sum_{m=0}^{[j/2]}
a_{j}\frac{j!}{m!(j-2m!)} \lambda^{j-2m} \bigl(\12 r_{\infty} \bigr)^{m}.
\]
We see that $\tilde{P}(\lambda)-P(\lambda)$ is of degree less than
$2n-1$. 
Applying Lemma \ref{int.3} to $\tilde{P}$ this yields
\[
\lim_{\Lambda\to \infty}\int g(\x):\!P(\phi_{\Lambda}(\x))\!:_{0}\d\x=
\lim_{\Lambda\to \infty}\int g(\x):\!\tilde{P}(\phi_{\Lambda}(\x))\!:
_{\beta}\d\x= \int g(\x):\!\tilde{P}(\phi(\x))\!:
_{\beta}\d\x,
\]
 which completes the proof of the proposition \qed .

\subsection{The spatially cutoff $\e^{\alpha \phi}\,\!_{2}$ interaction}
\label{int.sub11}
As in Subsection \ref{int.sub1} we set, for $|\alpha|<\sqrt{2\pi}$,
\[
V_{\Lambda}=\int g(\x):\!\e^{\alpha\phi_{\Lambda}(\x)}\!:_{0}\d \x 
\]
and
\[
V_{\Lambda, \beta}=\int g(\x):\!\e^{\alpha\phi_{\Lambda}(\x)}\!:
_{\beta}\d \x.
\]
Note that, as above, $V_{\Lambda}$ and $V_{\Lambda, \beta}$ are
affiliated to $\cU$.
\begin{lemma}
\label{int.5}
For $|\alpha|<\sqrt{2\pi}$ the family $\{ V_{\Lambda, \beta} \}$ is Cauchy in $L^{2}(Q, \Sigma_0, \mu)$ and
converges when $\Lambda\to \infty$ to a positive function $V_{\beta}\in L^{2}(Q, \Sigma_0, \mu)$. We set 
\[
V_{\beta}=:\int g(x):\!\e^{\alpha \phi(\x)}\!:_{\beta}\d\x.
\]
\end{lemma}
\proof The proof is completely similar to the
$0$-temperature case where $\rho=0$ (see e.g.\ \cite{Si1}, \cite{HK}). 
 For
completeness we will give an outline. 
Note first that by (\ref{wickdef}) $:
\!\e^{\alpha\phi_{\Lambda}(x)}\!:_\beta$ is a positive function on $Q$,
hence the same holds for $V_{\Lambda,\beta}$ as $g\geq 0$. We now
show that $V_{\Lambda, \beta}$ converges in $L^{2}(Q, \Sigma_0, \mu)$, and we
will identify $V_{\Lambda, \beta}$ with $V_{\Lambda,\beta}\Omega$.
We
have 
\[
\one_{\{n\}}(N)V_{\Lambda, \beta}= \frac{\alpha^{n}}{n!}\int g(\x):\!
\phi_{\Lambda}^{n}(\x)\!:\Omega \d\x=
\frac{\alpha^{n}}{(4\pi)^{n/2}\sqrt{n!}}\hat{g}(\sum_{1}^{n}k_{i})\prod_{1}^{n}\hat{\chi}\Bigl(\frac{k_{i}}{\Lambda}\Bigr)\frac{1}{\epsilon(k_{i})^{\12}}. 
\]
Hence
\[
\begin{array}{rl}
\|\one_{\{n\}}(N)V_{\Lambda,\beta}\|^{2}& =
\frac{1}{n!}\bigl(\frac{\alpha^{2}}{4\pi}\bigr)^{n}\int
|\hat{g}(\sum_{1}^{n}k_{i})|^{2}\prod_{1}^{n}\bigl|\hat{\chi}
\bigl(\frac{k_{i}}{\Lambda}\bigr)\bigr|^{2}\frac{1+2\rho(k_{i})}{\epsilon(k_{i})}\d
k_{1}\dots\d k_{n}\\
& \leq \frac{1}{n!}\bigl(\frac{\alpha^{2}}{4\pi}\bigr)^{n}\int
|\hat{g}(\sum_{1}^{n}k_{i})|^{2}\prod_{1}^{n}\frac{1+2\rho(k_{i})}{\epsilon(k_{i})}\d
k_{1}\dots\d k_{n}=: \epsilon_{n}. 
\end{array}
\]
Next we find 
\[
\epsilon_{n}= \frac{1}{n!} \Bigl(\frac{\alpha^{2}}{2\pi}\Bigr)^{n}\int g(\x)g({\rm y})K_{\beta}(\x-{\rm
y})^{n}\d\x\d{\rm y} 
\]
for
\[
 K_{\beta}(\x)=\12\int\e^{\i k\x}\frac{1+2\rho(k)}{\epsilon(k)}\d k.
\]
We claim now that
\beq
\label{int.e1}
\e^{\frac{\alpha^{2}}{2\pi}|K_{\beta}(\x)|}\in L^{1}(\rr)+L^{\infty}(\rr) 
\hbox{ for }|\alpha|<\sqrt{2\pi}.
\eeq
This implies that
\beq
\label{int.e2}
\sum_{n=0}^{\infty}\epsilon_{n}\leq \int g(\x)g({\rm
y})\e^{\frac{\alpha^{2}}{2\pi}|K_{\beta}|(\x-{\rm
y})}\d\x\d{\rm y}<\infty.
\eeq
If we set
\[
 K_{0}(\x)=\12\int\e^{\i k\x}\frac{1}{\epsilon(k)}\d k,
\]
then because of the rapid decay of $\rho(k)$ when $|k|\to \infty$, we
have $K_{0}-K_{\beta}\in L^{\infty}(\rr)$, and (see \cite[equ.
(2.4)]{HK}) $K_{0}(\x)\in O(1)$ in $|\x|\geq 1$, $K_{0}(\x)= -\ln(\x)+
O(1)$ in $|\x|\leq 1$. This implies (\ref{int.e1}).

Now by the arguments in the proof of Lemma \ref{int.3}, we see that 
\[
\lim_{\Lambda \to \infty}\one_{\{n\}}(N)V_{\Lambda,
\beta}=\frac{\alpha^{n}}{n!}\int g(\x):\!\phi(\x)^{n}\!:\Omega\d\x.
\]
Since $\one_{\{n\}}(N)V_{\Lambda, \beta}\to V_{n}$ in $L^{2}(Q, \Sigma_0, \mu)$
for each $n$ and $\sup_{\Lambda}\|\one_{\{n\}}(N)V_{\Lambda,
\beta}\|^{2}\leq \epsilon_{n}$ with $\sum \epsilon_{n}<\infty$, we see
that $V_{\Lambda, \beta}$ converges to some element $V\in L^{2}(Q, \Sigma_0, \mu)$, 
which is a.e.\ positive as a limit of positive functions  \qed .

\begin{proposition}
\label{int.6}
For $|\alpha|<\sqrt{2\pi}$, the family $\{ V_{\Lambda} \}$ is Cauchy in $L^{2}(Q, \Sigma_0, \mu)$ and
converges to a positive function $V\in L^{2}(Q, \Sigma_0, \mu)$. We set 
\[
V=:\int g(x):\!\e^{\alpha \phi(\x)}\!:_{0}\d\x.
\]
\end{proposition}
\proof By the Wick reordering identity (\ref{wickreordering2}) we
have 
\[
:\!\e^{\alpha \phi_{\Lambda, x}}\!:_{0}= :\!\e^{\alpha \phi_{\Lambda,
x}}\!:_{\beta}\e^{\frac{\alpha^{2}}{2}r_{\Lambda}},
\] 
Hence $V_{\Lambda}=\e^{\frac{\alpha^{2}}{2}r_{\Lambda}} V_{\Lambda,
\beta}$, which implies, using Lemma
\ref{int.5}, that $V_{\Lambda}$ converges in~$L^{2}(Q, \Sigma_0, \mu)$ to the
positive function $\e^{\frac{\alpha^{2}}{2}r_{\infty}}V_{\beta}$  \qed .

\subsection{The spatially cutoff $P(\varphi^*\varphi)_2$ interaction}
\label{chargedp}
We consider now the complex Klein-Gordon field in one space dimension which is described by
the Weyl algebra $\fW(X)$ for $X=\ch\oplus\ch$, $\ch=L^{2}(\rr, \d k)$. 
We recall that the Gibbs state at inverse temperature~$\beta$ is
given by $\omega(W(x))=\e^{\frac{1}{4}(x, (1+2\rho x))}$, where
$\rho=(\e^{\beta {\rm h}}-1)^{-1}$ and ${\rm
h}=\epsilon\oplus \epsilon$.

We set
\[
\varphi_{\Lambda}(\x)=\varphi_{\omega}(f_{\Lambda,\x}\oplus
f_{\Lambda, \x}),\:\varphi^{*}_{\Lambda}(\x)=\varphi^{*}_{\omega}(f_{\Lambda,\x}\oplus
f_{\Lambda, \x}),\:\x\in \rr.
\]
Note that $f_{\Lambda, \x}$ is invariant under the conjugation $h\to
\overline{h}$. This implies that $\varphi_{\Lambda}(x)$ is affiliated
to~$\cU$, since~$f_{\Lambda,\x}\oplus
f_{\Lambda, \x}\in X_{\kappa}$. Moreover, $\varphi_{\Lambda}^{*}(\x)\varphi_{\Lambda}(\x)= \12 \bigl(\phi^{2}_{\omega}(f_{\Lambda,\x}\oplus
f_{\Lambda, \x} )+\phi^{2}_{\omega}(\i f_{\Lambda,\x}\oplus
-\i f_{\Lambda, \x}) \bigr)$. 

For $P$ a real polynomial of degree $2n$, which is bounded from below,  and $g$ a
positive function in~$L^{1}(\rr)\cap L^{2}(\rr)$, we set  
\[
V_{\Lambda}=\int g(\x):\!P(\varphi_{\Lambda}^{*}(\x)\varphi_{\Lambda}(\x))\!:_{0}\d \x, 
\]
where $:\: :_{0}$ denotes Wick ordering  with respect to the $0$-temperature
covariance $c_{0}(x,x)=\12(x,x)$, and 
\[
V_{\Lambda, \beta}=\int g(\x):\!P(\varphi_{\Lambda}^{*}(\x)\varphi_{\Lambda}(\x))\!:_{\beta}\d \x,
\]
where $:\: :_{\beta}$ denotes Wick ordering with respect to the
covariance at inverse temperature~$\beta$ specified by $c_{\beta}(x,x)=\12(x, (1+2\rho)x)$. The
following two results can be shown by exactly the same methods as in
Subsection \ref{int.sub1}.
\begin{lemma}
\label{int.7}
The family $\{ V_{\Lambda,\beta} \}$ is Cauchy in all $L^{p}(Q, \Sigma_0, \mu)$ spaces 
and converges, when $\Lambda\to \infty$, to a function
$V_{\beta}\in L^{p}(Q, \Sigma_0, \mu)$, $1\leq p<\infty$,
which satisfies $\e^{-tV_{\beta}}\in L^{1}(Q, \Sigma_0, \mu)$ for all $t>0$.
We set

\[
V_{\beta}=:\int g(\x):\!P(\varphi^{*}(\x)\varphi(\x))\!:_{\beta}\d \x.
\]
\end{lemma}
\begin{proposition}
\label{int.8}
The family $\{V_{\Lambda} \}$ is Cauchy in all spaces $L^{p}(Q, \Sigma_0, \mu)$ and converges, when $\Lambda\to \infty$,
 to a function
$V\in L^{p}(Q, \Sigma_0, \mu)$, $1\leq p<\infty$, which satisfies
$\e^{-tV}\in L^{1}(Q, \Sigma_0, \mu)$ for all $t>0$.
We set
\[
V=:\int g(\x):\!P(\varphi^{*}(\x)\varphi(\x))\!:_{0}\d \x.
\]
\end{proposition}
\subsection{Scalar quantum fields at positive temperature with
spatially cutoff interactions}
\label{mainres}
To construct the space-cutoff $P(\phi)_{2}$ and $\e^{\alpha\phi}\,\!_{2}$
models at positive temperature, we apply the general results of
Subsection \ref{otot}. Note that by Subsections \ref{int.sub1} and
\ref{int.sub11}, the interactions terms $V=\int g(\x):\!P(\phi(\x))\!:
_{0}\d \x$ and $V= \int g(\x):\!\e^{\alpha \phi(\x)}\!:_{0}\d \x$ for
$|\alpha|<\sqrt{2\pi}$ satisfy all the hypotheses of Subsection
\ref{otot}. Consequently we obtain the following theorem:
\begin{theoreme}
Let $\bigl(\cW, \cW_\kappa, \tau^\circ,\omega \bigr)$ be the
quasi-free
$\beta$-KMS system describing the free neutral Klein-Gordon field in
one space dimension at temperature $\beta^{-1}$, described in
Subsection \ref{poskg}. Let $\cH, L, \Omega$ be the
associated GNS objects described in Subsection \ref{sec4.3}. 
Let $V$ be the selfadjoint operator on $\cH$
affiliated to $\cW_{\kappa}$ equal either to $\int g(\x):
\!P(\phi(\x))\!:_{0}\d \x$ or to  $\int g(\x):\!\e^{\alpha
\phi(\x)}\!:_{0}\d \x$. Then the following statements hold true:

\vskip .3cm
\halign{ \indent \indent \indent #  \hfil & \vtop { \parindent = 0pt \hsize=12cm
                            \strut # \strut} 
\cr 
(i)     & 
$L+V$ is essentially selfadjoint and $\Omega\in
\cD(\e^{-\frac{\beta}{2}H_{V}})$, where $H_{V}:= \overline{L+V}$.
\cr 
(ii)     & 
Let $\tau_{V}(t)$  be the $W^{*}$-dynamics generated by $H_{V}$
and $\omega_{V}$ be the vector state induced 
by~$\Omega_{V}=\|\e^{-\frac{\beta}{2}H_{V}}\Omega\|^{-1}\e^{-\frac{\beta}{2}H_{V}}\Omega$.
Then $\tau_{V}$ is a group of $^{*}$-automorphisms of $\cW$,
 continuous for the strong operator topology such that
$\bigl(\cW, \cW_\kappa, \tau_{V},\omega_{V} \bigr)$ is a
stochastically positive $\beta$-KMS system.
\cr 
(iii)     & 
The generalized path space associated to $\bigl(\cW, \cW_\kappa,
\tau_{V},\omega_{V} \bigr)$ satisfies the Markov property.
\cr 
(iv)     & 
Let $L_{V}$, $J_{V}$ be the perturbed Liouvillean and modular
conjugation associated to $\bigl(\cW, \cW_\kappa, \tau_{V},\omega_{V}
\bigr)$. Then $J_{V}= J$ and $L_{V}= \overline{H_{V}-JVJ}$. \cr}
\end{theoreme}
Finally we state the corresponding result for the charged Klein-Gordon
field:
\begin{theoreme}
Let $\bigl(\cW, \cW_\kappa, \tau^\circ,\omega \bigr)$ be the
quasi-free
$\beta$-KMS system describing the free charged  Klein-Gordon field in
one space dimension at temperature $\beta^{-1}$ and zero chemical
potential, described in Subsection \ref{poskg}. Let $\cH, L, \Omega$ be the
associated GNS objects described in Subsection \ref{sec4.3}. 
Let $V$ be the selfadjoint operator on $\cH$
affiliated to $\cW_{\kappa}$ equal  to \\ $\int g(\x):
\!P(\overline{\varphi}(\x)\varphi(\x))\!:_{0}\d \x$. Then the following statements hold true:

\vskip .3cm
\halign{ \indent \indent \indent #  \hfil & \vtop { \parindent = 0pt \hsize=12cm
                            \strut # \strut} 
\cr 
(i)     & 
$L+V$ is essentially selfadjoint and $\Omega\in
\cD(\e^{-\frac{\beta}{2}H_{V}})$, where $H_{V}:= \overline{L+V}$.
\cr 
(ii)     & 
Let $\tau_{V}(t)$  be the $W^{*}$-dynamics generated by $H_{V}$
and $\omega_{V}$ be the vector state induced  
by~$\Omega_{V}=\|\e^{-\frac{\beta}{2}H_{V}}\Omega\|^{-1}\e^{-\frac{\beta}{2}H_{V}}\Omega$.
Then $\tau_{V}$ is a group of $^{*}$-automorphisms of $\cW$,
 continuous for the strong operator topology such that
$\bigl(\cW, \cW_\kappa, \tau_{V},\omega_{V} \bigr)$ is a
stochastically positive $\beta$-KMS system.
\cr 
(iii)     & 
The generalized path space associated to $\bigl(\cW, \cW_\kappa,
\tau_{V},\omega_{V} \bigr)$ satisfies the Markov property.
\cr 
(iv)     & 
Let $L_{V}$, $J_{V}$ be the perturbed Liouvillean and modular
conjugation associated to $\bigl(\cW, \cW_\kappa, \tau_{V},\omega_{V}
\bigr)$. Then $J_{V}= J$ and $L_{V}= \overline{H_{V}-JVJ}$. \cr}
\end{theoreme}

\end{document}